\documentclass[12pt,preprint]{aastex}
\usepackage{graphicx}
\let\Newcommand=\newcommand
\def\newcommand{\providecommand}
\usepackage[pdftitle={Galaxy Rotation Curves Without Non-Baryonic Dark Matter}, pdfauthor={J. R. Brownstein and J. W. Moffat},
pdfsubject={\tt astrophysics, Galaxies, Gravitation, Quantum Gravity, RG Flow}, pdfkeywords={rotation curves, galaxies, quantum gravity, mond, dark matter}, bookmarks, bookmarksopen, 
colorlinks=true, linkcolor=blue, citecolor=blue, anchorcolor=blue, urlcolor=blue]{hyperref}
\let\newcommand=\Newcommand

\newcommand{\etal}{{\em et al.}\ }

\shorttitle{Galaxy Rotation Curves Without Non-Bar. D.M.}
\shortauthors{Brownstein and Moffat}

\begin{document}


\title{Galaxy Rotation Curves Without Non-Baryonic Dark Matter}


\author{ J. R. Brownstein\altaffilmark{1} and J. W. Moffat\altaffilmark{2}}
\affil{The Perimeter Institute for Theoretical Physics, Waterloo, Ontario, N2J 2W9, Canada}
\and
\affil{Department of Physics, University of Waterloo, Waterloo, Ontario N2Y 2L5, Canada}
\altaffiltext{1}{\href{mailto:jbrownstein@perimeterinstitute.ca}{\tt jbrownstein@perimeterinstitute.ca}}
\altaffiltext{2}{\href{mailto:john.moffat@utoronto.ca}{\tt john.moffat@utoronto.ca}}



\begin{abstract}
We apply the modified acceleration law obtained from Einstein gravity coupled to a massive skew
symmetric field $F_{\mu\nu\lambda}$ to the problem of explaining galaxy rotation curves without exotic dark matter.  Our
sample of galaxies includes low surface brightness (LSB) and high surface brightness (HSB) galaxies, and an elliptical
galaxy. In those cases where photometric data are available, a best fit via the single parameter $(M/L)_{\rm{stars}}$ to
the luminosity of the gaseous (HI plus He) and luminous stellar disks is obtained.  Additionally, a best fit to the
rotation curves of galaxies is obtained in terms of a parametric mass distribution (independent of luminosity
observations) -- a two
parameter fit to the total galactic mass, (or mass-to-light ratio $M/L$), and a core radius associated with a model of
the galaxy cores using a nonlinear least-squares fitting routine including estimated errors. The fits are compared to
those obtained using Milgrom's phenomenological MOND model and to the predictions of the Newtonian-Kepler acceleration
law.
\end{abstract}



\keywords{dark matter --- galaxies: general --- galaxies: kinematics and dynamics --- galaxies: fundamental parameters}


\section{Introduction}

In spite of intensive searches for the components of non-baryonic
dark matter, no candidate particles have been observed. This leads
one to search for a gravity theory that can explain galaxy
dynamics without the need for exotic dark matter. Such a possible
model was introduced in~\citet{mil83}, and for more than twenty
years since its publication, attempts to form a consistent
relativistic theory containing Milgrom's phenomenological model
have been pursued~\citep{bek04,san05}. A generalization of
Einstein's general relativity (GR) based on a pseudo-Riemannian
metric tensor and a skew symmetric rank three tensor field
$F_{\mu\nu\lambda}$, called metric-skew-tensor-gravity (MSTG),
leads to a modified acceleration law that can explain the flat
rotation curves of galaxies and cluster lensing without
postulating exotic dark matter~\citep{mof05a}.  A relativistic
scalar-tensor-vector gravity (STVG) theory has also been developed
which yields the same modified acceleration law as MSTG, and
provides an effective description of the running of the $G$,
$\gamma_{c}$ and $\mu$ with space and time~\citep{mof05b}.

A cosmological model obtained from the field equations and a
running of the effective gravitational coupling constant $G$ can
also explain the growth of large scale structure formation without
invoking cold dark matter. The running of the cosmological
constant would produce a quintessence-like dark energy that could
account for the acceleration of the expansion of the
universe~\citep{per97,rie98,rie04,gar98,spe03,ben03}.

A renormalization group (RG) framework~\citep{reu04a,reu04b} for MSTG was developed to describe the running of the
effective gravitational coupling constant $G$, and the effective coupling constant $\gamma_{c}$ that measures the strength
of the coupling of the $F_{\mu\nu\lambda}$ field to matter. A momentum cutoff identification $k=k(x)$ associates the RG
scales to points in spacetime. For the static, spherically symmetric solution, the RG flow equations allow a running
with momentum $k$ and proper length $\ell(r)=1/k$ and provides for a spatial variation of the effective Newton's
coupling constant, $G=G(r)$; together with a variation of the skew field coupling constant, $\gamma_{c}(r)$, to matter;
and the effective mass of the skew field $\mu=\mu(r)=1/r_{0}$ where $r$ denotes the
radial coordinate. Such a ``running'' of coupling constants and mass are well-known from particle physics (e.g.\,quantum
chromodynamics) and condensed matter and are applied here to the behavior of the infrared fixed point of the
renormalization group method of quantum gravity. The form of $G(r)$ as a function of $r$, obtained from the modified
Newtonian acceleration law, leads to agreement with solar system observations, terrestrial gravitational experiments and
the binary pulsar PSR 1913+16 observations, while strong renormalization effects in the infrared regime at large
distances lead to fits to galaxy rotation curves.

In this paper, a fit to both low surface brightness (LSB) and high surface brightness (HSB) galaxies over the range from
dwarf galaxies to giant galaxies is achieved in terms of the total galaxy mass $M$ (or $M/L$) without exotic dark
matter. A satisfactory fit is achieved to the rotational velocity data generic to the elliptical galaxy NGC 3379.  A
model of the modified acceleration law that includes a description of radial velocity curves in the core of galaxies as
well as in the outer regions of the galaxy is shown to yield good fits to rotational velocity data.  The significance of
this result is made explicit by the depth of the galaxy rotation curve study, here.  We performed photometric fits to 58
HSB and LSB and Dwarf galaxies utilizing a single parameter -- the $M/L$ ratio of the stellar disk.  29 of these
galaxies were based upon $B-$band and the remaining 29 galaxies were based upon $K-$band.  The $K-$band data is the more
modern data set and is a more precise tracer of the mean radial distribution of the dominant stellar population -- these
were the photometric results of Ursa Major.  Since 2001, the Ursa Major data has been revised since the estimated
distance to the cluster is now taken as 18.6 Mpc as opposed to the original publication's use of 15.5 Mpc.  We updated
the original analysis of \citet{san98} using that group's Groningen Image Processing System
(GIPSY)\footnote{\url{http://www.astro.rug.nl/~gipsy/}}.

In addition, we performed 101 galaxy rotation curve fits to HSB and LSB and Dwarf galaxies (including the 58 galaxies
that were fit photometrically) using a parametric mass distribution.  These fits were necessary for those galaxies for
which photometric data of the HI gas (corrected for He component)  and stellar disk was unavailable.  Although the model
contains two parameters as opposed to one parameter for the photometric fits, it makes no assumptions on the constancy
of the $M/L$ ratio within a galaxy although it does not distinguish between gas or stellar disk.

A comparison of the fits to the rotational velocity curve data obtained from MSTG and Milgrom's MOND reveals that the
results from both models are qualitatively similar for almost all of the galaxy data fitted, although predictions of the
galactic masses differ.

\section{Modified Acceleration Law}

From the derivation of the radial acceleration experienced by a test particle in a static, spherically symmetric
gravitational field due to a point source, we obtain~\citep{mof05a,mof05b}:
\begin{equation}
\label{accelerationlaw} a(r)=-\frac{G_{\infty}M}{r^2}+\sigma\frac{\exp(-r/r_0)}{r^2}
\biggl(1+\frac{r}{r_0}\biggr).
\end{equation}
Here, $G_{\infty}$ is defined to be the {\it effective} gravitational constant at infinity
\begin{equation}
\label{renormG} G_{\infty}=G_0\biggl(1+\sqrt{\frac{M_0}{M}}\biggr),
\end{equation}
where $G_0$ is Newton's ``bare'' gravitational constant. This conforms with our definition of $G$ in the RG flow
formalism, which requires that the effective $G$ be renormalized in order to guarantee that equation
(\ref{accelerationlaw}) reduces to the Newtonian acceleration
\begin{equation}
\label{Newtonianacceleration} a_{\rm Newton}=-\frac{G_0M}{r^2}
\end{equation}
at small distances $r\ll r_0$. We model the coupling constant $\sigma$ for 
the repulsive Yukawa force contribution in equation (\ref{accelerationlaw}) by 
\begin{equation}
\sigma=G_{0}\sqrt{M_{0} M},
\end{equation}
 where $M_{0}$ is a coupling parameter.   We obtain the acceleration on a point particle
\begin{equation}
\label{accelerationlaw2} a(r)=-\frac{G_{\infty}M}{r^2}+G_0\sqrt{MM_0}\frac{\exp(-r/r_0)}{r^2}
\biggl(1+\frac{r}{r_0}\biggr).
\end{equation}
By using equation (\ref{renormG}), we can express the modified acceleration in the form
\begin{equation}
\label{accelerationlaw3} a(r)=-\frac{G_0M}{r^2}
\biggl\{1+\sqrt\frac{M_0}{M}\biggl[1-\exp(-r/r_0)\biggl(1+\frac{r}{r_0}\biggr)
\biggr]\biggr\}.
\end{equation}

We can rewrite equation (\ref{accelerationlaw3}) as
\begin{equation}
\label{runG} a(r)=-\frac{G(r)M}{r^2},
\end{equation}
where
\begin{equation}
\label{runningG} G(r)=G_0\biggl\{1+\sqrt\frac{M_0}{M}\biggl[1-\exp(-r/r_0)\biggl(1+\frac{r}{r_0}\biggr)
\biggr]\biggr\}.
\end{equation}
Thus, $G(r)$ describes the running with distance of the effective gravitational constant in the RG flow scenario.

The gravitational potential for a point source obtained from the modified acceleration formula (\ref{accelerationlaw3})
is given by
\begin{equation}
\label{Phi}
\Phi(r)=\frac{G_0 M}{r} \biggl[ 1+\sqrt{\frac{M_0}{M}} \bigl( 1-\exp(-r/r_0) \bigr) \biggr].
\end{equation}

We apply equation (\ref{accelerationlaw3}) to explain the flatness of rotation curves of galaxies, as well as the
approximate Tully-Fisher law~\citep{tul77}, $v^4\propto G_0M\propto L$, where $v$ is the rotational velocity of a
galaxy, $M$ is the galaxy mass
\begin{equation}
\label{Mass}  M=M_{\rm stars}+M_{HI}+M_{DB}+M_f,
\end{equation}
and $L$ is the galaxy luminosity. Here, $M_{\rm stars},M_{HI}$, $M_{DB}$ and $M _f$ denote the visible mass, the mass of
neutral hydrogen, possible dark baryon mass and gas, and the mass from the skew field energy density, respectively.

The rotational velocity of a star $v$ obtained from $v^2(r)/r=a(r)$ is given by
\begin{equation}
\label{rotvelocity} v(r)=\sqrt{\frac{G_0M}{r}}\biggl\{1+\sqrt{\frac{M_0}{M}}\biggl[1-\exp(-r/r_0)
\biggl(1+\frac{r}{r_0}\biggr)\biggr]\biggr\}^{1/2}.
\end{equation}

Let us postulate that the parameters $M_0$ and $r_0$ give the magnitude of the constant acceleration
\begin{equation}
\label{specialacceleration} a_0=\frac{G_0M_0}{r^2_0}.
\end{equation}
We assume that for galaxies and clusters of galaxies this acceleration is determined by
\begin{equation}
\label{Hubbleacceleration} a_0=cH_0.
\end{equation}
Here, $H_0$ is the current measured Hubble constant $H_0=100\, h\, {\rm km}\, s^{-1}\,{\rm Mpc}^{-1}$ where $h=(0.71 \pm
0.07)$~\citep{eid04}\footnote{\url{http://pdg.lbl.gov}}. This gives
\begin{equation}
\label{speciala} a_0=6.90\times 10^{-8}\,{\rm cm}\, s^{-2}.
\end{equation}
We note that $a_0=cH_0\sim (\sqrt{\Lambda/3})c^2$, so there is an interesting connection between the parameters $M_0$,
$r_0$ and the cosmological constant $\Lambda$.

Let us now describe a model of a spherically symmetric galaxy such that the density of visible matter, $\rho(r)$,
contains an inner core at radius $r = r_c$. The acceleration law of equation (\ref{accelerationlaw3}) takes the form
\begin{equation}
\label{coreacceleration} a(r)=-\frac{G_0{\cal
M}(r)}{r^2}\biggl\{1+\sqrt{\frac{M_0}{M}}\biggl[1-\exp(-r/r_0)\biggl(1+\frac{r}{r_0}
\biggr)\biggr]\biggr\},
\end{equation}
where
\begin{equation}
{\cal M}(r)=4\pi \int_0^rdr'r'^2\rho(r')
\end{equation}
is the total amount of ordinary matter within a sphere of radius $r$. A simple model for ${\cal M}(r)$ is
\begin{equation}
\label{coremodel} {\cal M}(r)=M\biggl(\frac{r}{r_c+r}\biggr)^{3\beta},
\end{equation}
where
\begin{equation}
\label{surfaceBrightnessBeta}
\beta = \left\{
\begin{array}{ll} 1 & \mbox{for HSB galaxies,} \\
2 & \mbox{for LSB \& Dwarf galaxies.}
\end{array} \right.
\end{equation}
The constituents of $M$ in equation (\ref{coreacceleration}) and
(\ref{coremodel}) are determined by equation (\ref{Mass}). The
density of ordinary matter is given by
\begin{equation}
\frac{d{\cal M}(r)}{dr}=4\pi r^2\rho(r)\equiv 3\beta {\cal M}(r) \biggl[\frac{r_c}{r(r+r_c)}\biggr],
\end{equation}
so that we have
\begin{equation}
\label{density}
\rho(r)=\frac{3}{4\pi r^{3}} \beta {\cal M}(r) \biggl[\frac{r_c}{r+r_c}\biggr].
\end{equation}

Well inside the core radius, where $r \ll r_c$, the density $\rho(r) \propto $ constant for HSB galaxies with $\beta =
1$.  Although for LSB galaxies, it is the case for $\beta = 2$ that $\rho(r) \propto \left(r/r_{c}\right)^{3}$ for $r
\ll r_c$.  This behavior is quickly suppressed and the density profile shows no rapid change for $r \ll r_c$. 
Indeed the choice of $\beta = 2$ for LSB galaxies was found to be favorable because it resulted in a more slowly rising
mass profile (the integrated density distribution) for small $r$.  This critical distinction between high and low
surface brightness galaxies has been studied in the context of cold dark matter, with the common conclusion that the
rotation curves of LSB and Dwarf galaxies rise more slowly than those of HSB galaxies (or elliptical
galaxies)~\citep{deb96, deb03, bai05}. However, the ordinary matter
density of equation (\ref{density}) does not exhibit the divergent or cuspy behavior for either HSB or LSB galaxies,
contrary to the observations in these cold dark matter studies. Moreover, provided $r_{c} \ll r_{0}$ (which is the
result for all the galaxies fitted), then we see that the exponential factor in the acceleration law of equation
(\ref{coreacceleration}) has a damping effect well within the core radius such that the dynamics for $r \ll r_c$ is
described by Newtonian theory.  Thus the high resolution rotation curves for the LSB galaxies provide a clean testing
ground for any theory of galaxy rotation curves. Without the distinction we have implemented by equation
(\ref{surfaceBrightnessBeta}), the quality of our rotation curve fits would not be as good.

Well outside the core radius, where $r \gg r_{c}$, equation (\ref{coremodel}) implies that
\begin{equation}
\lim_{r \gg r_{c}} {\cal M}(r) = M,
\end{equation}
and the acceleration is described by equation (\ref{accelerationlaw3}).

For those galaxies for which there is photometric data available, the observed luminosity of the gaseous disk (HI plus
He), and the stellar components (the disk and separately the bulge) can be used to provide a more precise fit to the
velocity curve data, which would include the kinkiness of the rotation curves.  By assuming a constant $M/L$ ratio, it
is possible to invert the Poisson equation for the potential, yielding the mass distributions of the individual
constituents.  This results in a single parameter fit -- $(M/L)_{\rm stars}$.  By including the bulge, a more precise
fit is possible, in principle.  However this would necessitate a second parameter $(M/L)_{\rm bulge}$ which may be seen
as a weakness to this approach.  We choose to provide the single parameter fit -- which we present for comparison. We
note that the assumption of constant ratio $M/L$ forces an averaging over the radial distances thereby working to reduce
the accuracy of the photometric fits.  Our parametric mass model requires no assumptions on the ratio of $M/L$. The
photometric fits also include the assumption that the HI gas is on stable circular orbits around the centers of the
galaxies. This assumption may be violated for interacting galaxies, or in galaxies with strong bars. The photometric
fits also utilize a correction of the mass for the Helium fraction, which is based on the results from Big-Bang
nucleosynthesis and is not well known.

The rotational velocity derived from the acceleration law equation (\ref{coreacceleration}) is
\begin{equation}
\label{rotvelocity2} v(r)=\sqrt{\frac{G_0M}{r}}\biggl(\frac{r}{r_c+r}\biggr)^{\frac{3}{2}\beta}
\biggl\{1+\sqrt{\frac{M_0}{M}}\biggl[1-\exp(-r/r_0)
\biggl(1+\frac{r}{r_0}\biggr)\biggr]\biggr\}^{1/2}.
\end{equation}
The modified acceleration law equation (\ref{coreacceleration}) can be compared to the Newtonian law using equation
(\ref{coremodel}):
\begin{equation}
\label{coreNewton}
a_{\rm Newton}(r)=-\frac{G_0{\cal M}(r)}{r^2}.
\end{equation}

The gravitational acceleration described by Milgrom's phenomenological MOND model~\citep{mil83,san02} is given by
\begin{equation}
\label{milgromacc} a \mu\biggl(\frac{a}{{a_0}_{\rm Milgrom}}\biggr)=a_{\rm Newton},
\end{equation}
where $\mu(x)$ is a function that interpolates between the Newtonian regime, $\mu(x)=1$, when $x\gg 1$ and the MOND
regime, $\mu(x)=x$, when $x\ll 1$. The function normally used for galaxy fitting is
\begin{equation}
\mu(x)=\frac{x}{\sqrt{1+x^2}}.
\end{equation}

\section{Galaxy Rotational Velocity Curves}

A good fit to LSB and HSB galaxy data is obtained with the parameters
\begin{equation}
\label{parameters} M_0=9.60\times 10^{11}M_{\sun},\quad r_0=13.92\,{\rm kpc}=4.30\times 10^{22}\,{\rm cm},
\end{equation}
where we have substituted the value of $a_{0}$ from equation (\ref{speciala}) into equation (\ref{specialacceleration}),
relating the parameters $M_{0}$ and $r_{0}$. Thus, the modified acceleration law contains only a single parameter,
$M_{0}$ or $r_{0}$, which once set as in equation (\ref{parameters}) is universal for galaxies and is no longer a free
parameter for fitting the galaxy rotation curves.

We allow for the smaller scale Dwarf galaxies, for which the outermost observed radial position, $r_{\rm out} \lesssim
12\,\mbox{kpc}$, by rescaling equation (\ref{parameters}) such that the values for $M_{0}$ and $r_{0}$ continue to
satisfy the conditions (\ref{specialacceleration}) and (\ref{Hubbleacceleration}):
\begin{equation}
\label{DwarfParameters} M_0=2.40\times 10^{11}M_{\sun},\quad r_0=6.96\,{\rm kpc}=2.15\times 10^{22}\,{\rm cm}.
\end{equation}

The RG flow equations do indeed require that the parameters $M_0$ and $r_0$ are scale parameters and thus it is expected
that while we may obtain reasonable fits assuming they are roughly constant within a set population, this is an
approximation.  We are able to fit all HSB galaxies with a fixed $M_0$ and $r_0$.  We are also able to fit all LSB and Dwarf
galaxies by rescaling the HSB fixed values of $M_0$ and $r_0$ by a factor of 4 and 2, respectively.  This is
done to keep the number of parameters to an absolute minimum while respecting  the fact that the parameters $r_0$ and $M_0$
are determined by the scale of the system and should be different for HSB vs.\,LSB and Dwarf galaxies.

The fits to the galaxy rotation curves, $v$, in km/s versus the galaxy radius, $r$, in kpc are shown in
Figs.\,\ref{photometricRotationCurves} through
\ref{NGC4010}. The acceleration law is given by equation (\ref{coreacceleration}); and the rotational velocity by
equation (\ref{rotvelocity2}).

\subsection{Photometric Velocity Curve Fits}
\begin{figure}
\figurenum{5}
\plottwo{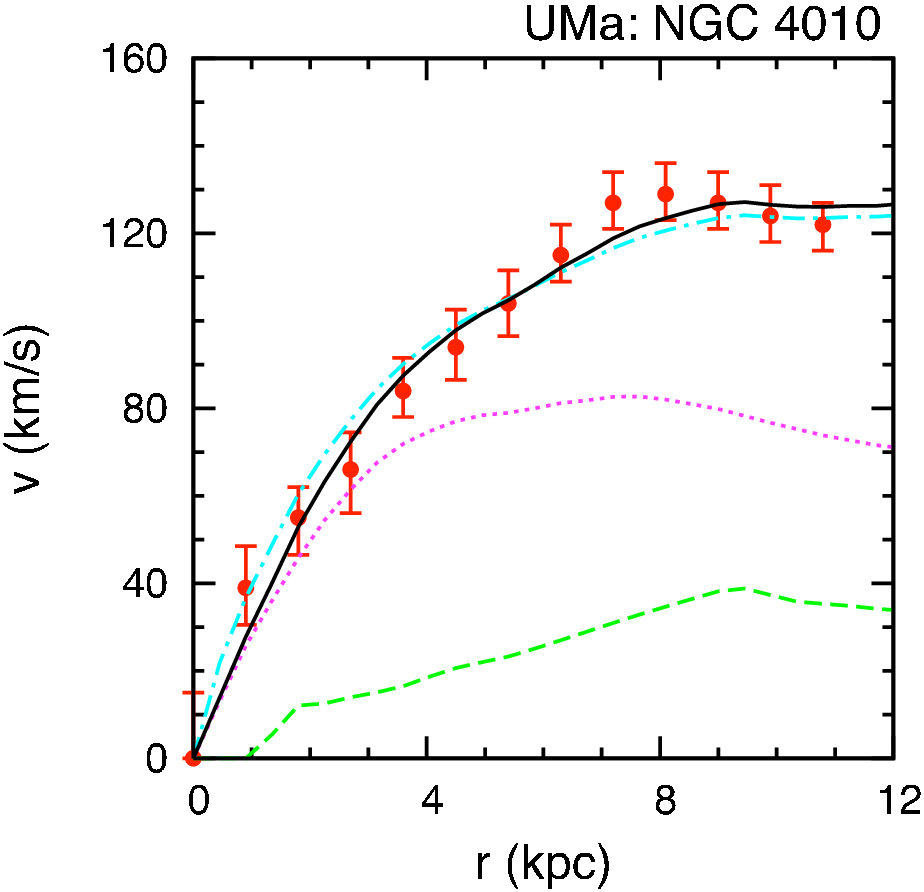}{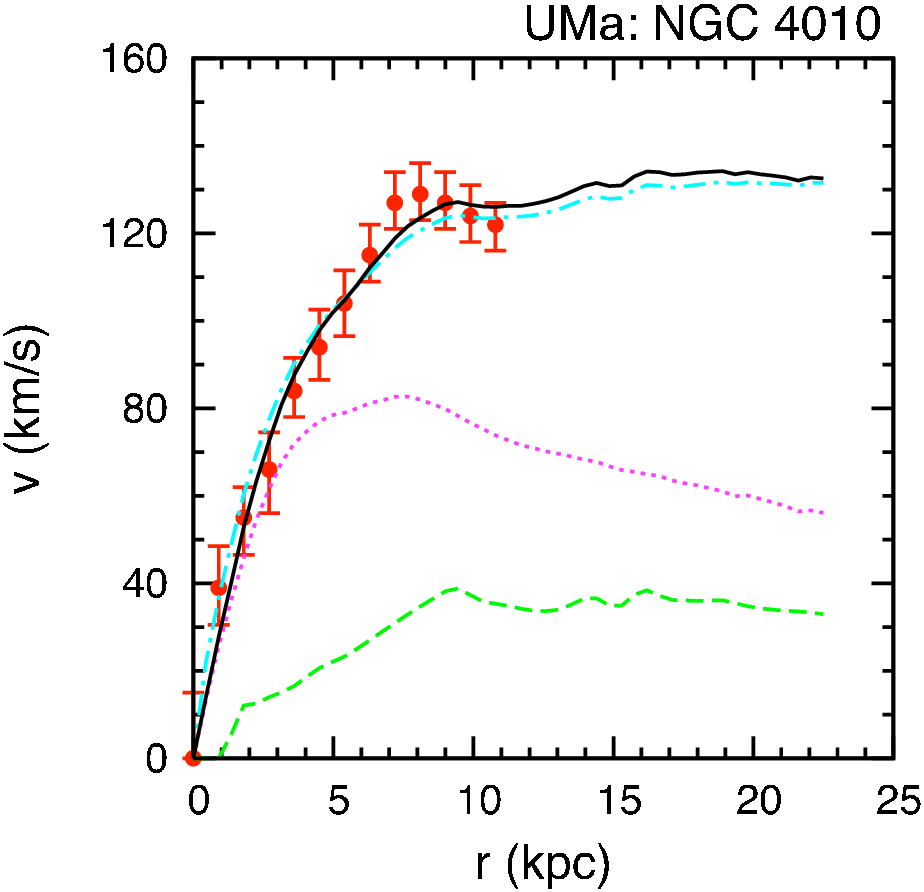}
\label{NGC4010}
\caption{Rotation curve for NGC 4010 in the Ursa Major cluster of galaxies reproduced from the original photometry of
~\citet{ver01a, ver01b} -- accounting for the revised distance estimate to UMa from 15.5 Mpc to 18.6 Mpc.
Both rotation curves are the same best fit via the single parameter $(M/L)_{\rm{stars}}$ based on the $K$-band photometric
observations of the gaseous (HI plus He) and luminous stellar disks for MSTG and MOND.   In every case,
the radial coordinate (horizontal axis) is given in kpc and the rotation velocity (vertical axis) in km/s.  The red
points with error bars are the observations, the solid black line is the rotation curve determined from MSTG, the
dash-dot cyan line is the rotation curve determined from MOND.  The other curves are the Newtonian rotation curves of
the various separate components: the long dashed green line is the rotation curve of the gaseous disk (HI plus He); the
dotted magenta curve is that of the luminous stellar disk. The first rotation curve
(leftmost) is plotted out to the edge of the visible stars in the galaxy at a distance of $r_{\rm out} \approx 10.8$ kpc.
 The second
rotation curve (rightmost) is plotted out to  $\approx 23$ kpc -- the extent of the HI and $K$-band data -- in order to
extend and
distinguish the MSTG and MOND ``flat rotation velocity'' predictions. The numerical results of the fit are presented in
Table~\ref{UMaPhotometricRotationCurveResults}.}
\end{figure}

The fits to the data in Fig.\,\ref{photometricRotationCurves} and \ref{NGC4010} are based on the photometric data of the
gaseous disk (HI plus He) component and luminous stellar disk component and only a single parameter $(M/L)_{stars}$ is
used to fit the rotation curves.  The data includes 29 galaxies -- both LSB and HSB galaxies -- obtained from
~\citet{deb98,deb01a,deb01b,beg91,mcg01,san02,san96} and
an additional 29 galaxies from the Ursa Major (UMa) cluster of galaxies derived from
~\citet{san98,ver01a,ver01b}.  For the UMa cluster of galaxies, available $K$-band data was used to reproduce the
velocity profiles of the gaseous disk (HI plus He) distribution and luminous stellar disks (via the
ROTMOD task of GIPSY\footnote{\url{http://www.astro.rug.nl/~gipsy/}}).   The gaseous disk (HI plus He) was assumed to be
infinitely thin and to have a total mass given by
\begin{equation}
\label{rotmodGas}
M_{gas} = \frac{4}{3} M_{HI}
\end{equation}
where the He correction factor of 4/3 is roughly determined from Big Bang nucleosynthesis, and the value of
\begin{equation}
\label{rotmodHI}
M_{HI} = 2.36 \times 10^{5} D^2 \int S dv\,[M_{\sun}],
\end{equation}
where $\int S dv$ is the integrated HI flux density in units of Jy km/s as measured from the global HI profile -- taken
from Column (15) of Table 2 of~\citet{ver01a}, and D is the distance in Mpc.  We accounted for
the revised distance estimate to UMa from $D=15.5\,\mbox{Mpc}$ to $D=18.6\,\mbox{Mpc}$ which has changed since the rotation
curves were originally presented in~\citet{san98}.  The luminous stellar disk was assumed to be described by the Van der
Kruit and Searle law, where the disk density distribution as a function of z (vertical height from the plane of the
disk) is given by
\begin{equation}
\label{rotmodDisk}
\Sigma(z)={\rm sech}^{2}(z/z_{0})/z_{0},
\end{equation}
where $z_{0}$ is the vertical scale height of the luminous stellar disk, and was assumed to be 20\% of the near infrared
exponential disk scale length according to Column (13) of Table 2 of~\citet{ver01a}.

According to~\citet{san98} the existence of $K$-band surface photometry is a great advantage since the near-infrared
emission, being relatively free of the
effects of dust absorption and less sensitive to recent star formation, is a more precise tracer of the mean radial
distribution of the dominant stellar population.  The principal advantages of using infrared luminosities is that
stellar mass-to-light ratios are less affected by population differences and extinction corrections are
minimal~\citep{ver01a}.  We focus on NGC 4010 in Fig.\,\ref{NGC4010} to study the effect of the extended HI and $K$-band
data (beyond the rotation curve data) on the quality fits in MSTG and MOND.  The numerical results of the UMa fits are
presented in Table\,\ref{UMaPhotometricRotationCurveResults}.

\subsection{Parametric Velocity Curve Fits}
\begin{figure}
\figurenum{3}
\plottwo{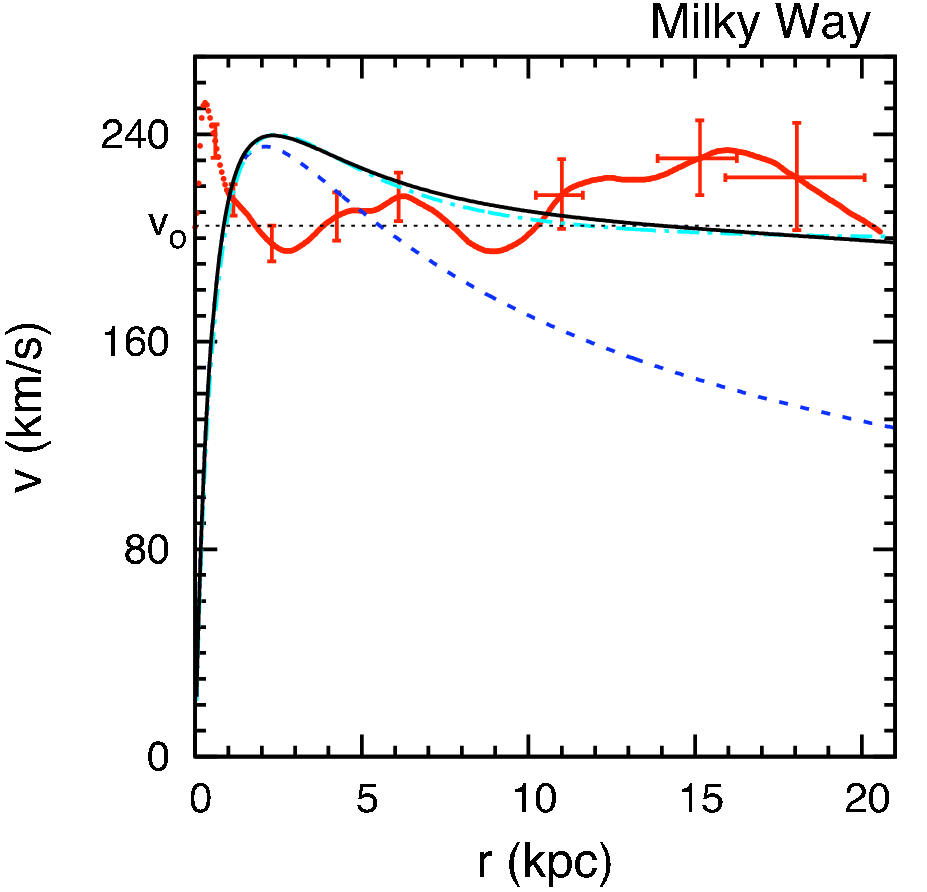}{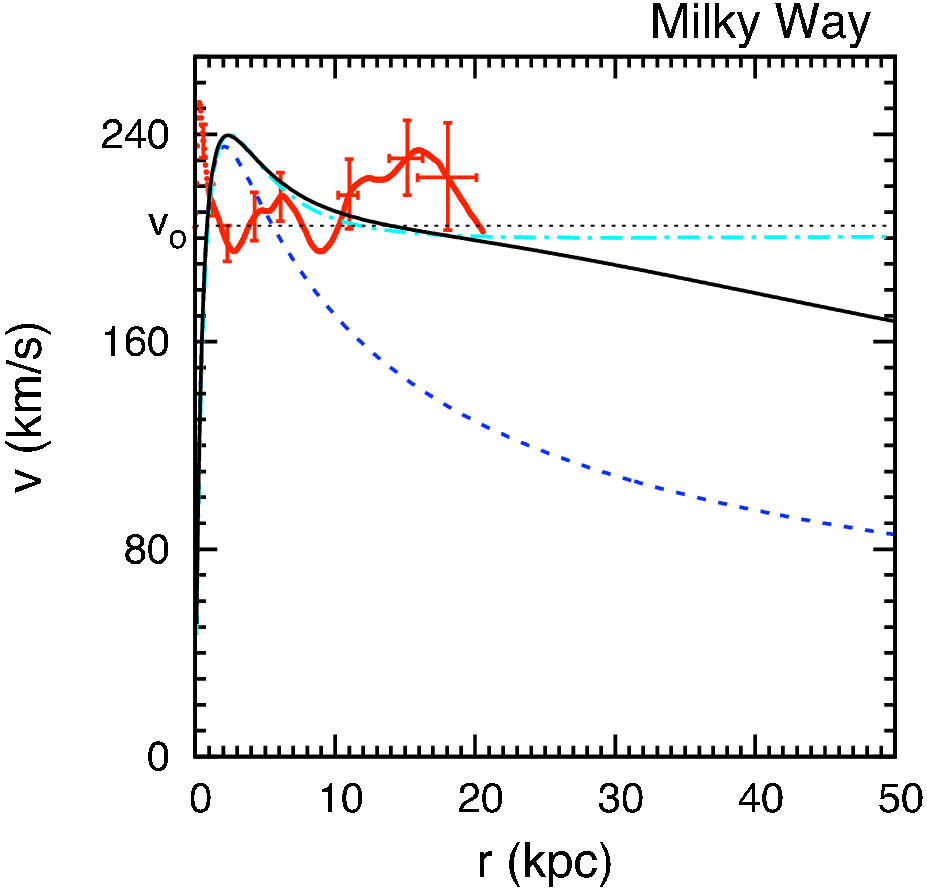}
\label{MilkyWay} \caption{High Resolution rotation curve for the
Milky Way.  Both rotation curves are the same best fit to a
parametric mass distribution (independent of luminosity
observations) -- a two parameter fit to the total galactic mass,
$M$, and a core radius, $r_{c}$.  The red points are the
observations -- error bars are shown as region specific. The solid
black line is the rotation curve determined from MSTG, the
dash-dot cyan line is the rotation curve determined from MOND.
The horizontal dotted black line is the MSTG predicted value of
the measured ``flat rotation velocity'', $v_{0}$ of equation
equation (\ref{v0}). The remaining curve -- the short dashed blue
line is the Newtonian galaxy rotation curve.  The first rotation
curve (leftmost) is plotted out to the edge of the visible stars
in the galaxy at a distance of $r_{\rm out} \approx 20.5$ kpc.
The second rotation curve (rightmost) is plotted out to 50 kpc in
order to distinguish the MSTG and MOND predictions. The numerical
results of the fit are presented in
Table~\ref{parametricRotationCurveResults}.}
\end{figure}

\begin{figure}
\figurenum{4}
\plottwo{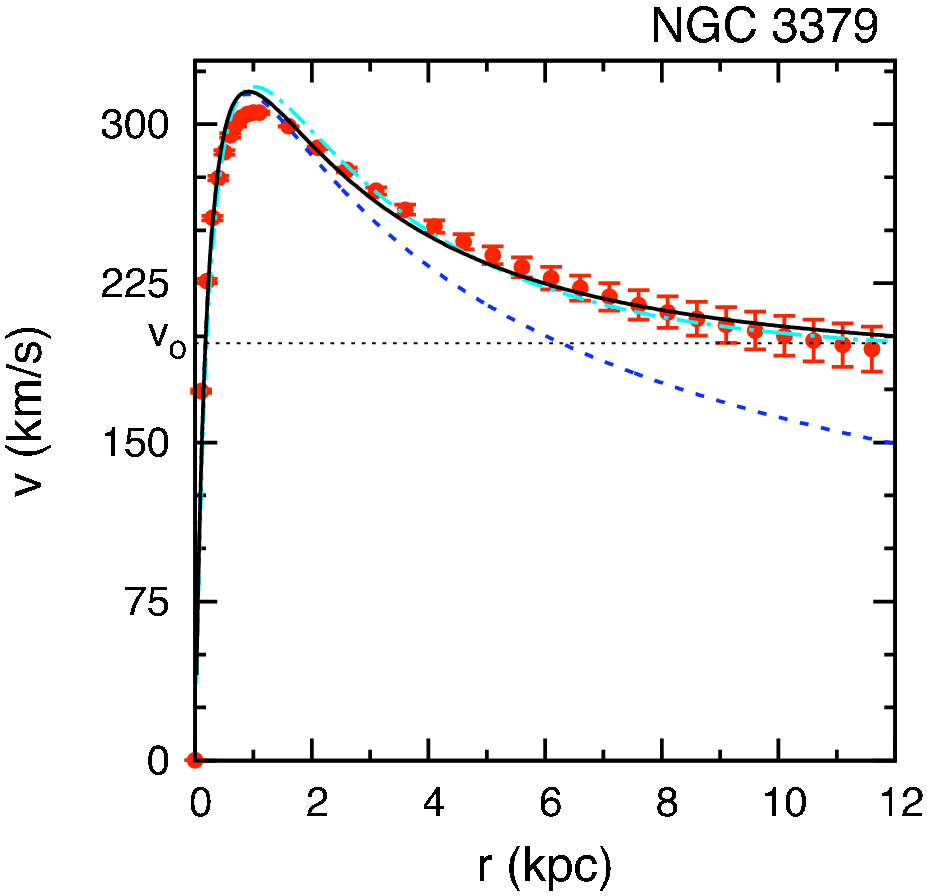}{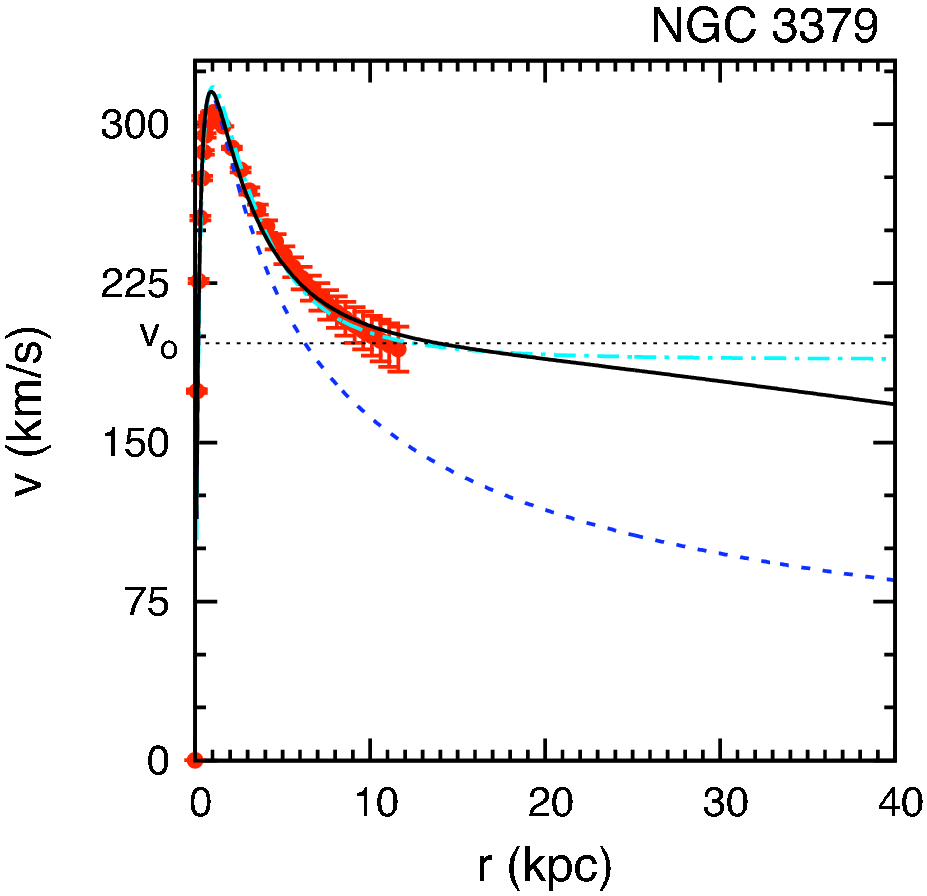}
\label{NGC3379} \caption{Rotation curve for the elliptical galaxy
NGC 3379.  Both rotation curves are the same best fit to a
parametric mass distribution (independent of luminosity
observations) -- a two parameter fit to the total galactic Mass,
$M$, and a core radius, $r_{c}$.  The red points (with error bars)
are samplings of the  circular velocity profile constrained by
orbit modeling according to~\citet{rom03,rom04}.  The solid black
line is the rotation curve determined from MSTG, the dash-dot cyan
line is the rotation curve determined from MOND.  The horizontal
dotted black line is the MSTG predicted value of the measured
``flat rotation velocity'', $v_{0}$ of equation equation
(\ref{v0}). The remaining curve -- the short dashed blue line is
the Newtonian galaxy rotation curve.  The first rotation curve
(leftmost) is plotted out to the edge of the visible stars in the
galaxy at a distance of $r_{\rm out} \approx 12$ kpc.  The second
rotation curve (rightmost) is plotted out to 40 kpc in order to
distinguish the MSTG and MOND predictions. The numerical results
of the fit are presented in
Table~\ref{parametricRotationCurveResults}.}
\end{figure}

The fits to the rotation curve data shown in Figs.\,\ref{parametricRotationCurves} through \ref{NGC3379} are based on
the parametric model of equations (\ref{coremodel}) and (\ref{surfaceBrightnessBeta}).  Since this model is independent of
the photometric data of the gaseous disk (HI plus He) component and luminous stellar disk component, a larger database
of galaxies is available including the high resolution rotation curves of~\citet{sof96} and the elliptical galaxy NGC
3379 of~\citet{rom03,rom04}.
This adds another 42 galaxies to the {\em complete sample} as described in Table\,\ref{CompleteSample}.  We focus on
the Milky Way in Fig.\,\ref{MilkyWay} and NGC 3379 in Fig.\,\ref{NGC3379} to clarify the
predictions of MOND and MSTG which are hard to distinguish in Fig.\,\ref{parametricRotationCurves}; but become apparent
at distances beyond the edge of the visible stars.    The numerical results of the fits to the {\em complete sample} are
presented in Table\,\ref{parametricRotationCurveResults}.

\subsection{Flat Rotation Velocity}
In Milgrom's phenomenological MOND model we have
\begin{equation}
\label{Milgromv} v_{c}^4=G_0 M {a_0}_{\rm Milgrom}.
\end{equation}
We see that equation (\ref{Milgromv}) predicts that the rotational velocity is constant out to an infinite range and the
rotational velocity does not depend on a distance scale, but on the magnitude of the acceleration ${a_0}_{\rm Milgrom}$.
In contrast, our modified acceleration formula does depend on the radius $r$ and the distance scale $r_0$, which for
galaxies is fixed by the formula equation (\ref{Hubbleacceleration}). The MSTG velocity curve asymptotically becomes the
same as the Newtonian-Kepler prediction as $r\rightarrow\infty$:
\begin{equation}
\label{asympvelocity} v\sim \sqrt{G_{\infty}M/r},
\end{equation}
where $G_{\infty}$ is the renormalized value of Newton's constant.

The flatness of the rotation curves arises due to an increased strength in the galactic gravitational potential due to
the running of Newton's constant.  By taking the first and second derivative of $G(r)$ from equation (\ref{runningG}):
\begin{eqnarray}
\label{GPrime} \frac{dG(r)}{dr} & = & \frac{G_0}{r_{0}^{2}}\sqrt\frac{M_0}{M}r\exp(-r/r_0), \\
\label{GDoublePrime} \frac{d^{2}G(r)}{dr^{2}} & = &
\frac{G_0}{r_{0}^{2}}\sqrt\frac{M_0}{M}
\left(1-\frac{r}{r_{0}}\right)\exp(-r/r_0),
\end{eqnarray}
and we see that $dG(r)/dr$ has a maximum at $r = r_{0}$. Therefore the rate of change of the running
of the effective gravitational constant has a maximum; and it is at this point where the tendency to return to the
Newtonian-Kepler behaviour of equation (\ref{asympvelocity}) is most opposed by the RG flow.  The effect of this on the
galaxy rotation curves is to produce an extended region where the curve seems flat.  The velocity at $r = r_{0}$ is
defined as $v_{0}$:
\begin{equation}
\label{v0} v_{0} \equiv v(r = r_{0}),
\end{equation}
is the MSTG equivalent of the flat rotational velocity; and is within the quoted experimental uncertainties to the MOND equivalent, $v_{c}$, in
those galaxies that MOND handles well.  The numerical results for $v_{0}$ and $v_{c}$ with calculated uncertainties are
included in Tables \ref{UMaPhotometricRotationCurveResults} and \ref{parametricRotationCurveResults}.

Using the Sloan Digital Sky Survey (SDSS),~\citet{pra03} have studied the velocities of satellites orbiting isolated
galaxies. They detected approximately 3000 satellites, and they found that the line-of-sight velocity dispersion of
satellites declines with distance to the primary. The velocity was observed to decline to a distance of $\sim 350$ kpc
for the available data. This result contradicts the constant velocity prediction equation (\ref{Milgromv}) of MOND, but
is consistent with the MSTG prediction equation (\ref{asympvelocity}). It also agrees with the cosmological models which
predict mass profiles of dark matter halos at large distances. During the last two decades of numerical modelling of
galaxy formation, they have produced a density profile of dark matter halos, $\rho\propto 1/r^3$ at large radii, which
does not depend on the nature of the dark matter~\citep{avi01,col02}. The results of~\citet{pra03} are consistent with
recent gravitational lensing results~\citep{she03}.

\section{The Tully-Fisher Relation}

Unlike MOND, the mass -- rotational velocity relationship is not absolute as in equation (\ref{Milgromv}); and indeed at
distances beyond the galaxy it is expected that Kepler's Law applies according to equation (\ref{asympvelocity}).

\begin{figure}
\figurenum{6}
\plottwo{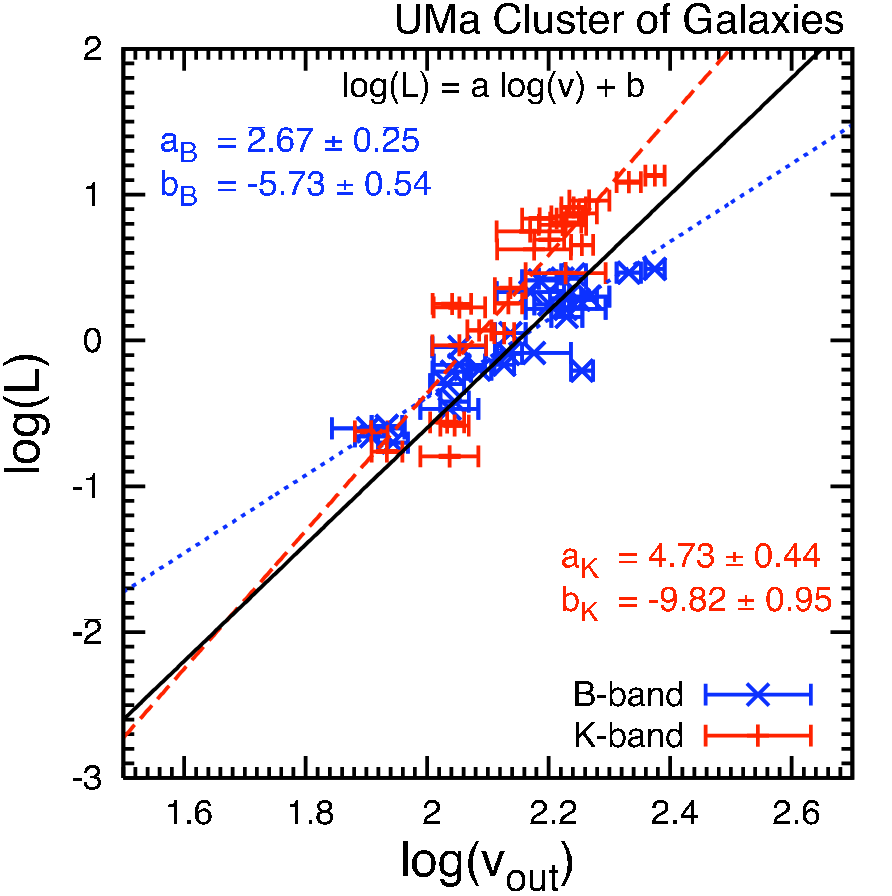}{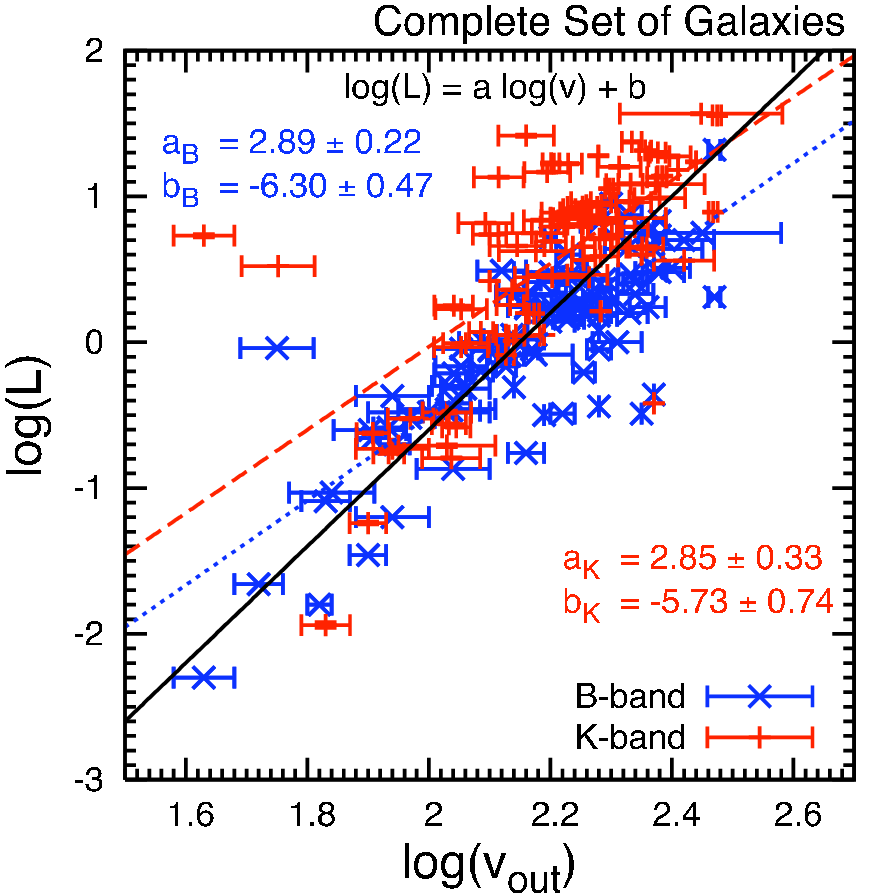}
\label{observedTullyFisher}
\caption{Observed $B$- and $K$-band Tully-Fisher relation for the UMa cluster of galaxies, and the {\em complete sample} of
galaxies. The vertical axis is the (base 10) logarithm of the observed galaxy luminosity (in $10^{10} L_{\sun}$); and the
horizontal axis is the
(base 10) logarithm of the observed rotational velocity (in km/s) at the maximum observed radius.  In both plots, the
blue $\times$-points are the
observed $B$-band luminosity data and the red $+$-points are the observed $K$-band luminosity data.  In all cases, the Tully-Fisher relation is parametrized by  $\log(L)
\equiv a log(v) + b$. In both plots, the blue
dotted lines are the best fit $B$-band Tully-Fisher relation and the red dashed lines are the best fit $K$-band Tully-Fisher
relation.  The best fit results
using a nonlinear least-squares fitting routine including estimated errors are summarized in Table
\ref{TullyFisherResults}. The solid black line is the MOND prediction with $\langle M/L \rangle \equiv 1$.}
\end{figure}

The observational Tully-Fisher relation implies a luminosity -- rotational velocity of the form~\citep{tul77}:
\begin{equation}
\label{TullyFisher} L \propto v_{\rm out}^{a} {\rm \ where\ } a \approx 4,
\end{equation}
where $L$ is the total observed luminosity of the galaxy (in units
of $10^{10} L_{\sun}$), and $v_{\rm out}$ is the observed velocity
at the outermost observed radial position (in km/s).
\citet{ver01a} considers an alternate definition of the ``flat
rotation velocity'', $v_{\rm flat}$, which categorizes galaxies
according to three kinds depending on the shape of the rotation
curve.  The behavior of $v_{out}$ is more closely related to the
asymptotic ``flat rotation velocity''. Taking the logarithm of
both sides of equation (\ref{TullyFisher}), we obtain
\begin{equation}
\label{TullyFisherLog} \log(L) = a \log(v_{\rm out}) + b.
\end{equation}
$B$-band luminosity data is available for practically all of the galaxies either from the
original references listed in Table \ref{parametricRotationCurveResults} or listed in~\citet{tul88}.  Moreover, the
majority of the galaxies in this study have been detected by 2MASS in the $K_{s}$-band.  In order to calculate the total $K$-band luminosity,
apparent $K$-band magnitudes from the 2MASS surver were used.  Given an apparent $K$-band magnitude it is possible to
calculate the $K$-band luminosity as
\begin{equation}
\label{luminosity}
\log_{10}(L_{K}) = 1.364 - \frac{2}{5} K_{T} + \log_{10}(1+z) + 2 \log_{10} D,
\end{equation}
where $L_{K}$ is the $K$-band luminosity in units of $10^{10} L_{\sun}$, $K_{T}$ is the $K$-band apparent magnitude and z
is the
redshift of the galaxy (determined from the NASA/IPAC Extragalactic Database), and D is the distance to the galaxy in
Mpc (from the original references).  The $\log_{10}(1+z)$ term is a first order $K$-correction.  We have plotted the observed Tully-Fisher relation for LSB and HSB galaxies in Fig.\,\ref{observedTullyFisher}.

As in~\citep{san98}, MOND is able to make predictions on both the slope and the intercept of the logarithmic
Tully-Fisher relation by assuming that the mass to luminosity ratio, $M/L$, is constant across all galaxies.
Although this is not the case, the assumption is enforced by using the mean mass to light ratio, $\langle
M/L\rangle$, to determine the values of the slope and intercept, $a$ and $b$, in equation
(\ref{TullyFisherLog}). Replacing the observed luminosity, $L$, in the logarithmic Tully-Fisher relation of equation
(\ref{TullyFisherLog}) with
\begin{equation}
\label{TullyFisherMean} L = \frac{M}{\langle M/L\rangle}.
\end{equation}
We obtain
\begin{equation}
\label{TullyFisherActual}
\log(M) = a \log(v_{\rm out}) + b - \log\left({\langle M/L\rangle}\right).
\end{equation}
Thus, the effect of $\langle M/L\rangle$ is to shift the intercept which vanishes when $\langle M/L\rangle = 1$.
We may quantify the predictions of MSTG and MOND by either computing the appropriate $\langle M/L\rangle$ values which depend on the particular band
of the luminosity measurements, or by considering the respective curve fits to the actual Tully-Fisher relation:

\begin{eqnarray}
\label{TullyFisherMSTG}
\log(M) = & a \log(v_{0}) + b & \mbox{MSTG}\\
\label{TullyFisherMOND}
\log(M) = & a \log(v_{c}) + b & \mbox{MOND}
\end{eqnarray}
where $v_{0}$ is the MSTG ``flat rotation velocity''  of equation (\ref{v0}); and $v_{c}$ is the MOND asymptotic
rotation velocity of equation (\ref{Milgromv}). Taking the logarithm (with respect to base 10) of both sides of equation
(\ref{Milgromv}), we have the MOND predictions
\begin{eqnarray}
\label{TullyFisherParameters} a &=& 4, \\
b &=& -8.21
\end{eqnarray}

\begin{figure}
\figurenum{7}
\plottwo{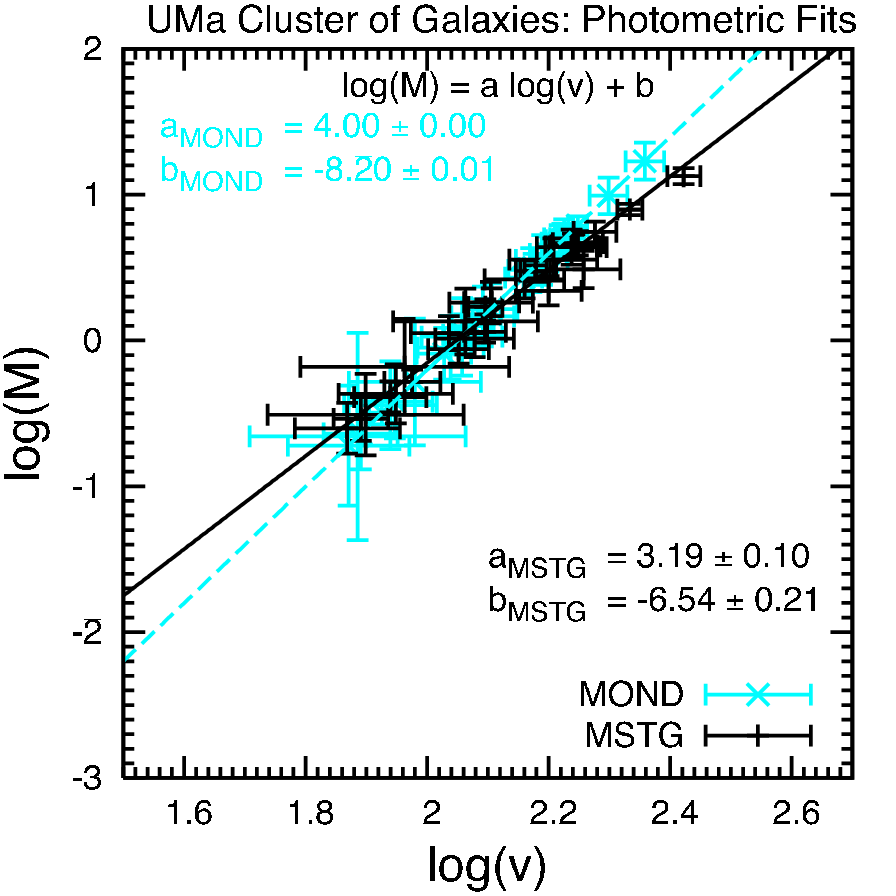}{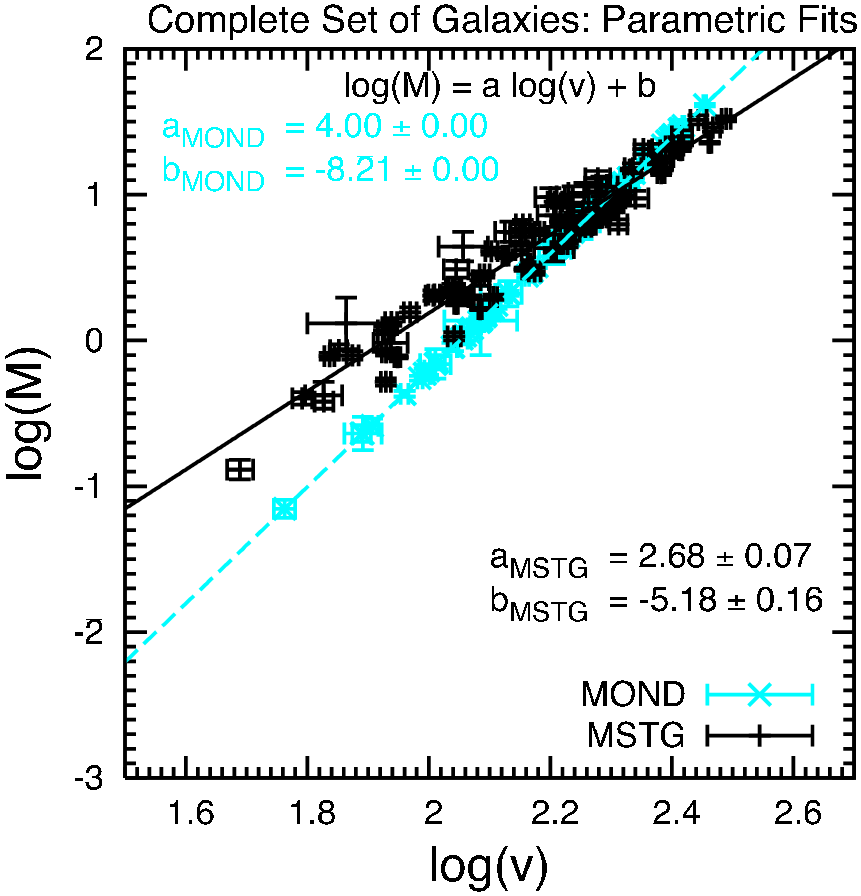}
\label{actualTullyFisher}
\caption{Actual Tully-Fisher relation for the UMa cluster of galaxies' photometry, and the {\em complete sample} of
galaxies resulting from the parametric fits.  The first curve
(leftmost) is the combined HSB, LSB and Dwarf galaxies fit to the photometric data of the UMa cluster of galaxies
corresponding to those galaxies of Fig.\,\ref{photometricRotationCurves} labeled by {\em UMa} and Table
\ref{UMaPhotometricRotationCurveResults}.  The second curve (rightmost) is the combined HSB, LSB and Dwarf
galaxies fit to the parametric data of the {\em complete sample} corresponding to Fig.\,\ref{parametricRotationCurves}
and Table \ref{parametricRotationCurveResults}.  The vertical axes are the
(base 10) logarithm of the total mass of the galaxy (in $10^{10} M_{\sun}$) resulting from the fits.  The horizontal axes
are the (base 10)
logarithm of the flat rotational velocity (in km/s) as determined from the fits -- $v_{0}$ in MSTG and $v_{c}$ in MOND.
In all plots, the cyan $\times$-points are the MOND results,
and the black $+$-points are the MSTG results.  In all cases, the Tully-Fisher relation is parametrized by  $\log(M)
\equiv a log(v) + b$; and the results of the fits are shown for MSTG and MOND.  The dashed cyan line is the best fit
solution for the MOND results, and the solid black line is the best fit solution for MSTG results.  The best fit results
using a nonlinear least-squares fitting routine including estimated errors are summarized in Table
\ref{TullyFisherResults}.}
\end{figure}

Fig.\,\ref{actualTullyFisher} shows the curve fits to equations (\ref{TullyFisherMSTG}) and (\ref{TullyFisherMOND}) for
both the photometric fits and the fits to our parametric model of equations (\ref{coremodel}) and
(\ref{surfaceBrightnessBeta}).  In all cases, we are able to combine the HSB, LSB and  Dwarf galaxy data (and the
elliptical
galaxy NGC 3379) for the fitting, implying consistent physics across galaxies.  The numerical results of the
respective fits are presented in Table \ref{TullyFisherResults}.

We see that a comparison of the MSTG and MOND results of the actual and observed Tully-Fisher relation show that the MSTG prediction is closer to the observational data for the {\em complete sample} and the UMa $B-$band photometry.  This is most likely the result of the implicit assumption within the MOND framework that M/L is constant within each galaxy which does not appear to be the case in either the $B-$ or $K-$ bands.

\section{Conclusions}

A gravity theory consisting of a metric-skew-tensor action that leads to a modified Newtonian acceleration
law~\citep{mof05a,mof05b} can be fitted to a large class of galaxy rotation curves. We have presented the predictions
for the galaxy rotation curves from a covariant and relativistic gravitational theory without postulating non-baryonic
dark matter. The only other relativistic gravity theories that have been published that have attempted to do this are
\citet{man90,man05} and \citet{bek04}.  The latter recent publication attempts to incorporate Milgrom's MOND into a covariant gravitational theory.  The fully
relativistic gravitational theory presented in \citet{mof05a,mof05b} is a consistent and viable example of a
gravitational theory that fits the galaxy rotation curves and galaxy cluster mass X-ray data without non-baryonic dark
matter~\citet{bro05}.  The large sample of galaxy data which fits our predicted MSTG acceleration law warrants taking
seriously the proposal that a modified gravity theory can explain the flat rotational velocity curves of galaxies
without (as yet, undetected) non-baryonic dark matter.  It represents an important foil in comparing modified
gravitational theory with dark matter.

It is interesting to note that we can fit the rotational velocity
data of galaxies in the distance range $0.02\,{\rm kpc} < r <
70\,{\rm kpc}$ and in the mass range $10^5\, M_{\sun}< M <
10^{11}\,M_{\sun}$ without exotic dark matter halos. The lensing
of clusters can also be explained by the theory without exotic
dark matter in cluster halos. An important prediction is that for
large enough distances from the galaxy cores, the rotational
velocity of stars declines as $v\propto 1/\sqrt{r}$ consistent
with a Newtonian-Kepler fall off. This is consistent with the
results of~\citet{pra03} and gravitational lensing results for
galaxies clusters~\citep{she03}.

In order to obtain a self-consistent description of solar dynamics, galaxies, clusters of galaxies and cosmology, it is
necessary to have the effective gravitational constant $G$, the MSTG coupling constant $\gamma_{c}$, the mass $\mu$ (range
$r_0$) of the skew field $F_{\mu\nu\lambda}$ and the cosmological constant $\Lambda$ run with distance (time). The RG
flow effective action description of MSTG quantum gravity allows for a running of these effective constants with
distance and time~\citep{mof05a,mof05b}. The RG flow framework for the theory is characterized by special RG
trajectories. On
the RG trajectory, we identify a regime of distance scales where solar system gravitational physics is well described by
GR, which is contained in MSTG as an approximate solution to the field equations. We are able to obtain agreement with
the observations in the solar system, terrestrial gravitational experiments and the binary pulsar PSR 1913+16. Strong
infrared renormalization effects become visible at the scale of galaxies and the modified Newtonian potential replaces
exotic dark matter as an explanation of flat rotation curves. Thus, gravity becomes a ``confining force'' that has
significant predictions for astrophysics and cosmology.

We have demonstrated that the RG flow running of $G$ and MSTG cosmology can lead to a description of the universe that
does not require dominant, exotic dark matter. Dark energy is described by an effective time dependent cosmological
constant. A detailed investigation of the MSTG cosmological scenario must be performed to establish that it can describe
the large scale structure of the universe, account for galaxy formation and big bang nucleosynthesis and be consistent
with the WMAP data.

The predictions for the galaxy rotation curves from MSTG and Milgrom's MOND agree remarkably for almost all
of the 101 galaxies fitted throughout the visible distance scales of the galaxies for LSBs as well as HSBs and the one
elliptical galaxy NGC 3379. In particular, for the fits using photometric data and only one parameter $\langle
M/L\rangle$ (once $r_0$ and $M_0$ are fixed), the agreement of the two models suggests that Milgrom's MOND critical
acceleration ${a_0}_{\rm Milgrom}=1.2\times 10^{-8}\,{\rm cm}/{\rm s}^2$ is closely related to the range parameter $r_0$ and the
distance scaling behavior of the MSTG modified acceleration law.

\acknowledgments
\section*{Acknowledgments}

This work was supported by the Natural Sciences and Engineering Research Council of Canada. JRB would like to thank the
Perimeter Institute for Theoretical Physics for additional funding. We thank Martin Green and Martin Reuter for helpful
discussions. We also thank Stacy McGaugh and Marc Verheijen for supplying data and for helpful discussions; and Erwin
de Blok for assisting in our deployment of the Groningen Image Processing System (GIPSY)\footnote{\url{http://www.astro.rug.nl/~gipsy/}} and the ROTMOD task on the Mac G4
platform.




\clearpage





\figcaption{Photometric Fits to Galaxy Rotation Curves: There are 58 galaxies presented here, each is  a best fit via
the single parameter $(M/L)_{\rm{stars}}$ based on the photometric data of the gaseous (HI plus He) and luminous stellar
disks.
 The 29
galaxies labeled {\em UMa} are members of the Ursa Major cluster of galaxies.  For the UMa subset, available $K$-band data
was used to reproduce the velocity profiles of the gaseous disk (HI plus He) distribution and luminous stellar disks
accounting for the revised distance estimate to UMa from 15.5 Mpc to 18.6 Mpc according to~\citet{ver01b}.
The numerical results of the UMa fits are presented in
Table\,\ref{UMaPhotometricRotationCurveResults}.  In all cases, the radial coordinate (horizontal axis) is given in kpc and the
rotation velocity (vertical axis) in km/s. The red points with error bars are the observations, the solid black line is
the rotation curve determined from MSTG, the dash-dot cyan line is the rotation curve determined from MOND.  The other
curves are the Newtonian rotation curves of the various separate components: the long dashed green line is the rotation
curve of the gaseous disk (HI plus He); the dotted magenta curve is that of the luminous stellar disk.
\label{photometricRotationCurves}}

\figcaption{Parametric Fits to Galaxy Rotation Curves: There are
101 galaxies presented here -- the {\em complete sample} of
galaxies of Table~\ref{CompleteSample}.  These rotation curves are
best fit to the parametric mass distribution (independent of
luminosity observations) of equations (\ref{coremodel}) and
(\ref{surfaceBrightnessBeta}) -- a two parameter fit to the total
galactic Mass, $M$, and a core radius, $r_{c}$.  In all cases, the
red points (with error bars) are the observations.  In those cases
where the high resolution observations connect together, error
bars if available are shown as region specific. The solid black
line is the rotation curve determined from MSTG, the dash-dot cyan
line is the rotation curve determined from MOND.  The horizontal
dotted black line is the MSTG predicted value of the measured
``flat rotation velocity'', $v_{0}$ of equation (\ref{v0}). The
remaining curve -- the short dashed blue line is the Newtonian
galaxy rotation curve.  The numerical results of the fits are
presented in Table\,\ref{parametricRotationCurveResults}.
\label{parametricRotationCurves}}
\begin{figure}[p]
\figurenum{1}
\begin{center}
\begin{tabular}{c}
\includegraphics{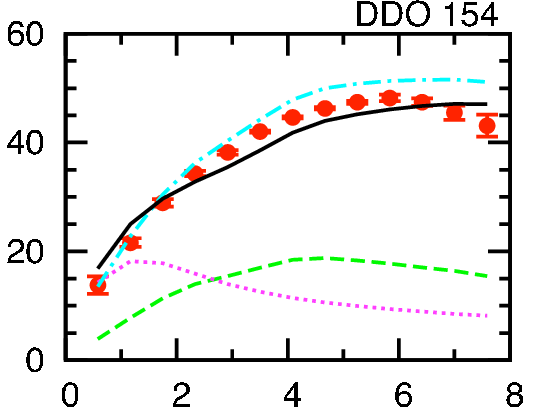}
\includegraphics{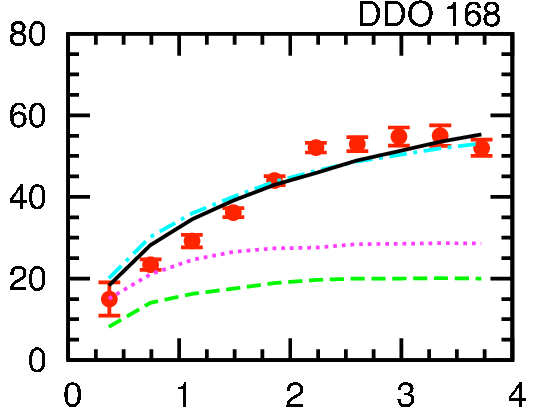}
\includegraphics{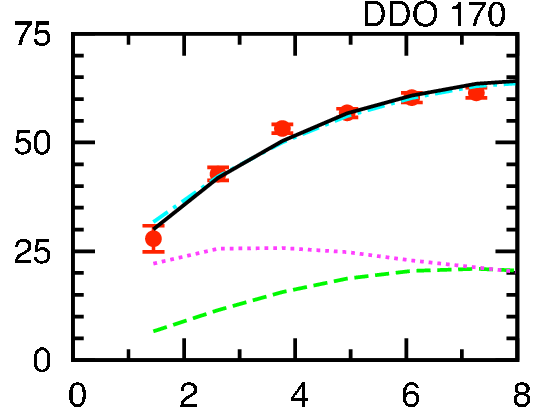} \\
\includegraphics{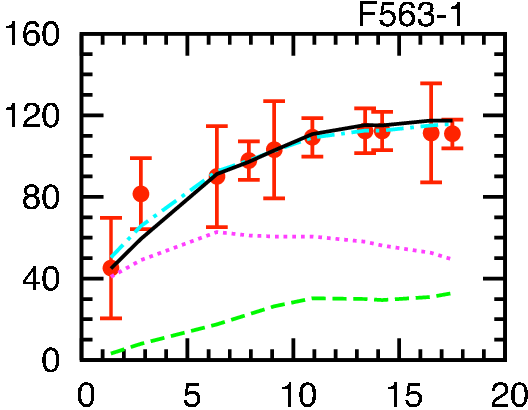}
\includegraphics{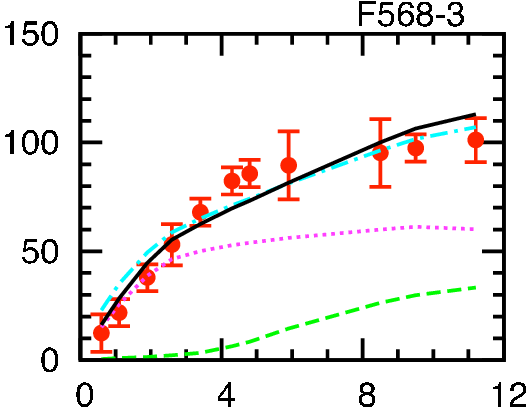}
\includegraphics{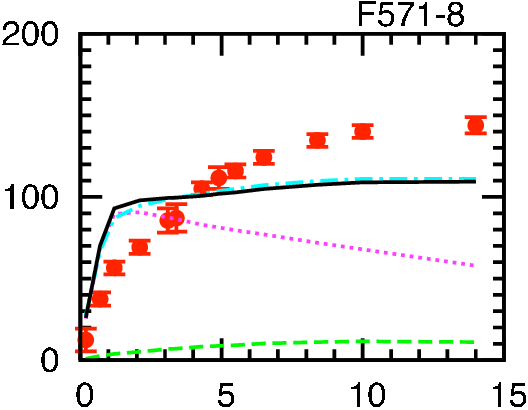} \\
\includegraphics{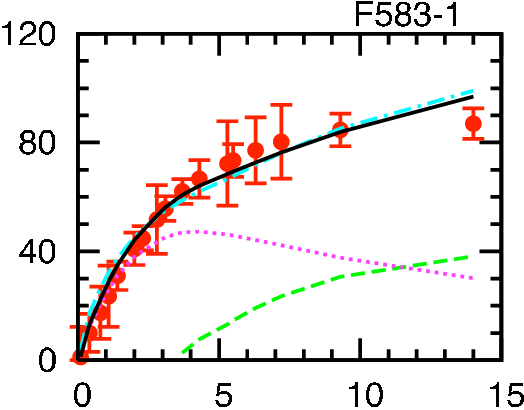}
\includegraphics{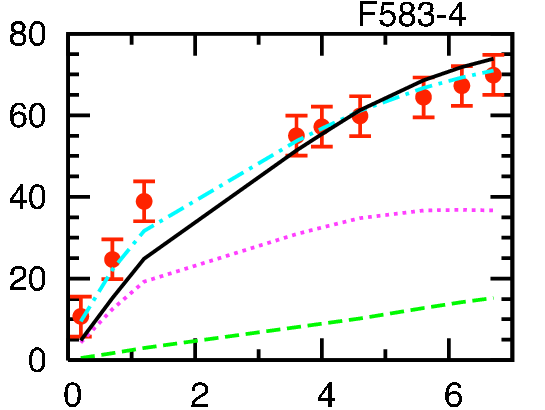}
\includegraphics{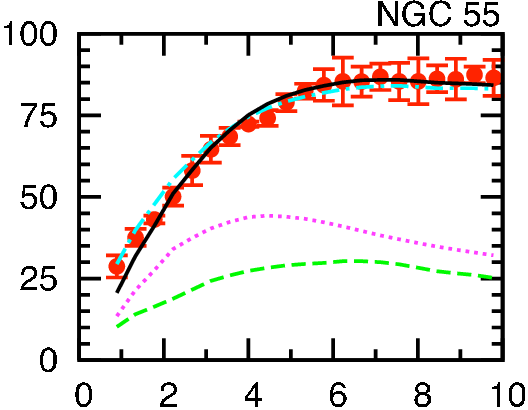} \\
\includegraphics{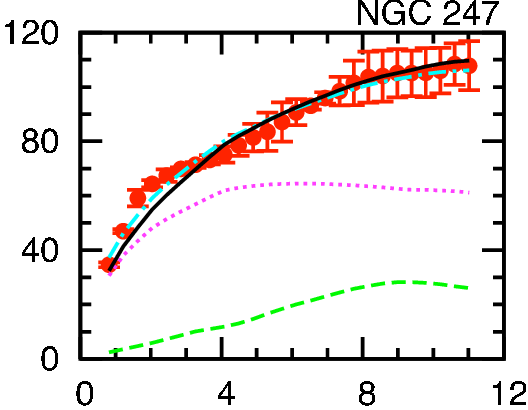}
\includegraphics{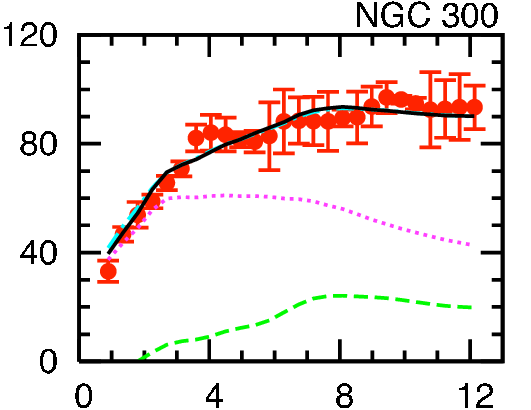}
\includegraphics{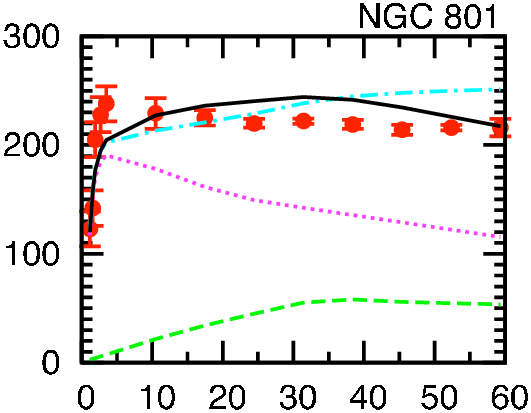} \\
\includegraphics{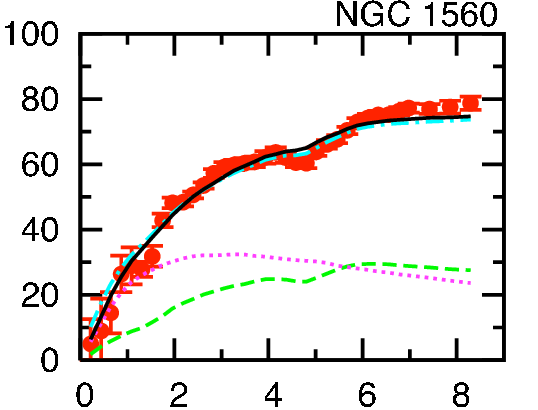}
\includegraphics{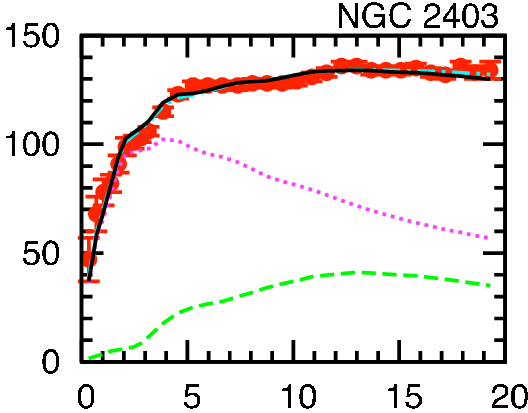}
\includegraphics{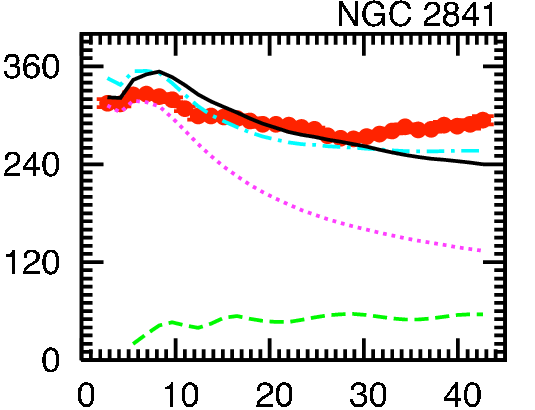}
\end{tabular}
\end{center}
\caption{Photometric Galaxy Rotation Curve Fits}
\end{figure}
\clearpage
\begin{figure}[p]
\figurenum{1 Continued}
\begin{center}
\begin{tabular}{c}
\includegraphics{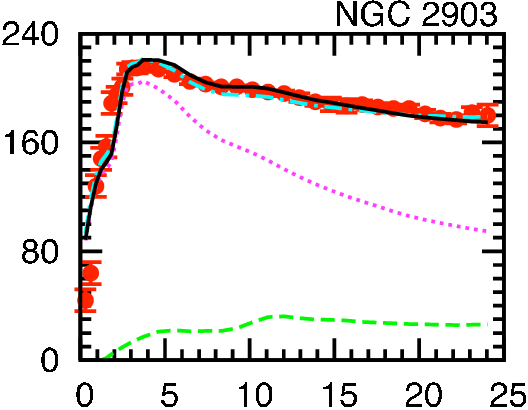}
\includegraphics{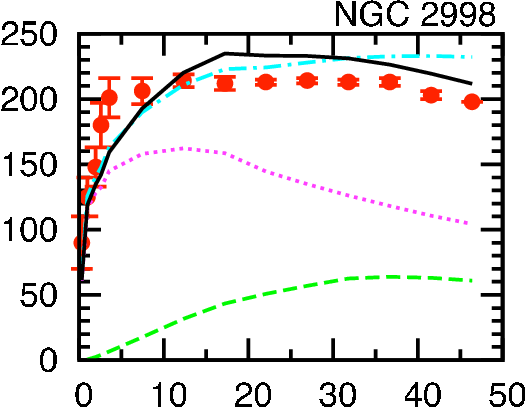}
\includegraphics{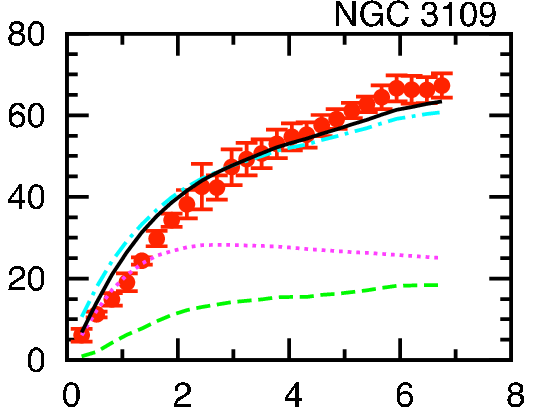} \\
\includegraphics{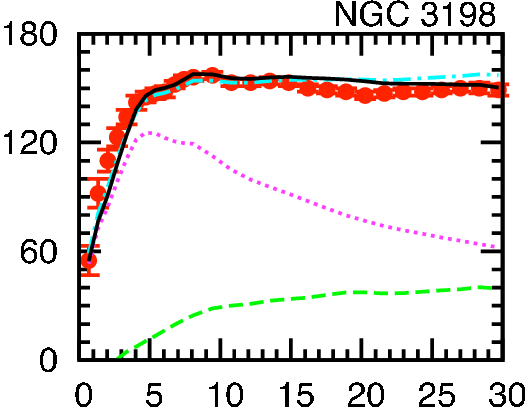}
\includegraphics{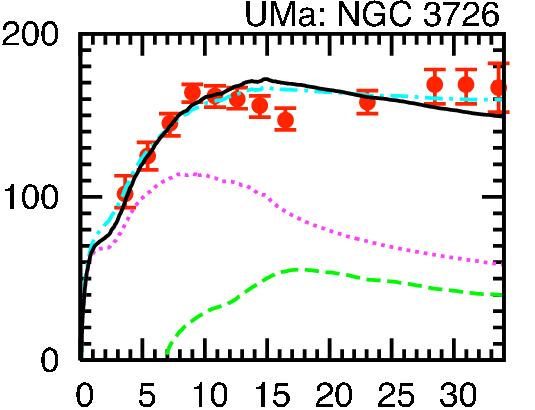}
\includegraphics{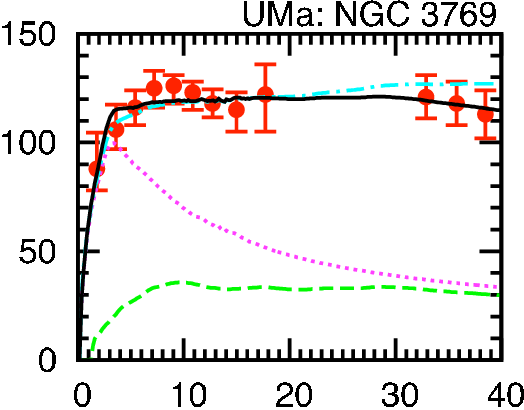} \\
\includegraphics{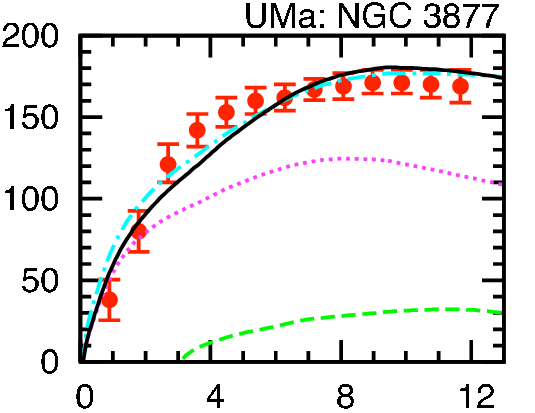}
\includegraphics{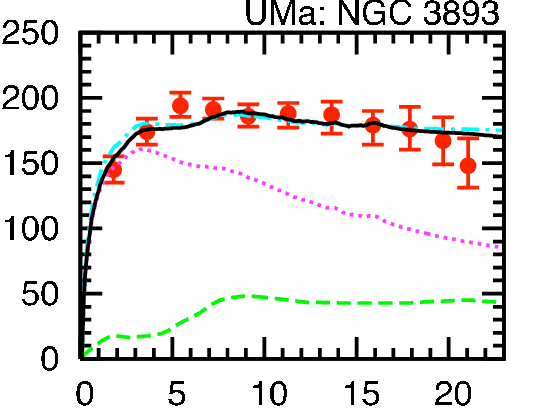}
\includegraphics{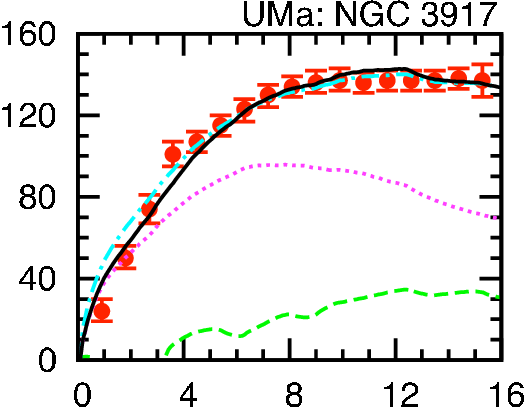} \\
\includegraphics{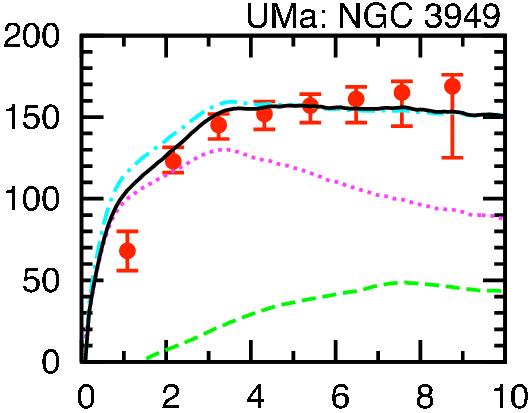}
\includegraphics{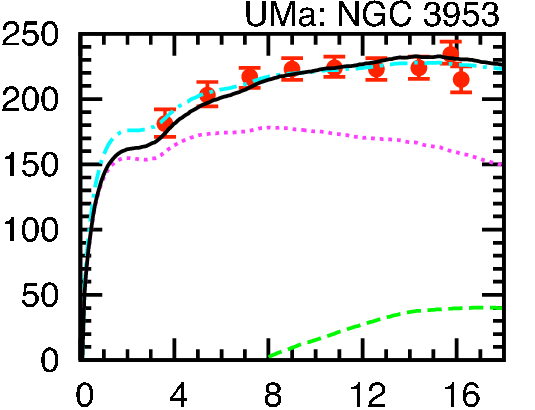}
\includegraphics{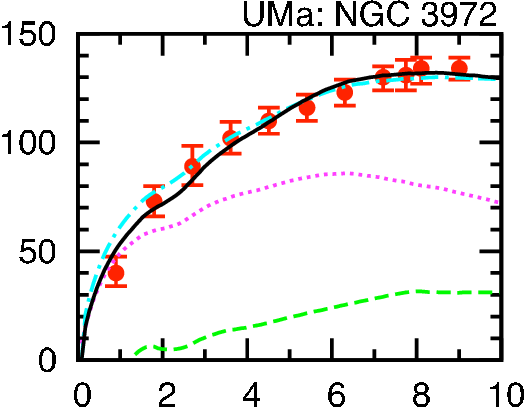} \\
\includegraphics{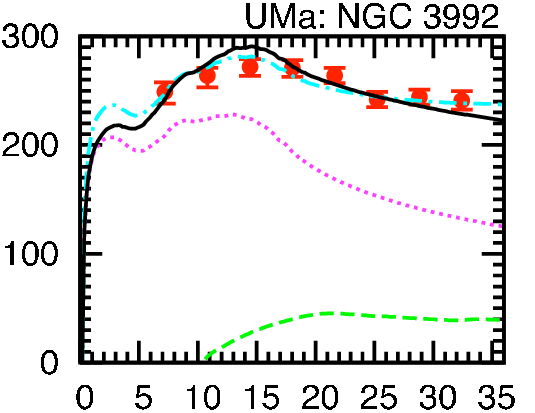}
\includegraphics{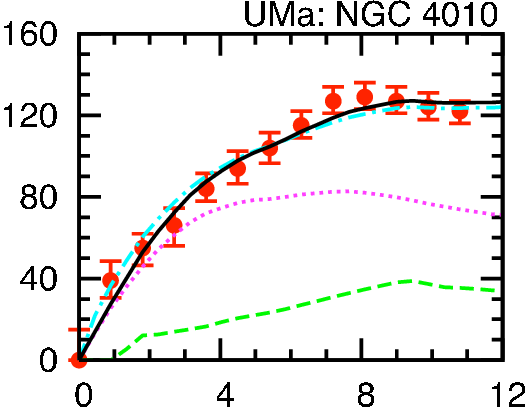}
\includegraphics{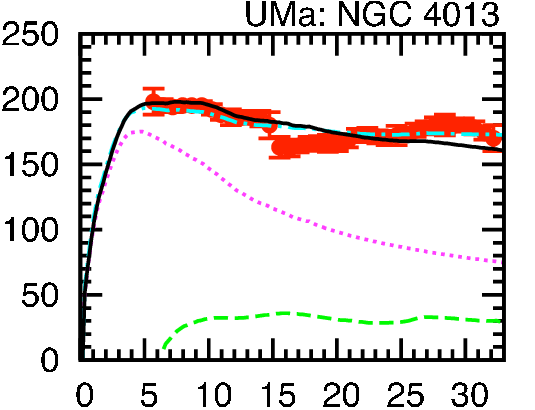}
\end{tabular}
\end{center}
\caption{Photometric Galaxy Rotation Curve Fits}
\end{figure}
\clearpage
\begin{figure}[p]
\figurenum{1 Continued}
\begin{center}
\begin{tabular}{c}
\includegraphics{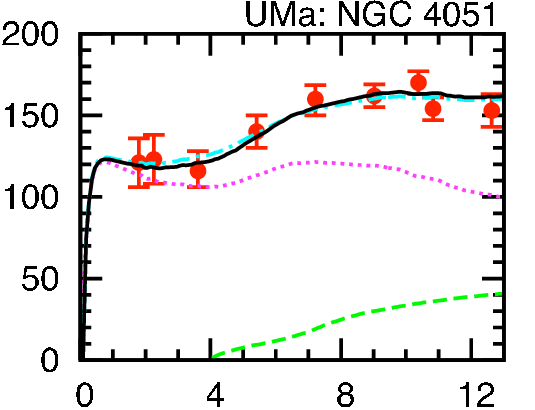}
\includegraphics{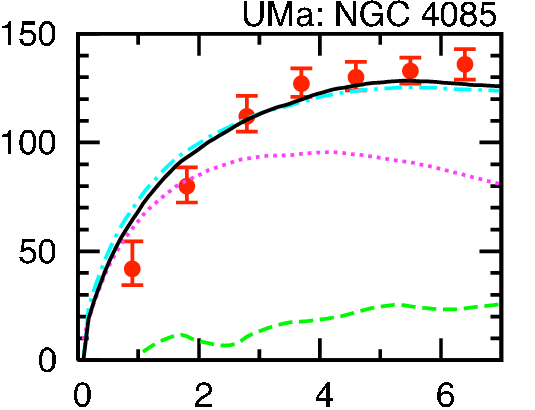}
\includegraphics{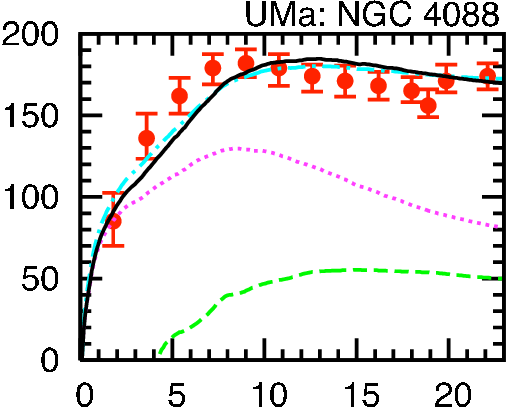} \\
\includegraphics{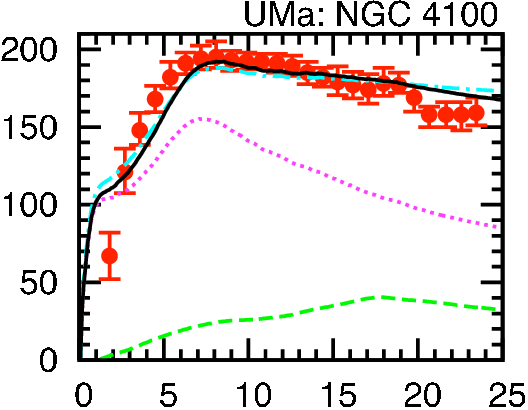}
\includegraphics{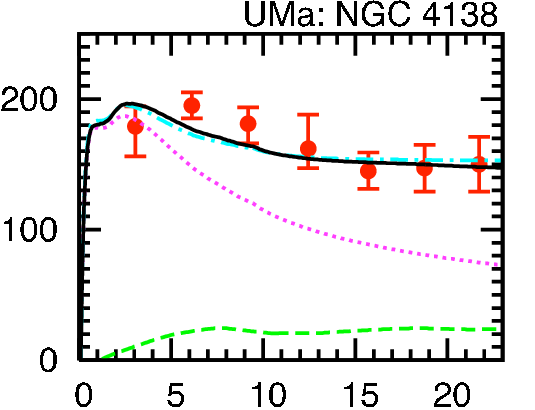}
\includegraphics{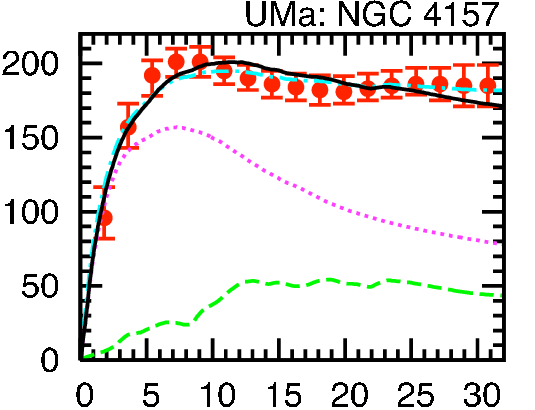} \\
\includegraphics{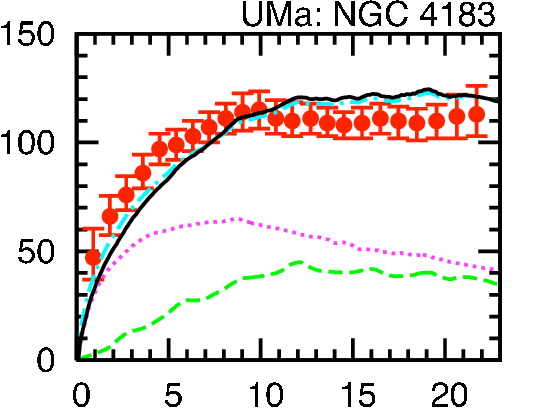}
\includegraphics{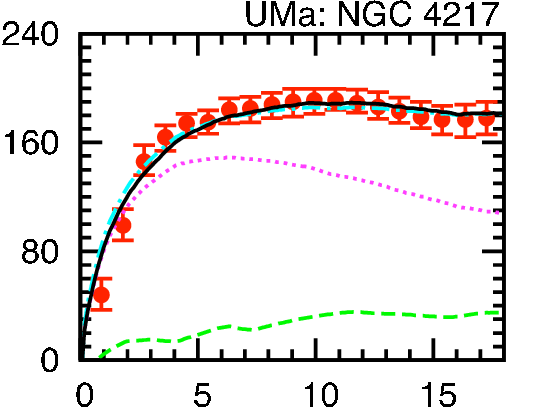}
\includegraphics{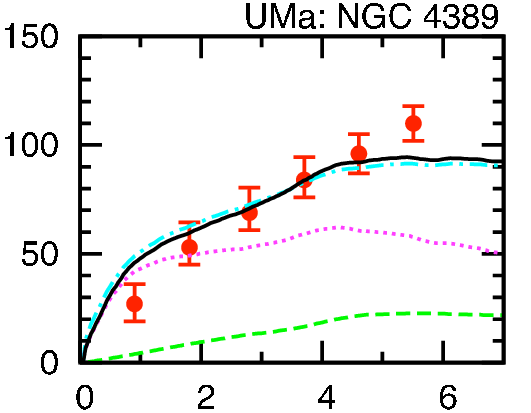} \\
\includegraphics{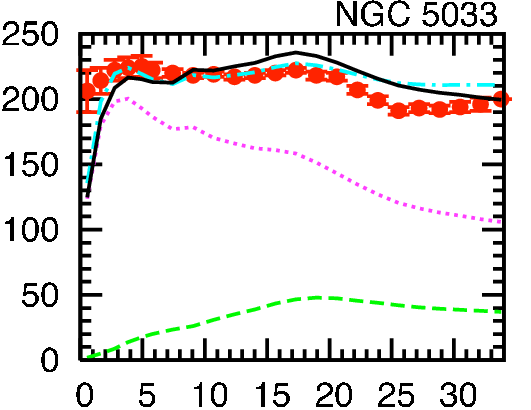}
\includegraphics{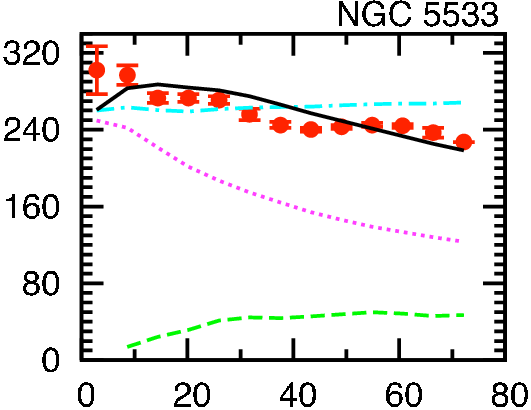}
\includegraphics{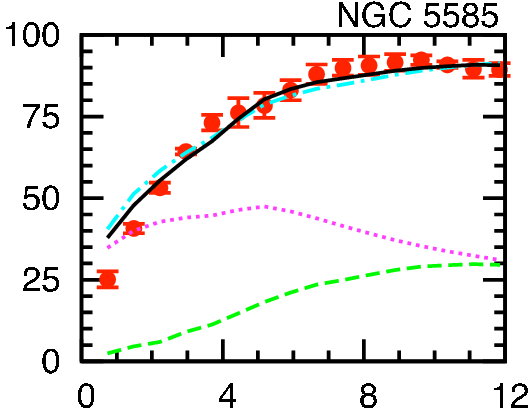}\\
\includegraphics{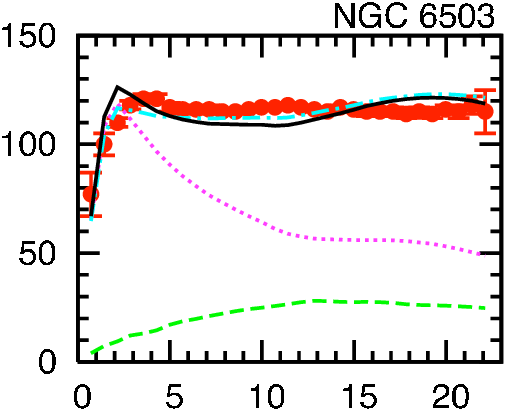}
\includegraphics{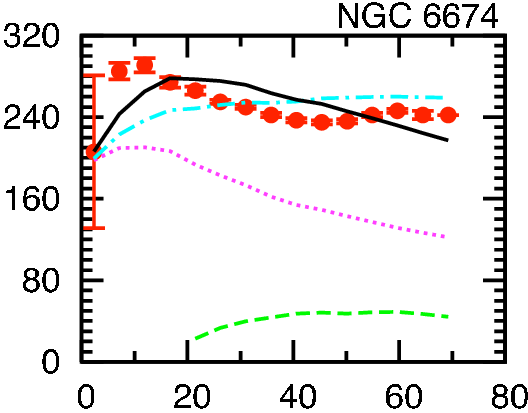}
\includegraphics{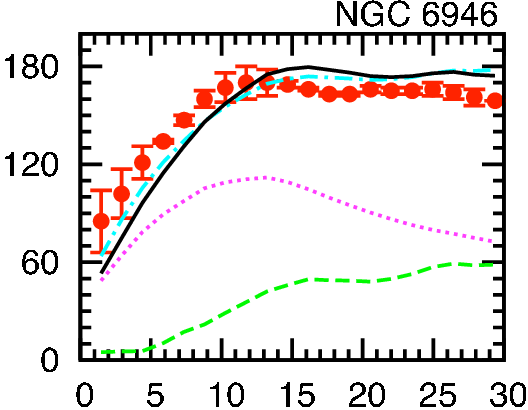}
\end{tabular}
\end{center}
\caption{Photometric Galaxy Rotation Curve Fits}
\end{figure}
\clearpage
\begin{figure}[p]
\figurenum{1 Continued}
\begin{center}
\begin{tabular}{c}
\includegraphics{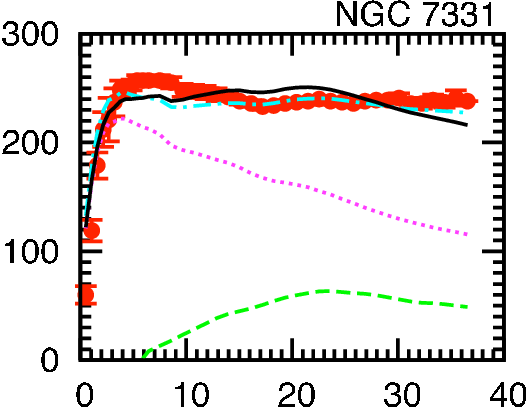}
\includegraphics{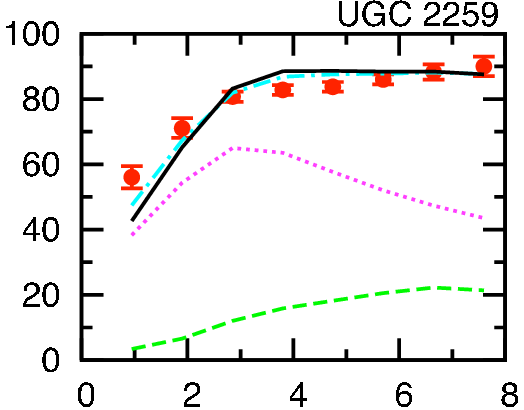}
\includegraphics{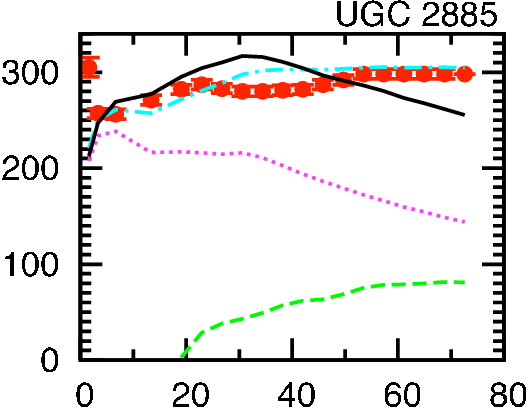} \\
\includegraphics{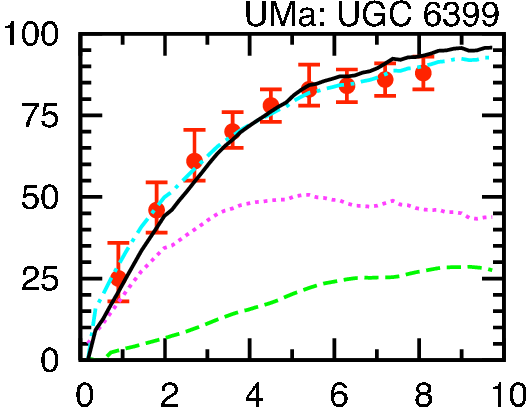}
\includegraphics{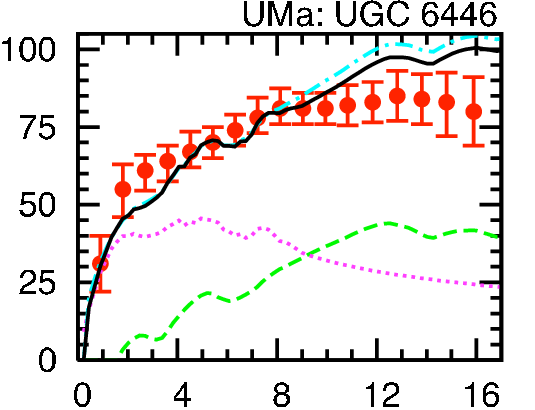}
\includegraphics{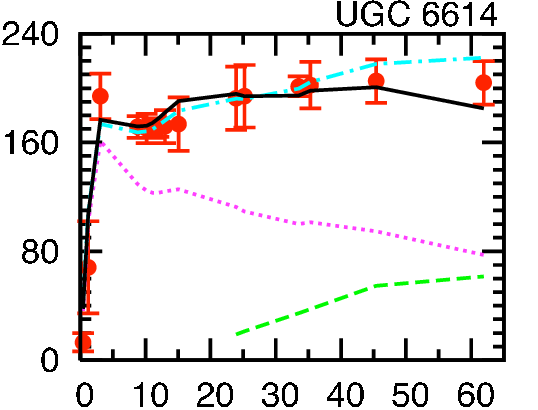} \\
\includegraphics{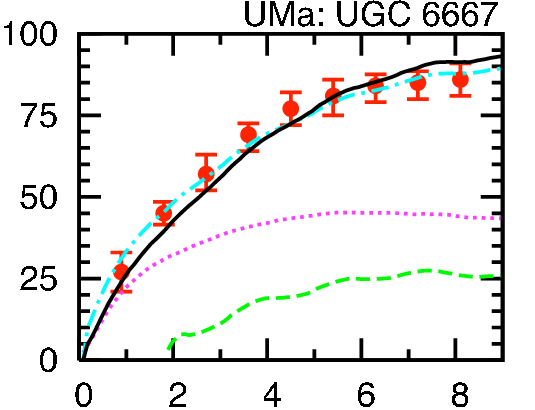}
\includegraphics{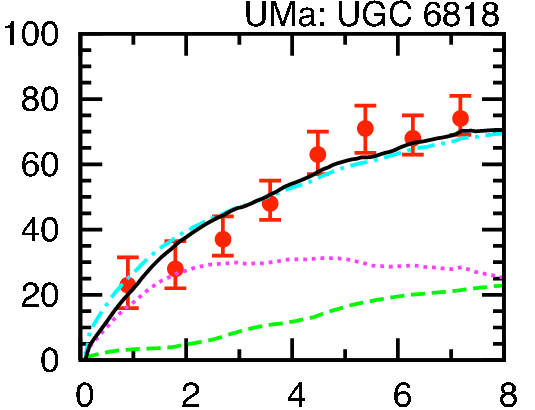}
\includegraphics{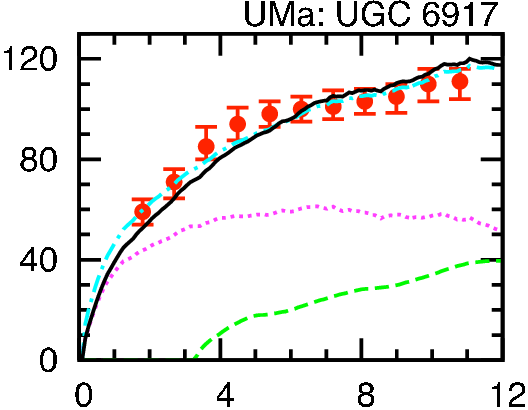} \\
\includegraphics{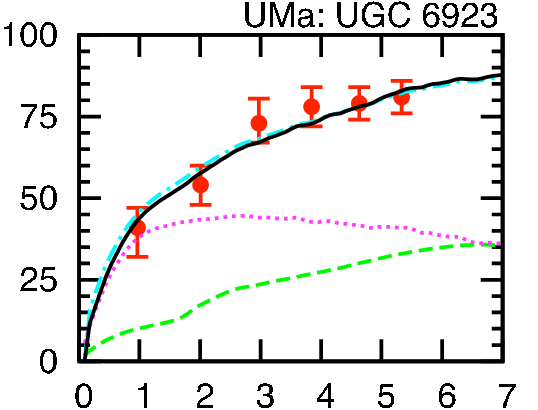}
\includegraphics{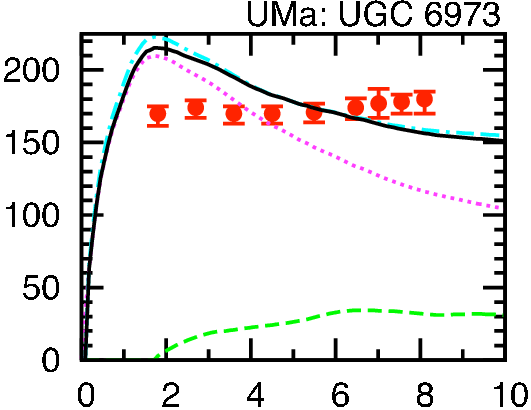}
\includegraphics{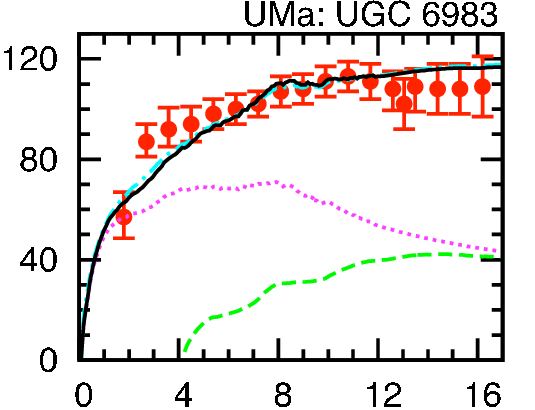}  \\
\includegraphics{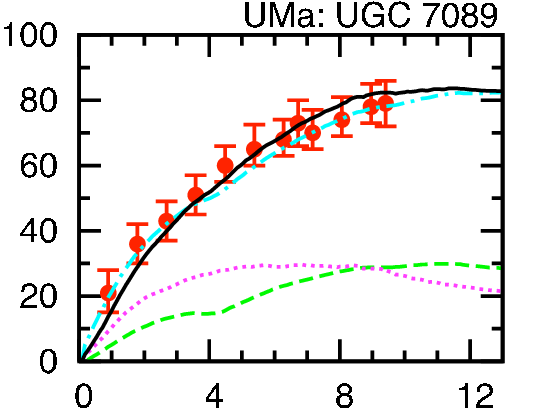}
\end{tabular}
\end{center}
\caption{Photometric Galaxy Rotation Curve Fits}
\end{figure}
\clearpage
\begin{figure}[p]
\figurenum{2}
\begin{center}
\begin{tabular}{c}
\includegraphics{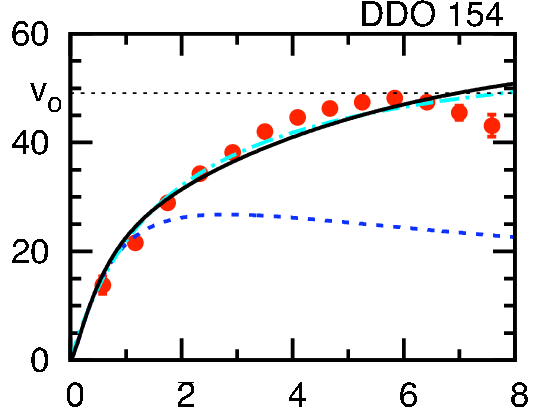}
\includegraphics{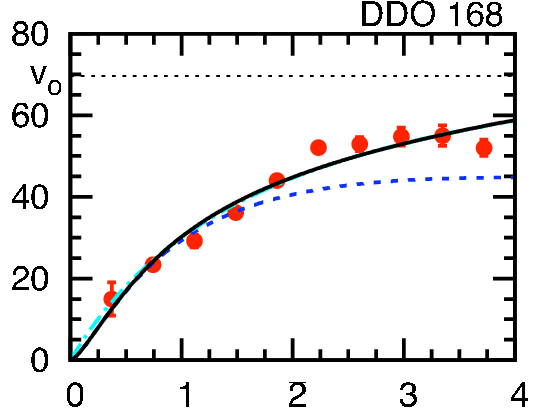}
\includegraphics{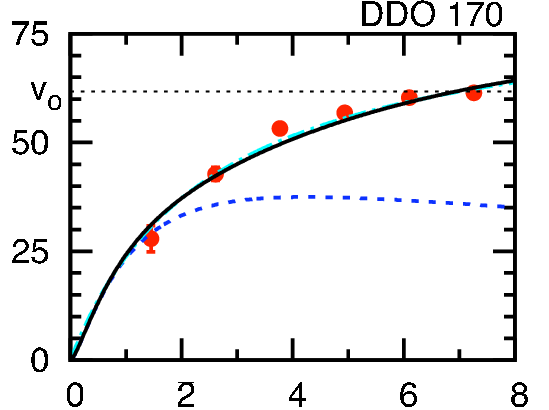} \\
\includegraphics{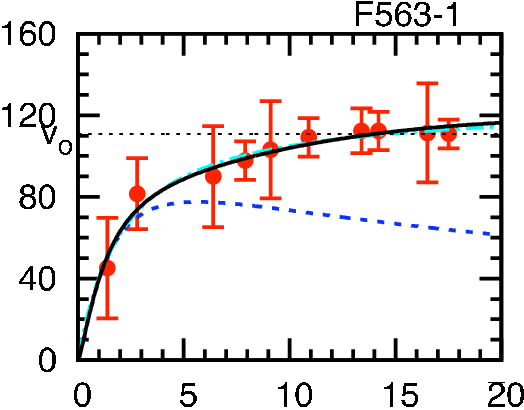}
\includegraphics{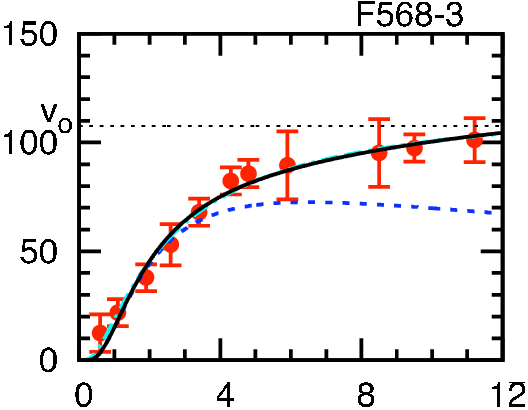}
\includegraphics{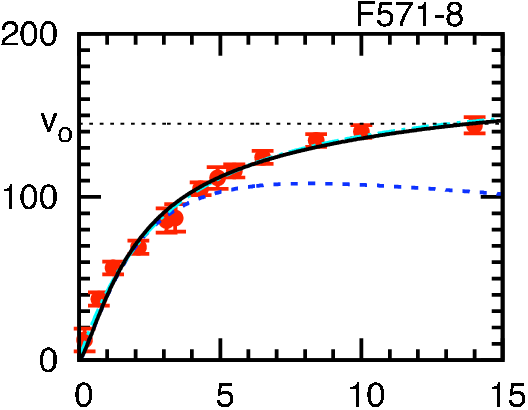} \\
\includegraphics{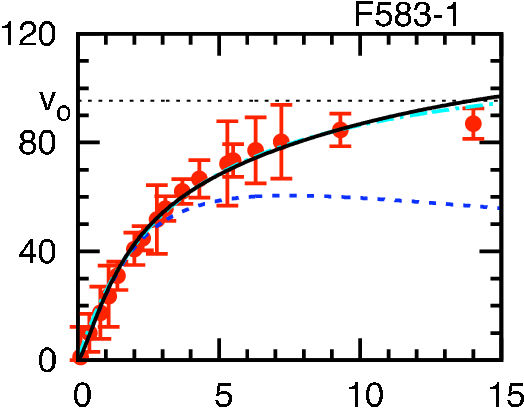}
\includegraphics{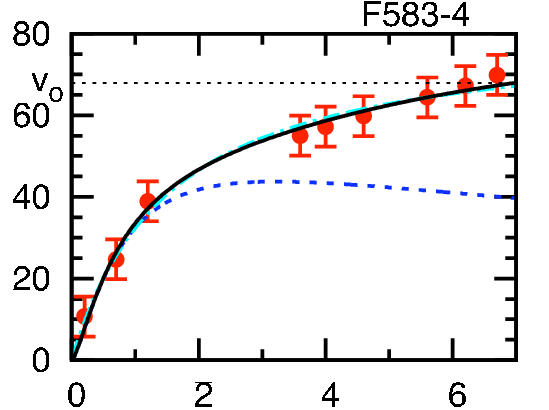}
\includegraphics{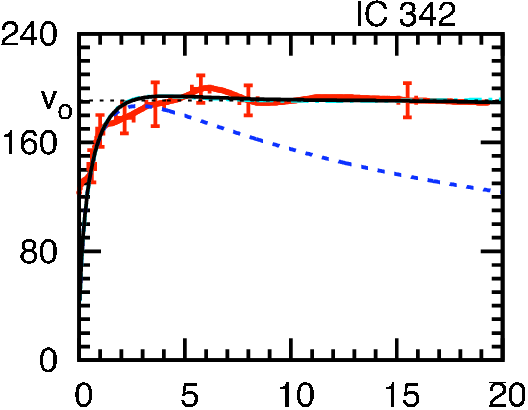}  \\
\includegraphics{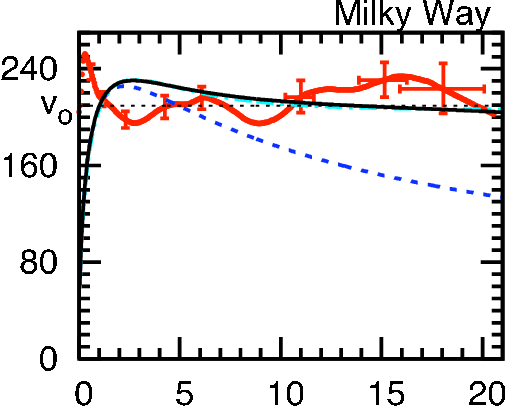}
\includegraphics{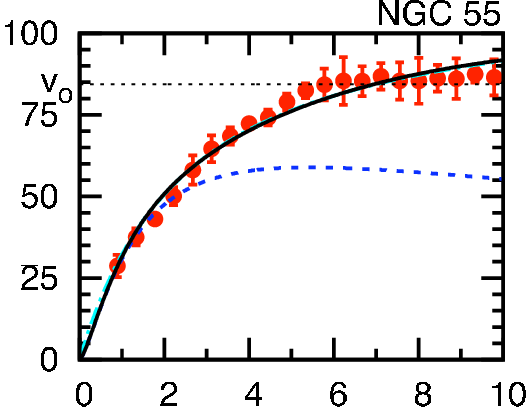}
\includegraphics{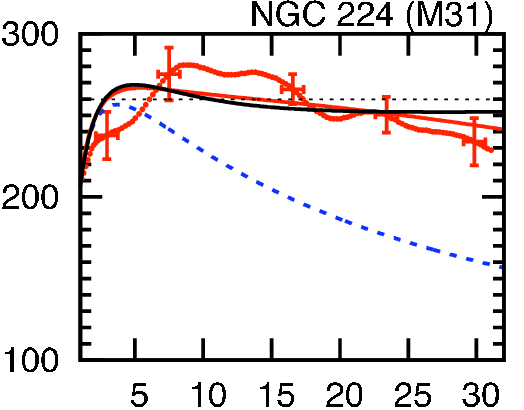}   \\
\includegraphics{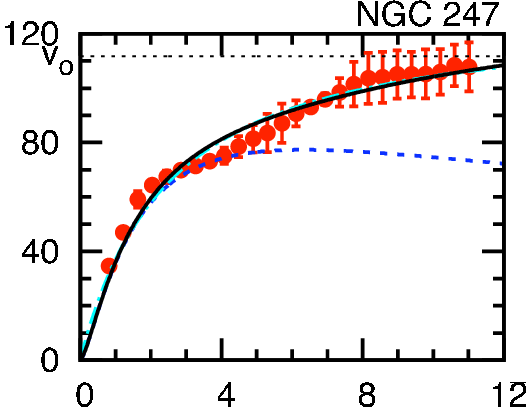}
\includegraphics{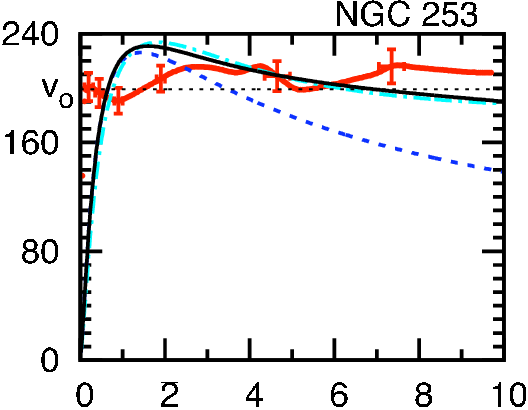}
\includegraphics{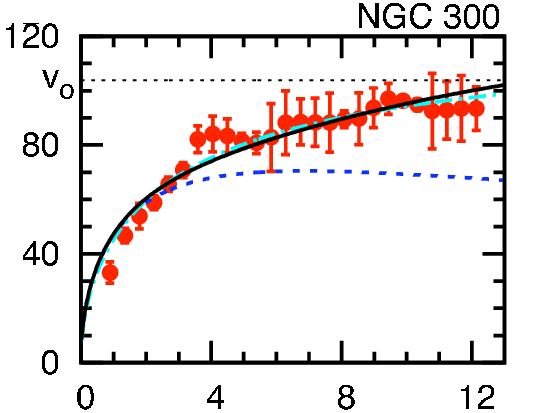}
\end{tabular}
\end{center}
\caption{Parametric Galaxy Rotation Curve Fits}
\end{figure}
\clearpage
\begin{figure}[p]
\figurenum{2 Continued}
\begin{center}
\begin{tabular}{c}
\includegraphics{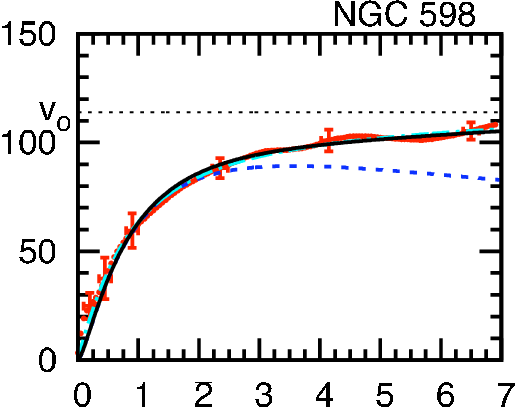}
\includegraphics{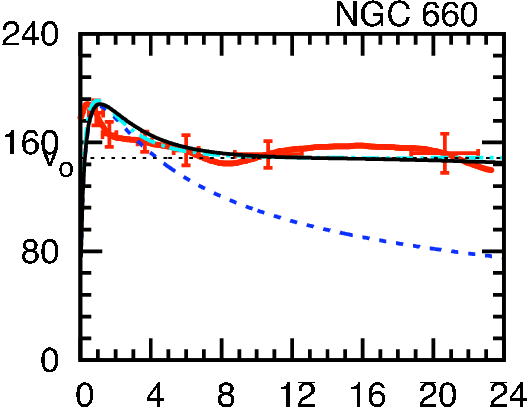}
\includegraphics{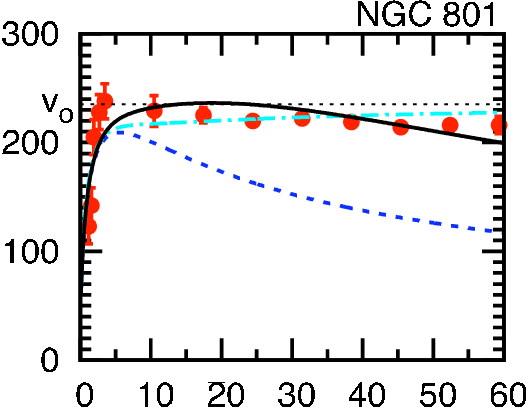}  \\
\includegraphics{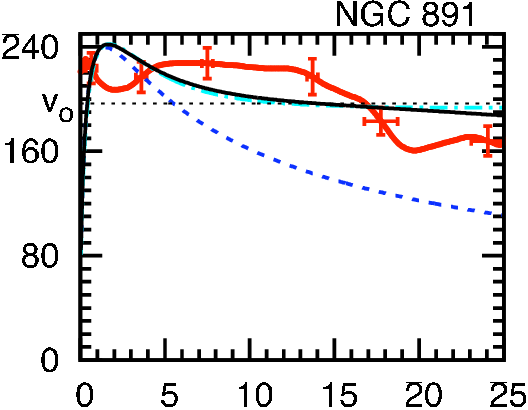}
\includegraphics{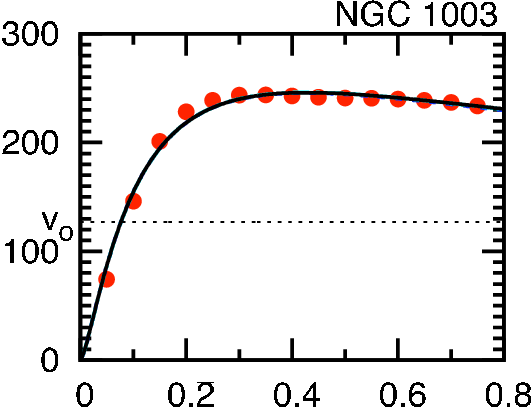}
\includegraphics{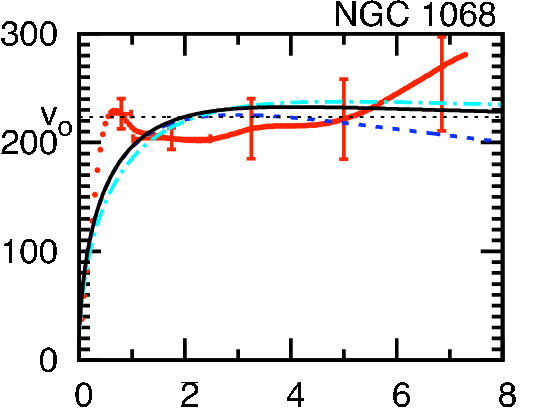}  \\
\includegraphics{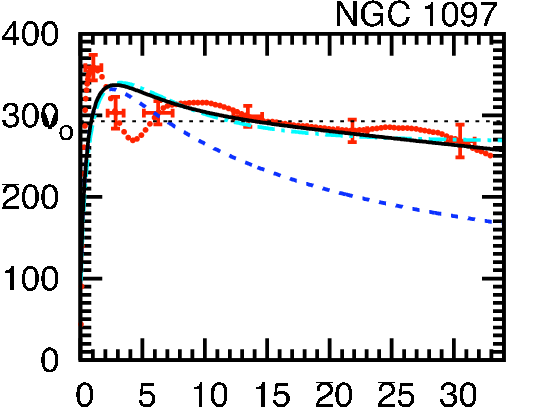}
\includegraphics{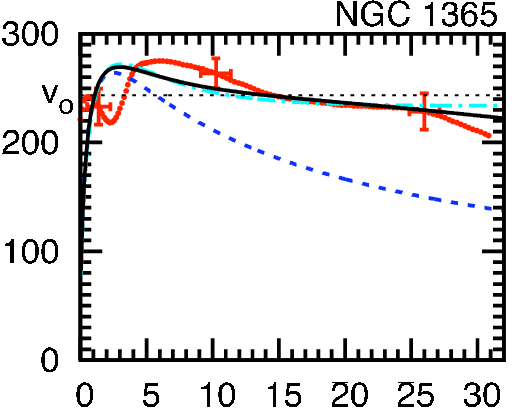}
\includegraphics{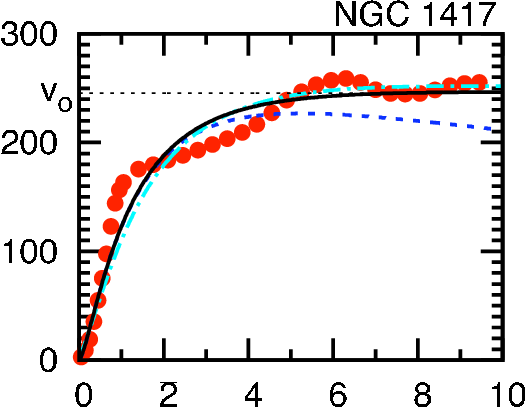}  \\
\includegraphics{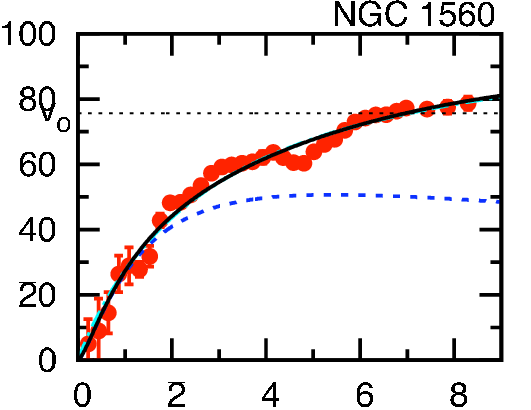}
\includegraphics{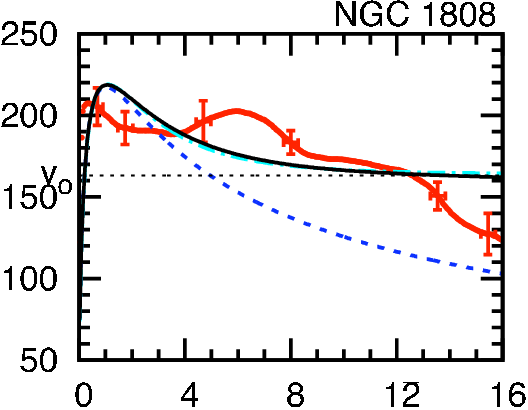}
\includegraphics{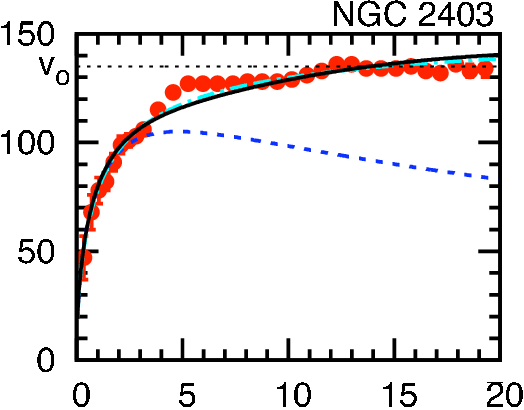}   \\
\includegraphics{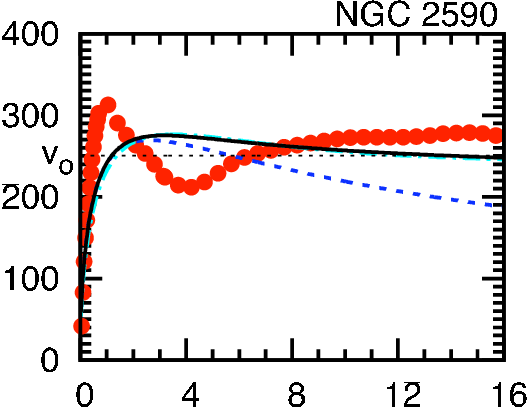}
\includegraphics{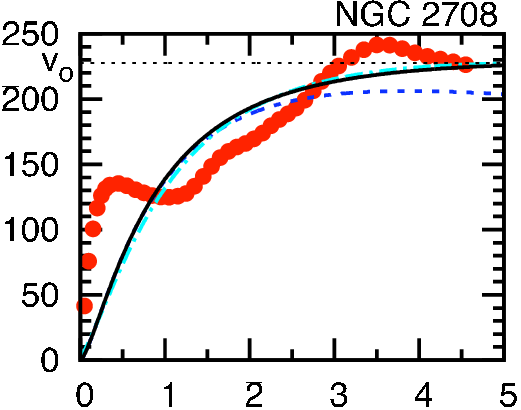}
\includegraphics{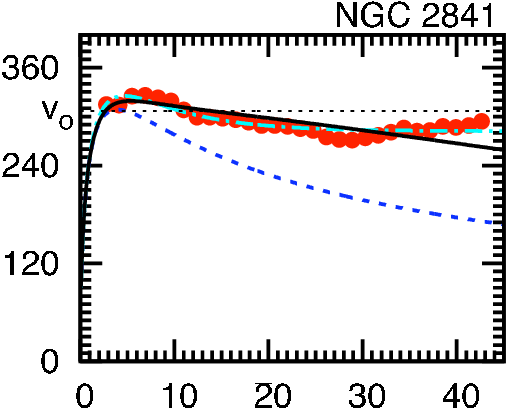}
\end{tabular}
\end{center}
\caption{Parametric Galaxy Rotation Curve Fits}
\end{figure}
\clearpage
\begin{figure}[p]
\figurenum{2 Continued}
\begin{center}
\begin{tabular}{c}
\includegraphics{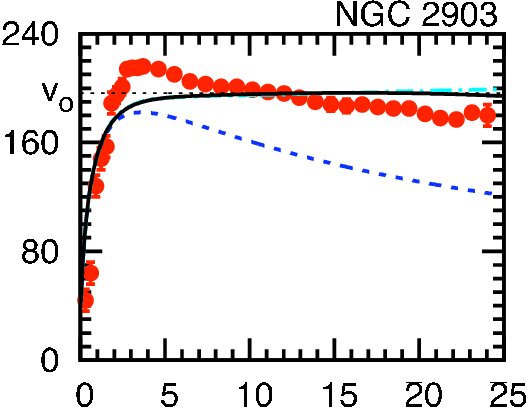}
\includegraphics{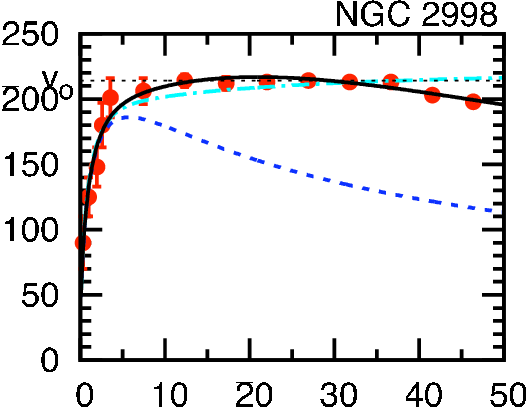}
\includegraphics{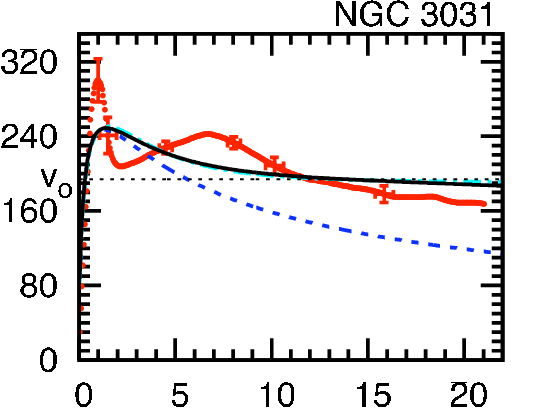}  \\
\includegraphics{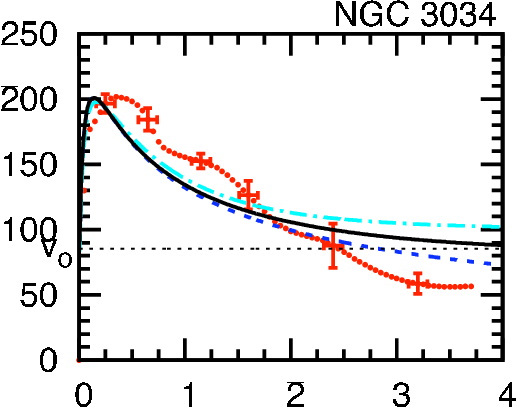}
\includegraphics{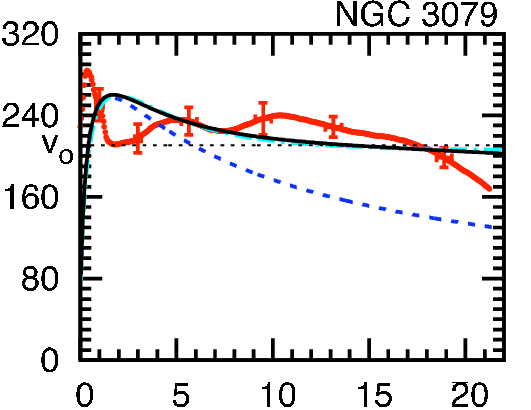}
\includegraphics{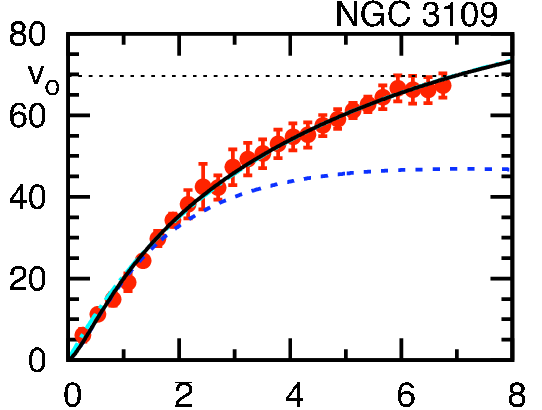}  \\
\includegraphics{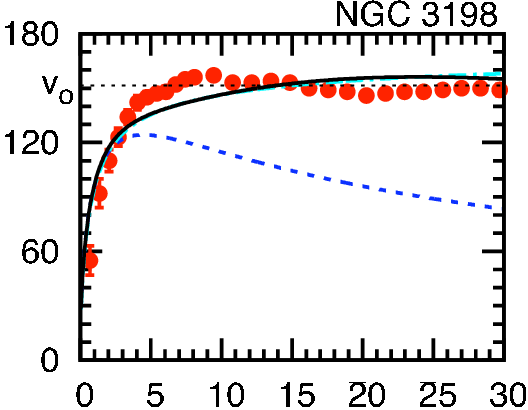}
\includegraphics{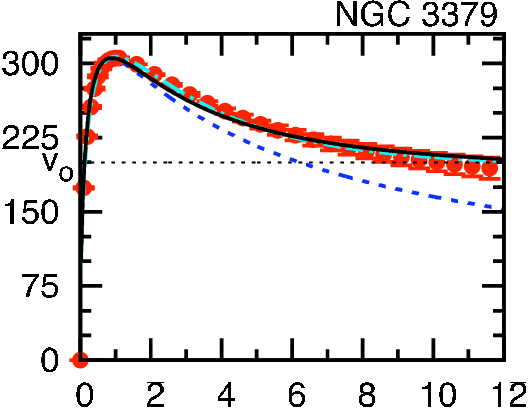}
\includegraphics{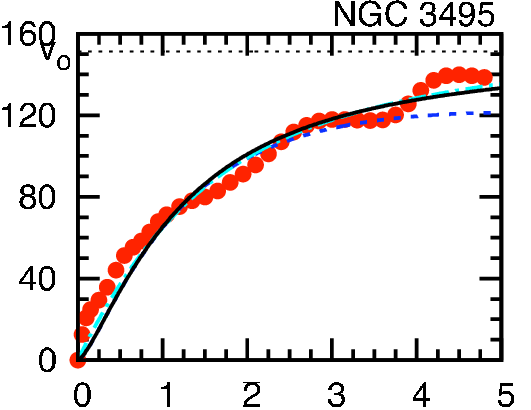}  \\
\includegraphics{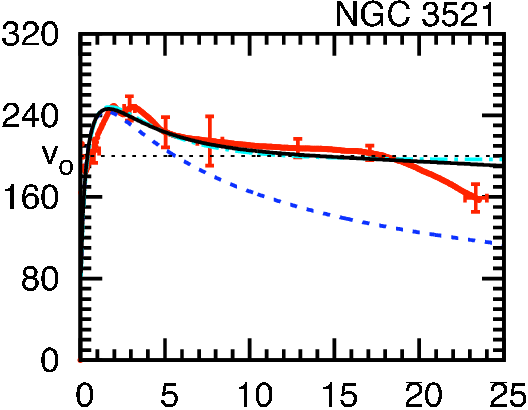}
\includegraphics{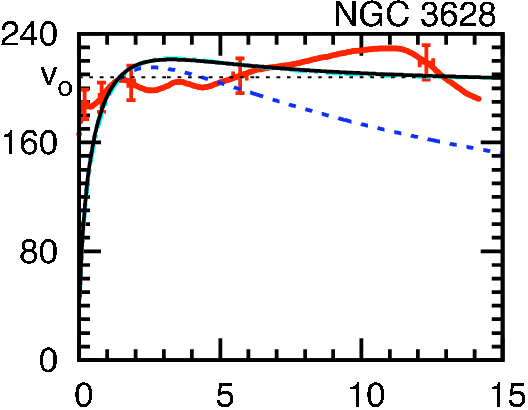}
\includegraphics{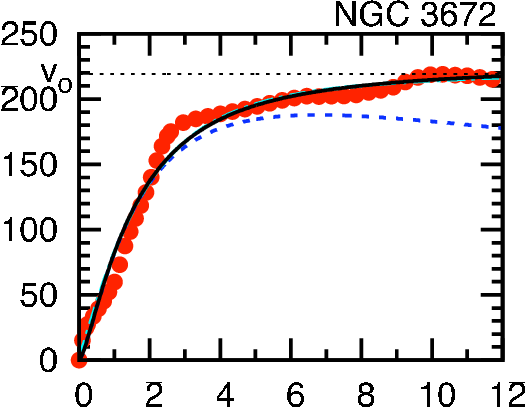}  \\
\includegraphics{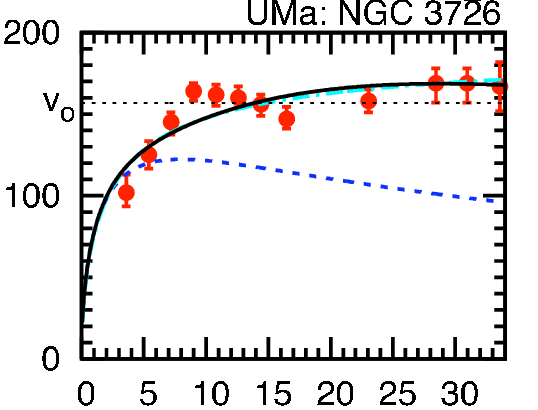}
\includegraphics{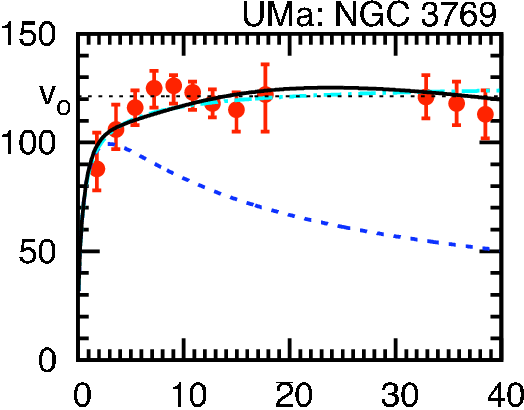}
\includegraphics{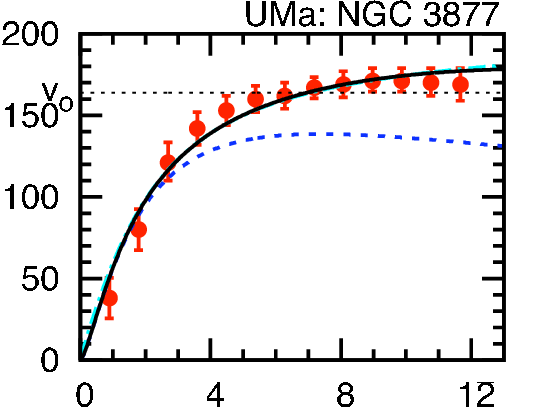}
\end{tabular}
\end{center}
\caption{Parametric Galaxy Rotation Curve Fits}
\end{figure}
\clearpage
\begin{figure}[p]
\figurenum{2 Continued}
\begin{center}
\begin{tabular}{c}
\includegraphics{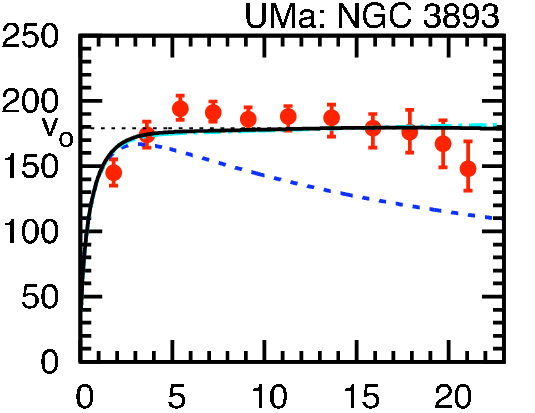}
\includegraphics{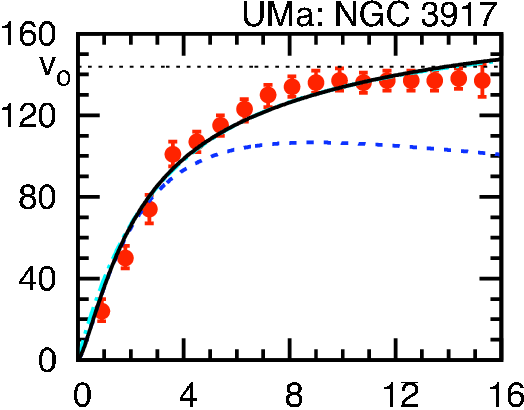}
\includegraphics{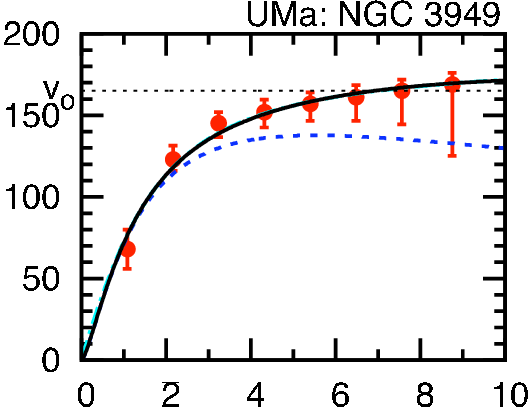} \\
\includegraphics{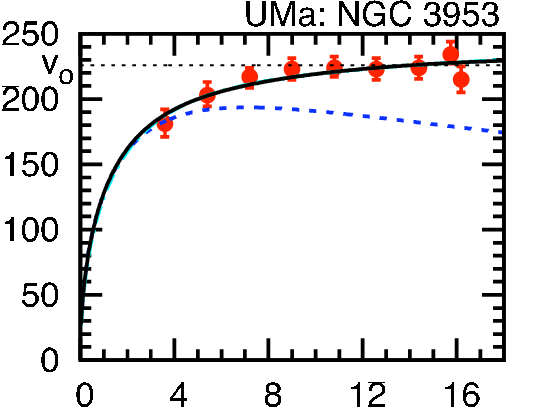}
\includegraphics{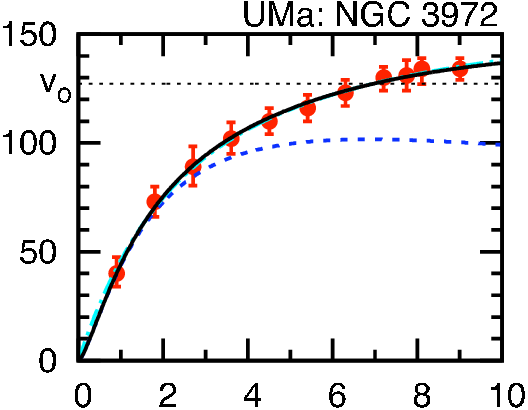}
\includegraphics{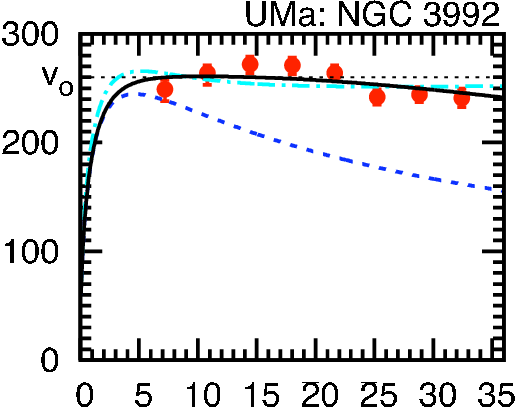} \\
\includegraphics{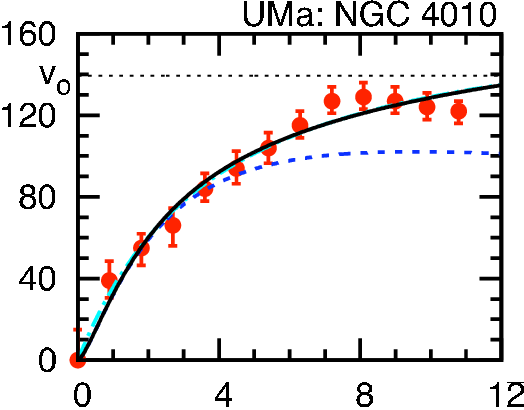}
\includegraphics{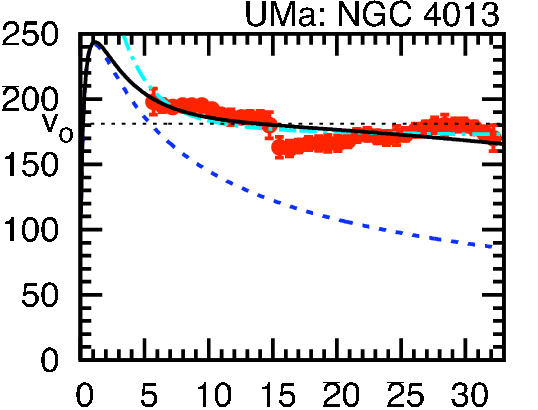}
\includegraphics{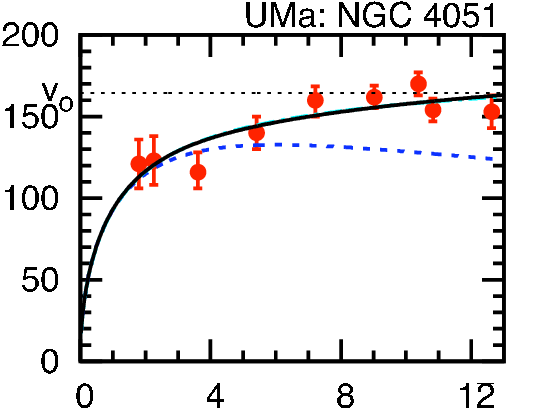} \\
\includegraphics{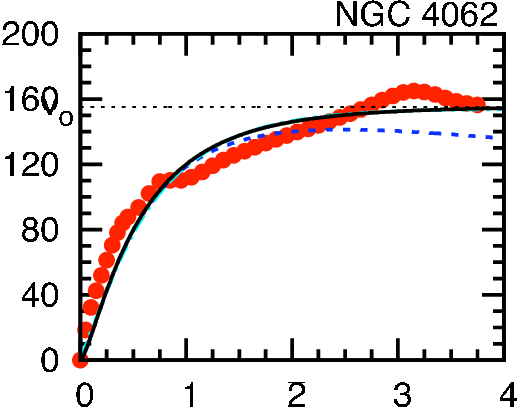}
\includegraphics{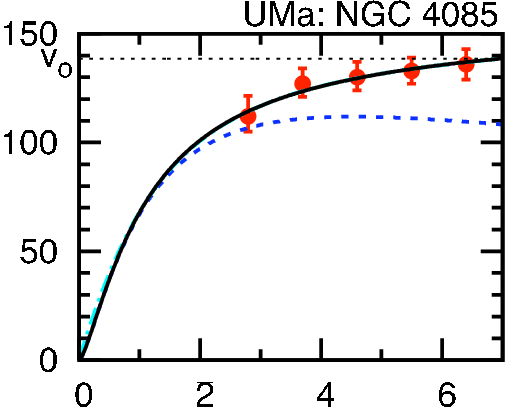}
\includegraphics{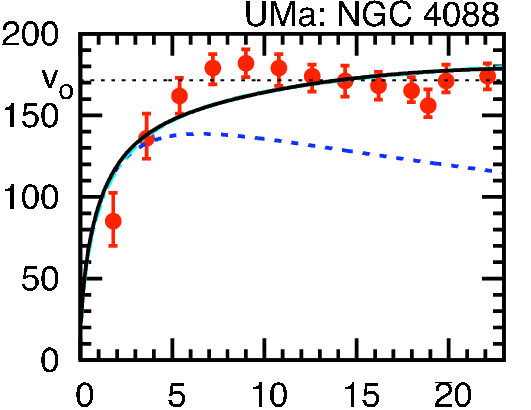}  \\
\includegraphics{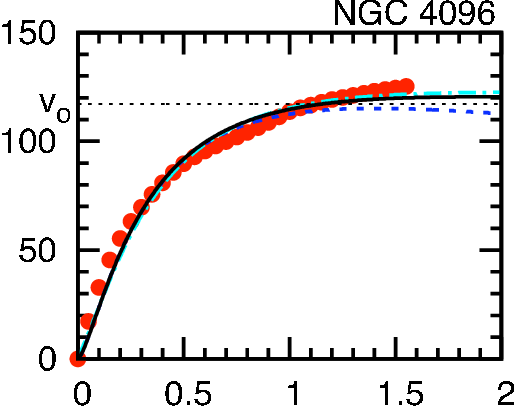}
\includegraphics{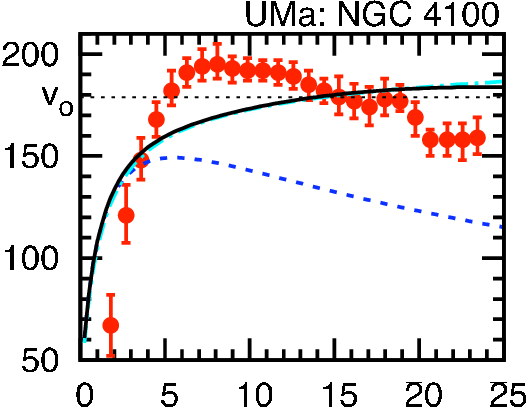}
\includegraphics{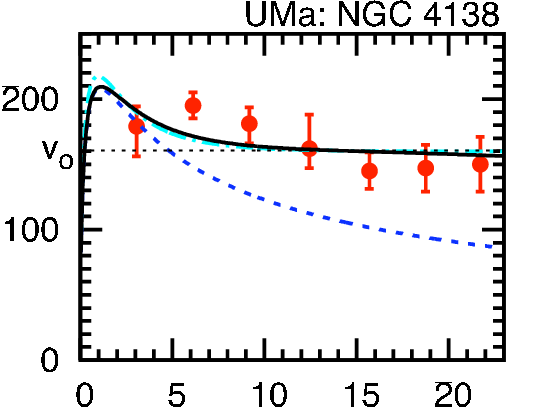}
\end{tabular}
\end{center}
\caption{Parametric Galaxy Rotation Curve Fits}
\end{figure}
\clearpage
\begin{figure}[p]
\figurenum{2 Continued}
\begin{center}
\begin{tabular}{c}
\includegraphics{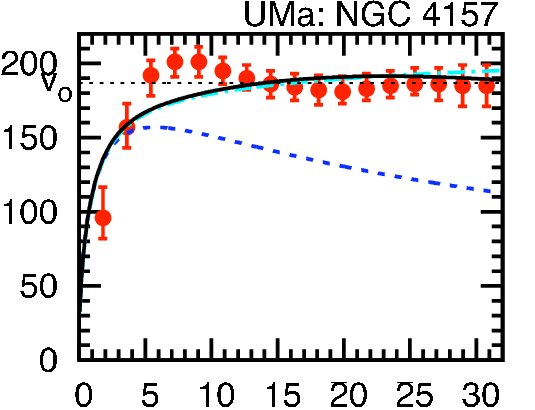}
\includegraphics{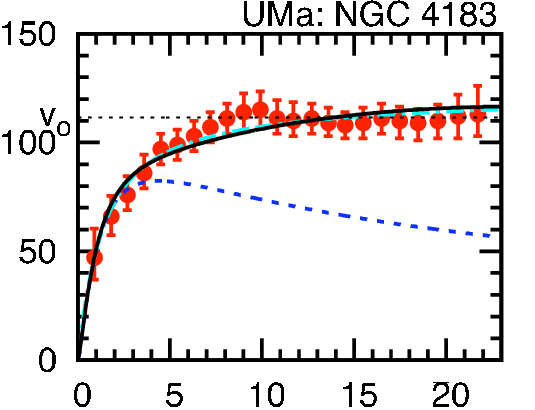}
\includegraphics{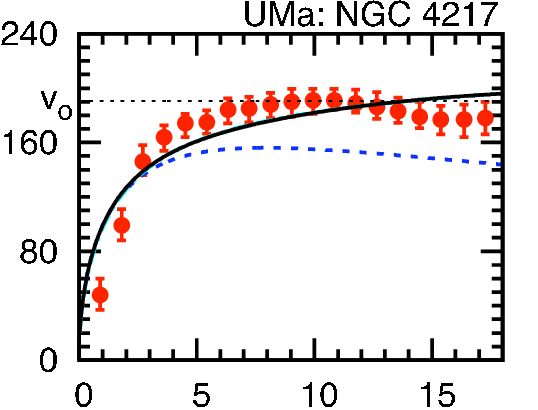}  \\
\includegraphics{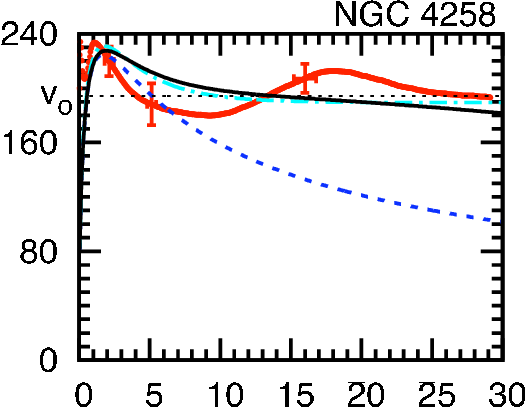}
\includegraphics{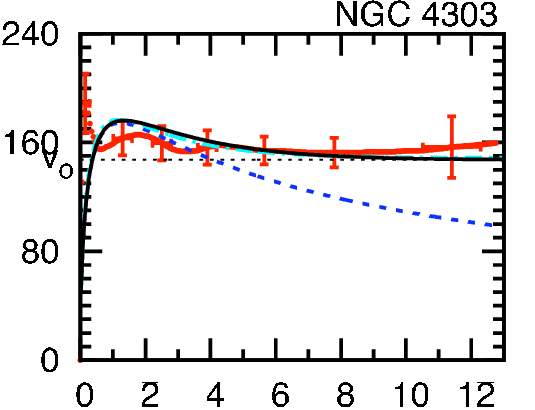}
\includegraphics{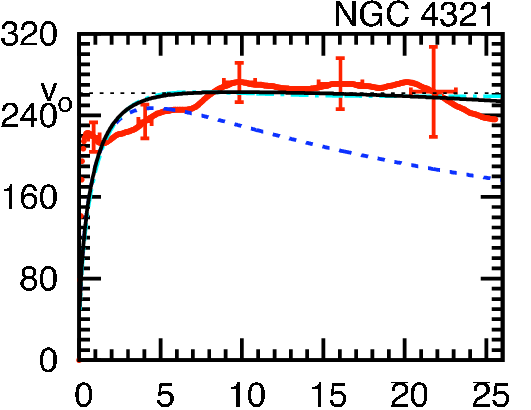}  \\
\includegraphics{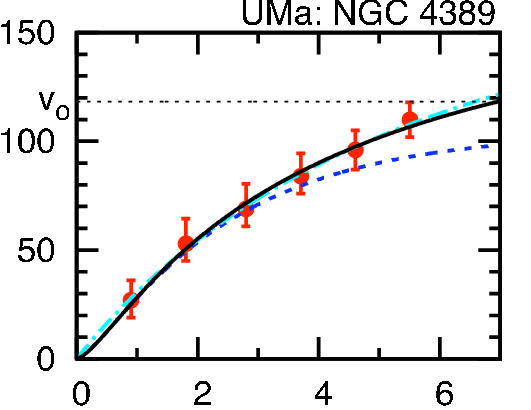}
\includegraphics{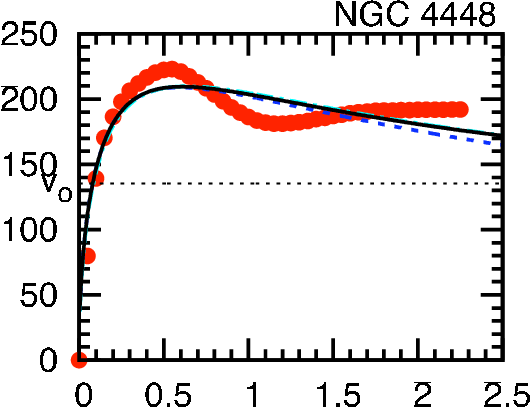}
\includegraphics{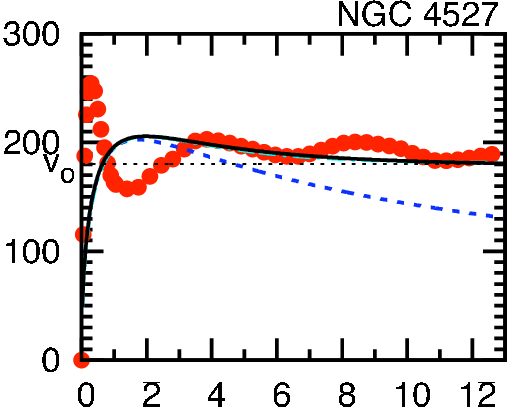}  \\
\includegraphics{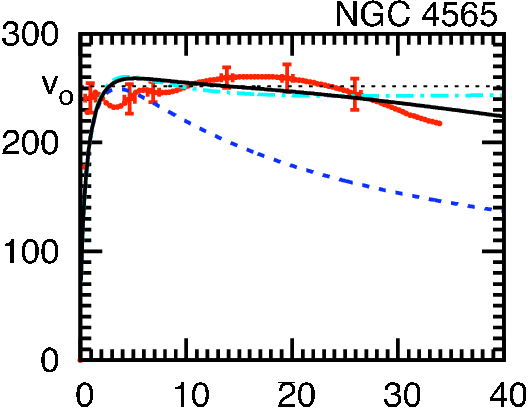}
\includegraphics{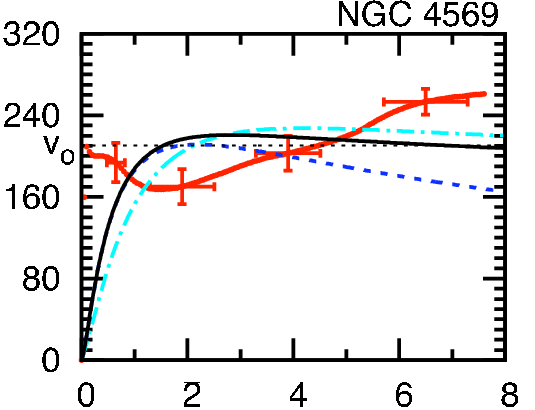}
\includegraphics{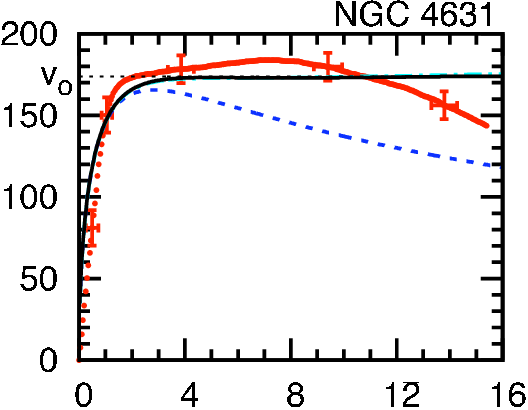}  \\
\includegraphics{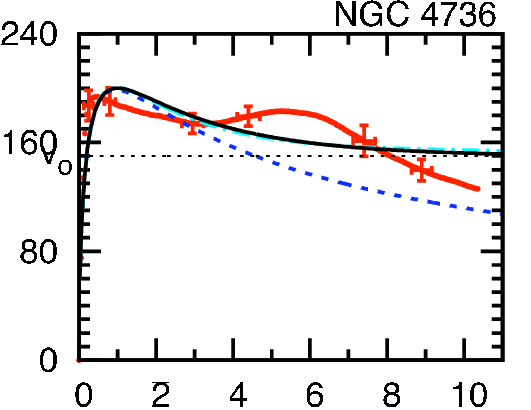}
\includegraphics{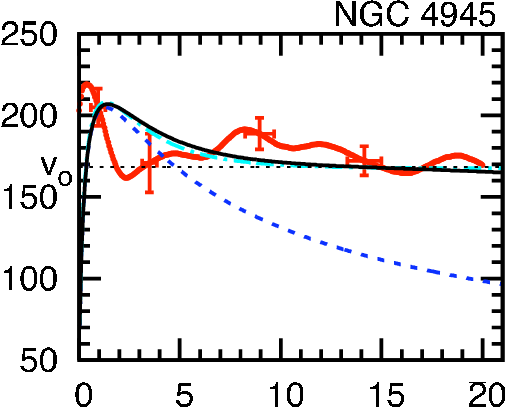}
\includegraphics{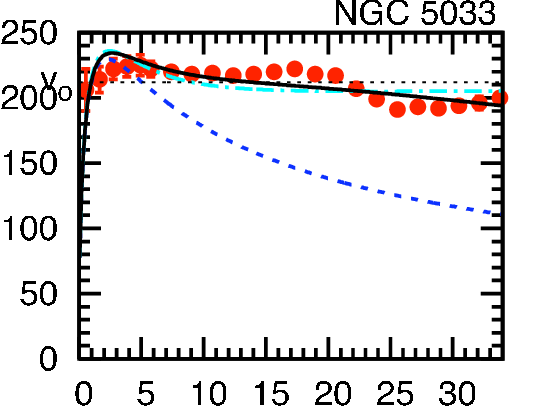}
\end{tabular}
\end{center}
\caption{Parametric Galaxy Rotation Curve Fits}
\end{figure}
\clearpage
\begin{figure}[p]
\figurenum{2 Continued}
\begin{center}
\begin{tabular}{c}
\includegraphics{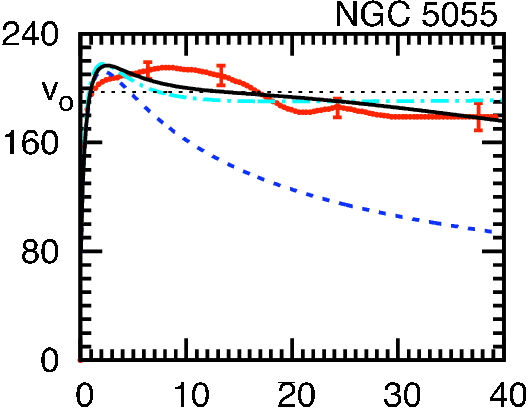}
\includegraphics{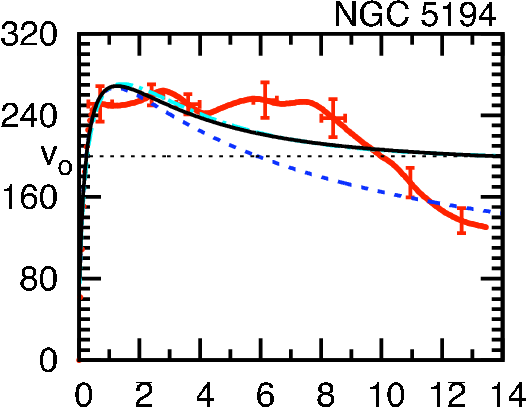}
\includegraphics{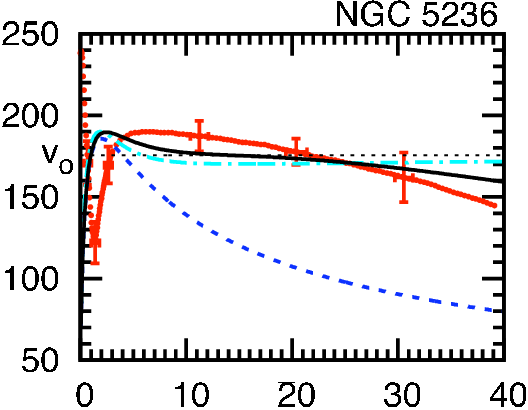}  \\
\includegraphics{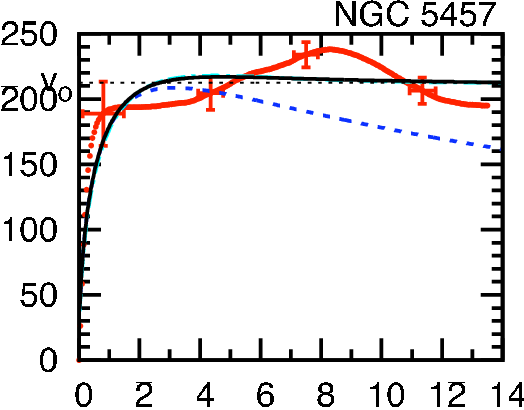}
\includegraphics{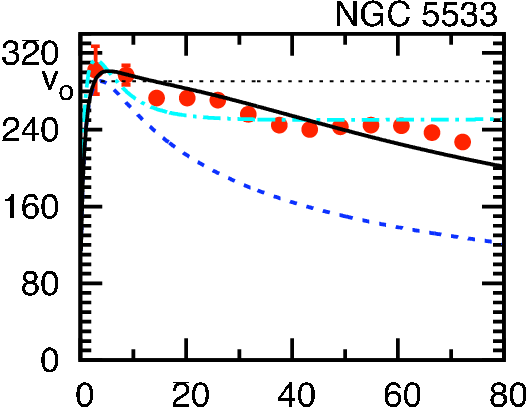}
\includegraphics{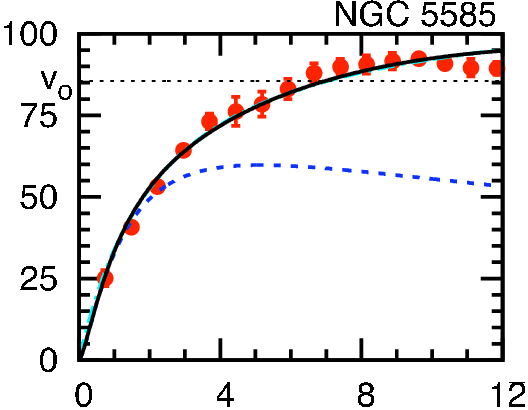}  \\
\includegraphics{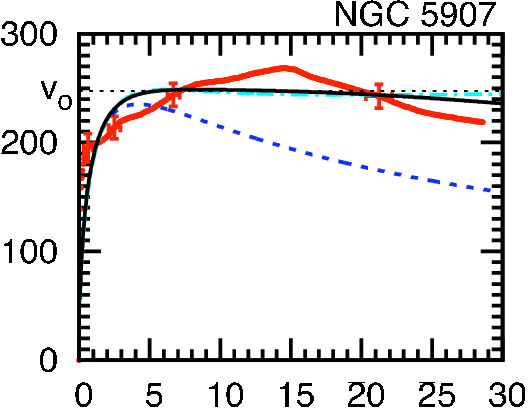}
\includegraphics{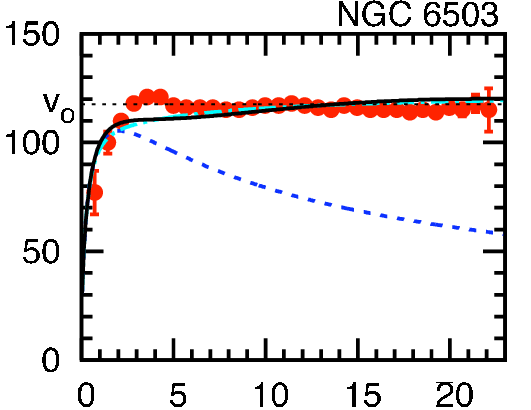}
\includegraphics{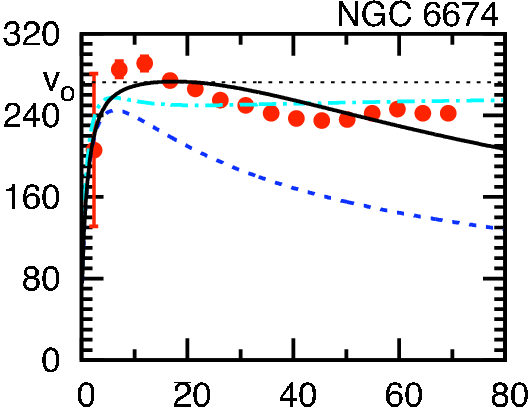}  \\
\includegraphics{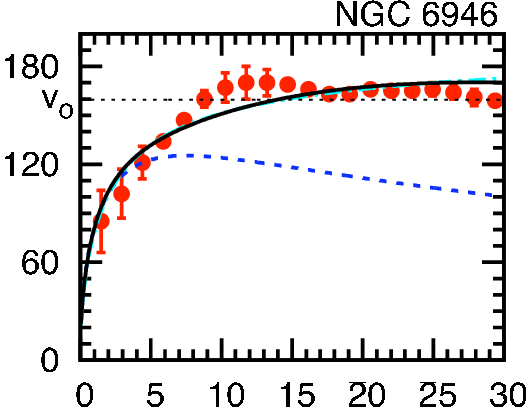}
\includegraphics{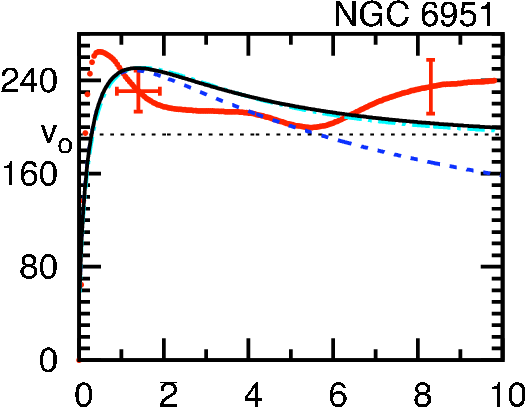}
\includegraphics{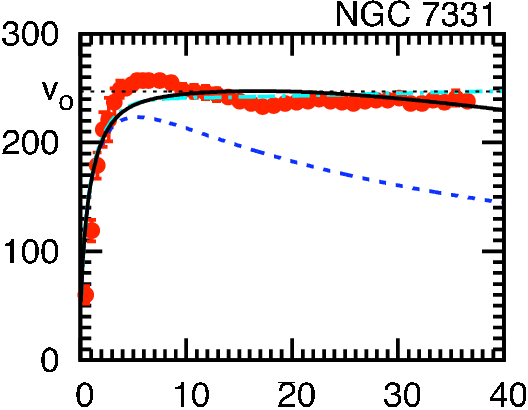}  \\
\includegraphics{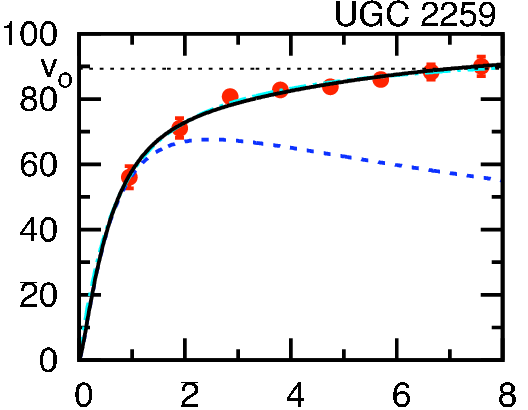}
\includegraphics{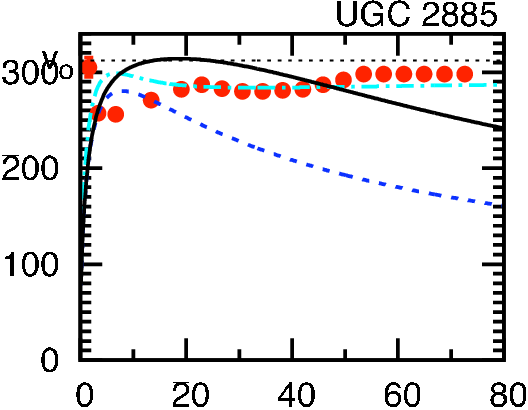}
\includegraphics{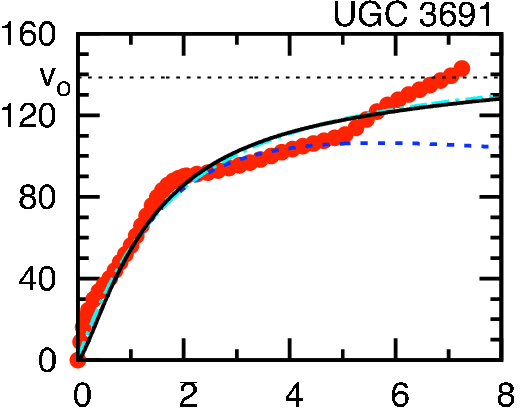}
\end{tabular}
\end{center}
\caption{Parametric Galaxy Rotation Curve Fits}
\end{figure}
\clearpage
\begin{figure}[p]
\figurenum{2 Continued}
\begin{center}
\begin{tabular}{c}
\includegraphics{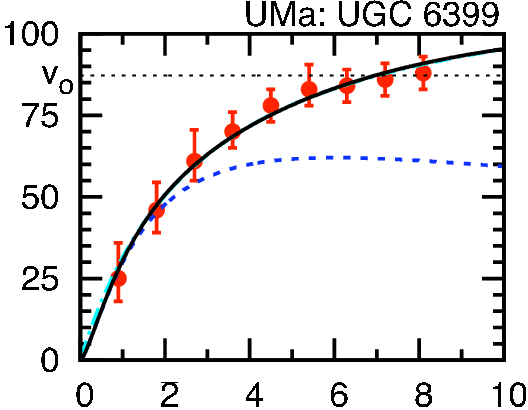}
\includegraphics{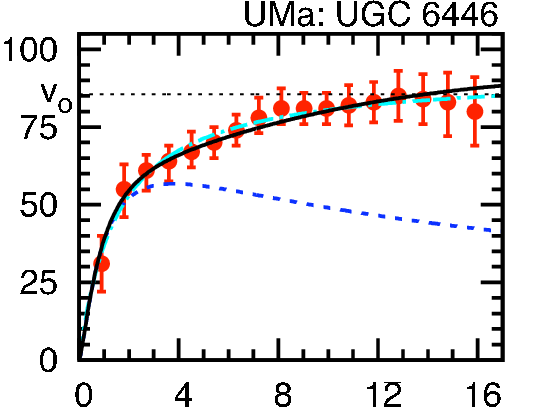}
\includegraphics{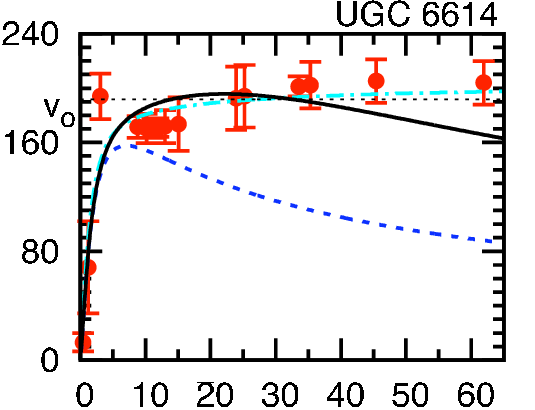}  \\
\includegraphics{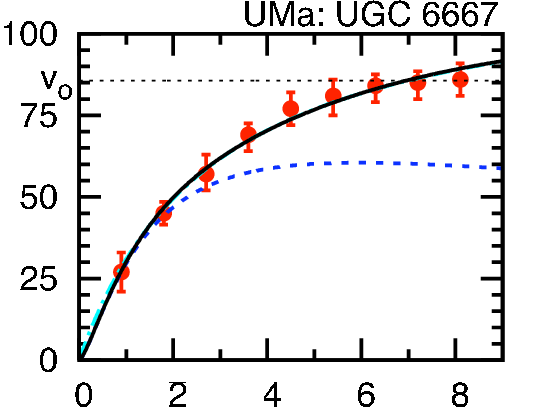}
\includegraphics{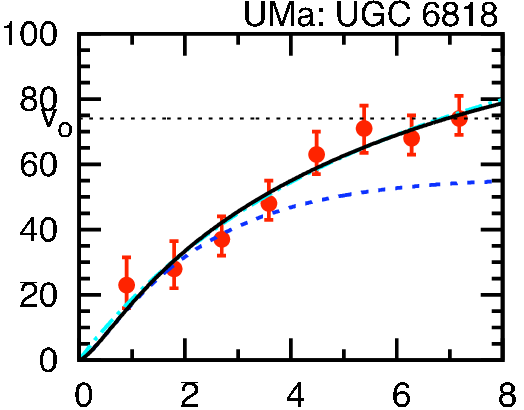}
\includegraphics{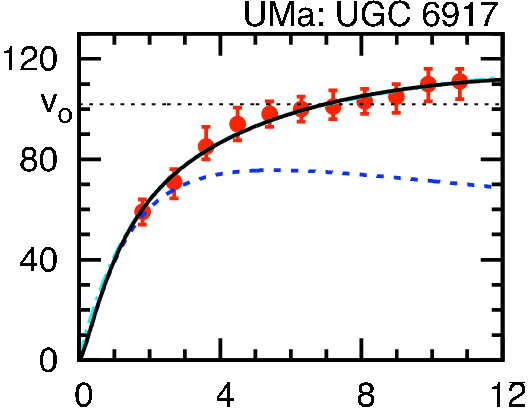} \\
\includegraphics{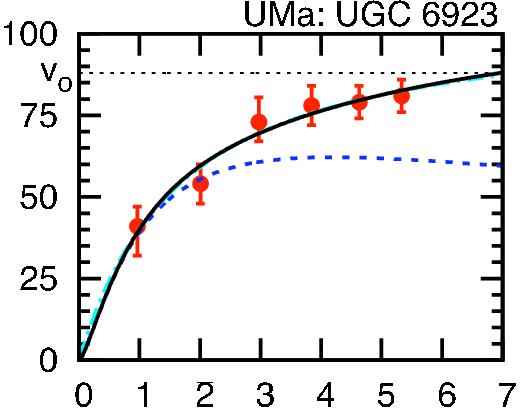}
\includegraphics{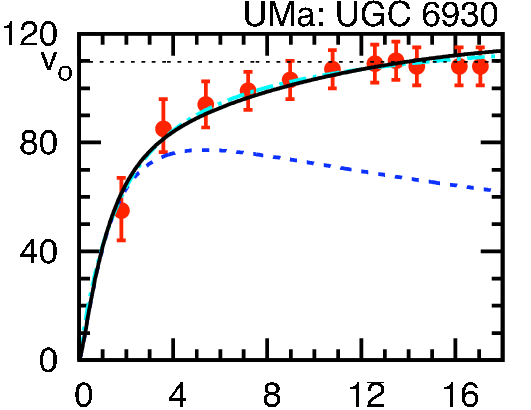}
\includegraphics{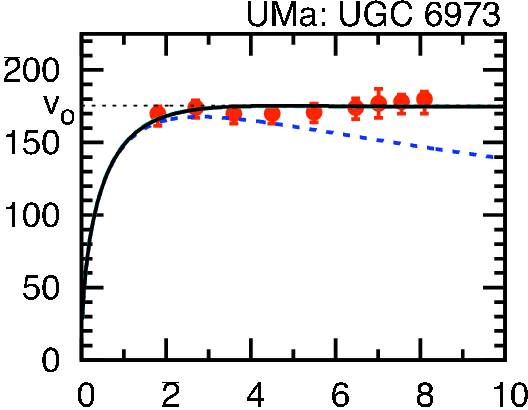}  \\
\includegraphics{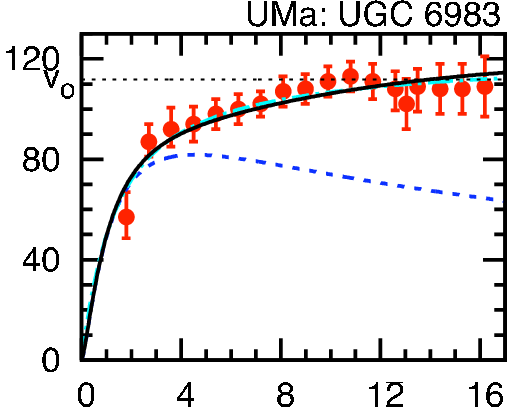}
\includegraphics{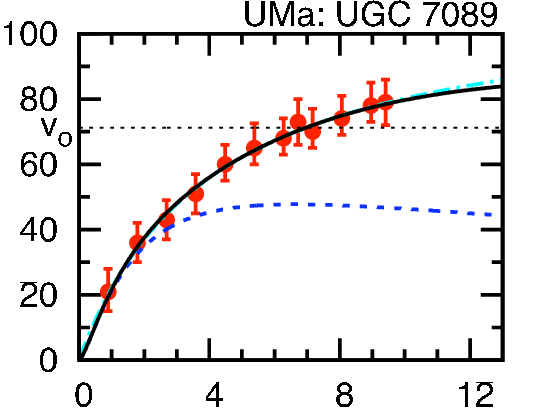}
\end{tabular}
\end{center}
\caption{Parametric Galaxy Rotation Curve Fits}
\end{figure}
\clearpage

\clearpage

\clearpage

\clearpage

\clearpage

\begin{deluxetable}{llcrrrrrrrrc}
\tabletypesize{\scriptsize}
\tablecaption{\sc Galaxy Properties of the {\em Complete Sample} \label{CompleteSample}}
\tablewidth{0pt}
\tablehead{\colhead{Galaxy}&\colhead{Type}&\colhead{Surface}&\colhead{$L_{B}$}&\colhead{$L_{K}$}&\colhead{$r_{\rm out}$}&\colhead{$v_{\rm out}$}&\colhead{Distance}&\colhead{Redshift}&\colhead{Ref}\\
&&\colhead{Brightness}&\colhead{\scriptsize ($10^{\tiny 10} L_{\sun})$}&\colhead{\scriptsize $(10^{\tiny 10} L_{\sun})$}&\colhead{\scriptsize (kpc)}&\colhead{\scriptsize (km s$^{\tiny -1}$)}&\colhead{\scriptsize (Mpc)}&&\\
\colhead{\scriptsize (1)}&\colhead{\scriptsize (2)}&\colhead{\scriptsize (3)}&\colhead{\scriptsize (4)}&\colhead{\scriptsize (5)}&\colhead{\scriptsize (6)}&\colhead{\scriptsize (7)}&\colhead{\scriptsize (8)}&\colhead{\scriptsize (9)}&\colhead{\scriptsize (10)}}
\tablecolumns{10}
\startdata
DDO~154&IB&LSB&0.005&5.392&7.6&$43.1\pm2.0$&4.0&0.000103&1\\
DDO~168&Irr&LSB&0.022&\nodata&3.7&$52.0\pm2.0$&3.8&0.000640&2\\
DDO~170&Im&\nodata&0.016&\nodata&9.6&$66.0\pm1.0$&12.0&0.003119&1\\
F563-1&Sm&LSB&0.135&0.100&17.5&$110.9\pm7.0$&45.0&\nodata&3\\
F568-3&Sd&LSB&0.331&0.272&11.2&$101.0\pm10.2$&77.0&\nodata&3\\
F571-8&\nodata&LSB&0.174&\nodata&14.0&$143.9\pm5.0$&48.0&\nodata&3\\
F583-1&Sm&LSB&0.063&0.054&14.0&$86.9\pm5.6$&32.0&\nodata&3\\
F583-4&Sc&LSB&0.093&0.072&6.7&$69.9\pm4.9$&49.0&\nodata&3\\
IC~342&Sc&HSB&6.607&5.126&19.3&$191.1\pm12.7$&3.9&0.000103&4\\
Milky Way&Sb&HSB&\nodata&\nodata&20.5&$202.6_{-20.3}^{+21.1}$&\nodata&\nodata&4\\
NGC~55&SBm&LSB&0.430&0.187&9.8&$86.5\pm5.5$&1.6&0.000430&2\\
NGC~224&Sb&HSB&1.738&4.443&31.3&$227.9\pm14.5$&0.7&-0.001001&4\\
NGC~247&SBc&LSB&0.350&0.193&11.0&$107.8\pm9.0$&2.8&0.000520&2\\
NGC~253&Sc&HSB&1.585&4.482&9.7&$211.5\pm12.5$&2.5&0.000811&4\\
NGC~300&Sc&HSB&0.300&0.300&12.1&$93.4\pm8.0$&2.2&0.000480&2\\
NGC~598&Sc&LSB&0.331&0.330&6.9&$108.1\pm4.0$&0.8&-0.000597&4\\
NGC~660&Sc&HSB&1.047&4.557&23.3&$139.6\pm14.5$&13.0&0.002835&4\\
NGC~801&Sc&HSB&7.400&23.640&59.0&$216.0\pm8.0$&80.0&0.019227&2\\
NGC~891&Sb&HSB&1.514&7.733&24.8&$164.6\pm11.5$&8.9&0.001761&4\\
NGC~1003&Scd&LSB&0.437&0.382&0.8&$233.6\tablenotemark{a}$&9.5&0.002088&4\\
NGC~1068&Sb&HSB&5.623&36.796&7.3&$280.8_{-86.5}^{+27.7}$&18.1&0.003793&4\\
NGC~1097&SBb&HSB&3.236&18.762&33.2&$249.1\pm20.0$&16.0&0.004240&4\\
NGC~1365&SBb&HSB&1.000&15.974&31.1&$205.7\pm16.7$&15.6&0.005457&4\\
NGC~1417&Sb&LSB&\nodata&15.200&9.6&$255.7\tablenotemark{a}$&54.1&0.013569&4\\
NGC~1560&Sd&LSB&0.035&0.058&8.3&$78.7\pm2.0$&3.0&-0.000120&1\\
NGC~1808&SBc&HSB&1.000&6.559&16.0&$123.8\pm12.7$&11.4&0.003319&4\\
NGC~2403&Sc&HSB&0.790&0.816&19.3&$134.0\pm4.0$&3.3&0.000437&1\\
NGC~2590&Sb&HSB&\nodata&17.278&15.8&$275.2\tablenotemark{a}$&64.5&0.016665&4\\
NGC~2708&Sb&HSB&0.324&3.952&4.6&$225.7\tablenotemark{a}$&24.6&0.006698&4\\
NGC~2841&Sb&HSB&2.050&7.797&42.7&$294.0\pm5.0$&9.5&0.002128&1\\
NGC~2903&Sc&HSB&1.530&3.654&24.0&$180.0\pm8.0$&6.4&0.001855&1\\
NGC~2998&SBc&HSB&9.000&11.288&46.4&$198.0\pm3.0$&67.0&0.015958&2\\
NGC~3031&Sb&HSB&0.324&7.167&21.1&$167.4\pm9.0$&3.3&-0.000113&4\\
NGC~3034&I&HSB&0.912&3.327&3.7&$56.5\pm7.8$&3.3&0.000677&4\\
NGC~3079&Sc&HSB&3.981&7.032&21.3&$167.9\pm9.6$&15.6&0.003723&4\\
NGC~3109&SBm&LSB&0.081&0.011&6.7&$67.3\pm3.0$&1.6&0.001344&1\\
NGC~3198&Sc&HSB&0.900&1.570&29.7&$149.0\pm3.0$&9.4&0.002212&1\\
NGC~3379&E&HSB&0.891&8.712&12.0&$192.5\pm11.0$&11.0&0.003039&6\\
NGC~3495&Sd&LSB&0.490&1.018&4.8&$138.5\tablenotemark{a}$&12.8&0.003789&4\\
NGC~3521&SBb&HSB&1.380&14.648&23.6&$156.8\pm13.6$&11.4&0.002672&4\\
NGC~3628&Sb/I&HSB&1.514&3.871&14.2&$192.0\pm12.7$&6.7&0.002812&4\\
NGC~3672&Sc&LSB&4.571&9.267&11.8&$214.7\tablenotemark{a}$&28.4&0.006211&4\\
NGC~3726&SBc&HSB&2.650&6.216&33.6&$167.0\pm15.0$&18.6&0.002887&5\\
NGC~3769&SBb&HSB&0.680&1.678&38.5&$113.0\pm11.0$&18.6&0.002459&5\\

NGC~3877&Sc&HSB&1.940&6.396&11.7&$169.0\pm10.0$&18.6&0.002987&5\\
NGC~3893&Sc&HSB&2.140&5.598&21.1&$148.0_{-17}^{+21}$&18.6&0.003226&5\\
NGC~3917&Scd&LSB&1.120&2.289&15.3&$137.0\pm8.0$&18.6&0.003218&5\\
NGC~3949&Sbc&HSB&1.650&2.901&8.8&$169.0_{-44}^{+7}$&18.6&0.002669&5\\
NGC~3953&SBbc&HSB&2.910&12.183&16.2&$215.0\pm10.0$&18.6&0.003510&5\\
NGC~3972&Sbc&HSB&0.680&1.124&9.0&$134.0\pm5.0$&18.6&0.002843&5\\
NGC~3992&SBbc&HSB&3.100&13.482&36.0&$237.0_{-10}^{+7}$&18.6&0.003496&5\\
NGC~4010&SBd&LSB&0.630&1.169&10.8&$122.0_{-6}^{+5}$&18.6&0.003008&5\\
NGC~4013&Sb&HSB&1.450&7.090&32.2&$170.0\pm10.0$&18.6&0.002773&5\\
NGC~4051&SBbc&HSB&2.580&6.856&12.6&$153.0\pm10.0$&18.6&0.002336&5\\
NGC~4062&Sc&HSB&0.316&1.131&3.8&$156.3\tablenotemark{a}$&9.7&0.002565&4\\
NGC~4085&Sc&HSB&0.810&1.797&6.4&$136.0\pm7.0$&18.6&0.002487&5\\
NGC~4088&SBc&HSB&2.830&8.176&22.1&$174.0\pm8.0$&18.6&0.002524&5\\
NGC~4096&Sc&HSB&0.891&2.610&1.6&$125.3\tablenotemark{a}$&12.2&0.001888&4\\
NGC~4100&Sbc&HSB&1.770&4.909&23.5&$159.0_{-8}^{+10}$&18.6&0.003584&5\\
NGC~4138&Sa&HSB&0.820&4.203&21.7&$150.0\pm21.0$&18.6&0.002962&5\\
NGC~4157&Sb&HSB&2.000&9.098&30.8&$185.0\pm14.0$&18.6&0.002583&5\\
NGC~4183&Scd&LSB&0.900&0.924&21.7&$113.0_{-10}^{+13}$&18.6&0.003102&5\\
NGC~4217&Sb&HSB&1.900&7.442&17.3&$178.0\pm12.0$&18.6&0.003426&5\\
NGC~4258&SBc&HSB&2.692&6.579&29.2&$193.1\pm10.6$&6.6&0.001494&4\\
NGC~4303&Sc&HSB&3.020&2.793&12.8&$159.6\pm22.5$&8.1&0.005224&4\\
NGC~4321&Sc&HSB&4.365&12.102&25.6&$236.1\pm44.0$&15.0&0.005240&4\\
NGC~4389&SBbc&HSB&0.610&1.782&5.5&$110.0\pm8.0$&18.6&0.002396&5\\
NGC~4448&SBab&HSB&0.363&1.634&2.3&$192.0\tablenotemark{a}$&9.7&0.002205&4\\
NGC~4527&Sb&HSB&1.202&18.974&12.8&$190.0\tablenotemark{a}$&22.0&0.005791&4\\
NGC~4565&Sb&HSB&2.138&9.099&34.1&$217.2\pm14.4$&10.2&0.004103&4\\
NGC~4569&Sab&HSB&5.012&3.621&7.6&$260.9\pm12.5$&8.2&-0.000784&4\\
NGC~4631&Sc/I&HSB&1.738&1.111&15.4&$143.5\pm8.5$&4.3&0.002021&4\\
NGC~4736&Sab&HSB&0.871&5.460&10.4&$125.8\pm7.9$&5.1&0.001027&4\\
NGC~4945&Sc/I&HSB&2.818&16.740&20.0&$169.6\pm9.0$&6.7&0.001878&4\\
NGC~5033&Sc&HSB&1.900&5.420&33.8&$200.0\pm5.0$&11.9&0.002919&2\\
NGC~5055&SBc&HSB&1.778&8.467&39.4&$179.0\pm10.0$&8.0&0.001681&4\\
NGC~5194&Sc&HSB&3.090&13.515&13.5&$130.4\pm12.3$&9.6&0.001544&4\\
NGC~5236&SBc&HSB&2.089&26.057&39.3&$144.4\pm15.2$&8.9&0.001711&4\\
NGC~5457&Sc&HSB&2.344&7.012&13.5&$195.0\pm10.2$&7.2&0.000804&4\\
NGC~5533&Sab&HSB&5.600&20.604&72.2&$227.0\pm5.0$&54.0&0.012896&2\\
NGC~5585&SBcd&HSB&0.240&0.212&11.9&$89.4\pm2.0$&7.6&0.001017&2\\
NGC~5907&Sc&HSB&2.570&6.181&28.6&$218.5\pm10.8$&11.6&0.002225&4\\
NGC~6503&Sc&HSB&0.480&0.985&22.1&$115.0\pm10.0$&5.9&0.000200&1\\
NGC~6674&SBb&HSB&6.800&13.757&69.2&$242.0\pm4.0$&49.0&0.011438&2\\
NGC~6946&SABcd&HSB&5.300&16.792&29.4&$159.0\pm5.0$&10.1&0.000160&2\\
NGC~6951&Sbc&HSB&3.020&9.741&9.8&$239.7\pm23.1$&18.0&0.004750&4\\
NGC~7331&Sb&HSB&5.400&19.969&36.0&$238.0\pm7.0$&14.9&0.002722&1\\
UGC~2259&SBcd&HSB&0.100&0.008&7.6&$90.0\pm3.0$&9.8&0.001945&1\\
UGC~2885&SBc&HSB&21.000&35.808&72.0&$298.0\pm5.0$&79.0&0.019353&2\\
UGC~3691&Scd&HSB&1.698&1.573&7.2&$144.1\tablenotemark{a}$&30.0&0.007348&4\\
UGC~6399&Sm&HSB&0.200&\nodata&8.1&$88.0\pm5.0$&18.6&0.002640&5\\
UGC~6446&Sd&LSB&0.250&\nodata&15.9&$80.0\pm11.0$&18.6&0.002149&5\\
UGC~6614&\nodata&LSB&\nodata&7.157&61.9&$203.9\pm16.0$&85.0&0.021185&3\\
UGC~6667&Scd&LSB&0.260&0.173&8.1&$86.0\pm5.0$&18.6&0.003246&5\\
UGC~6818&Sd&LSB&0.180&\nodata&7.2&$74.0_{-5}^{+7}$&18.6&0.002696&5\\
UGC~6917&SBd&LSB&0.380&0.260&10.8&$111.0_{-7}^{+5}$&18.6&0.003038&5\\
UGC~6923&Sdm&LSB&0.220&0.237&5.3&$81.0\pm5.0$&18.6&0.003556&5\\
UGC~6930&Sd&LSB&0.500&0.275&17.1&$108.0\pm7.0$&18.6&0.002592&5\\
UGC~6973&Sab&HSB&0.620&4.513&8.1&$180.0_{-10}^{+5}$&18.6&0.002337&5\\
UGC~6983&SBcd&LSB&0.340&0.160&16.2&$109.0\pm12.0$&18.6&0.003609&5\\
UGC~7089&Sdm&LSB&0.440&\nodata&9.4&$79.0\pm7.0$&18.6&0.002568&5\\
\enddata
\tablecomments{Relevant galaxy properties of the {\em complete sample}:  Column (1) is
the NGC/UGC galaxy number.  Column (2) is the galaxy morphological type.  Column (3) is the surface brightness --
denoted LSB if the galaxy central surface brightness is low ($\mu_{0} \gtrsim 23\ \rm{mag/arcsec}^{2}$) or HSB
if the galaxy central surface brightness is otherwise high.  Column (4) is $B$-band luminosity data taken from the
original references except for~\citet{sof96} and~\citet{rom03}
which are taken from~\citet{tul88}; and Column (5) is the $K$-band luminosity data converted from the 2MASS $K$-band apparent
magnitude via equation
(\ref{luminosity}) except for the Schombert F-type galaxies, which are taken from the original reference.   Column (6)
is the outermost observed radial position in the rotation velocity data; and Column (7) is the
observed velocity at the outermost observed radial position.  Column (8) is the distance to the galaxy adopted from
the original references; and Column (9) is the observed redshift taken from the NASA/IPAC Extragalactic Database.
Column (10) provides the primary original references.}
\tablenotetext{a}{Error bars were not part of the available high resolution rotation curve data for this galaxy.}
\tablerefs{ (1) Begeman, Broeils \& Sanders 1991; (2) Sanders 1996; (3) de Blok \& McGaugh 1998;  (4)
Verheijen \& Sancisi 2001; (5) Sofue 1996; (6) Romanowsky 2003.}
\end{deluxetable}
\clearpage

\begin{deluxetable}{crrrrrrrr}
\tabletypesize{\footnotesize}
\rotate
\tablecaption{\sc UMa Photometric Rotation Curve Fit Results \label{UMaPhotometricRotationCurveResults}}
\tablewidth{0pt}
\tablehead{&&&\colhead{\vector(-1,0){25}}&\colhead{MSTG}&\colhead{\vector(1,0){25}}&\colhead{\vector(-1,0){25}}&\colhead{MOND}&\colhead{\vector(1,0){25}}\\
\colhead{Galaxy}&\colhead{$M_{gas}$}&\colhead{$z_{0}$}&\colhead{$M_{disk}$}&\colhead{$M$}&\colhead{$v_{0}$}&\colhead{$M_{disk}$}&\colhead{$M$}&\colhead{$v_{c}$}\\
&\colhead{\scriptsize $(10^{\tiny 10} M_{\sun})$}&\colhead{\scriptsize (kpc)}&\colhead{\scriptsize $(10^{\tiny 10} M_{\sun})$}&\colhead{\scriptsize $(10^{\tiny 10} M_{\sun})$}&\colhead{\scriptsize (km s$^{\tiny -1}$)}&\colhead{\scriptsize $(10^{\tiny 10} M_{\sun})$}&\colhead{\scriptsize $(10^{\tiny 10} M_{\sun})$}&\colhead{\scriptsize (km s$^{\tiny -1}$)}\\
\colhead{\scriptsize (1)}&\colhead{\scriptsize (2)}&\colhead{\scriptsize (3)}&\colhead{\scriptsize (4)}&\colhead{\scriptsize (5)}&\colhead{\scriptsize (6)}&\colhead{\scriptsize (7)}&\colhead{\scriptsize (8)}&\colhead{\scriptsize (9)}}
\tablecolumns{9}
\startdata
\cutinhead{Dwarf (LSB \& HSB) Galaxies}
NGC~3877&0.21&0.56&$2.86\pm0.80$&$3.07\pm0.90$&$180.9\pm25.4$&$3.71\pm1.00$&$3.92\pm1.10$&$158.3\pm11.0$\\
NGC~3949&0.49&0.35&$1.70\pm0.42$&$2.19\pm0.50$&$158.6\pm19.7$&$2.22\pm0.80$&$2.70\pm0.75$&$144.2\pm10.0$\\
NGC~3972&0.18&0.39&$0.97\pm0.15$&$1.15\pm0.20$&$124.6\pm9.9$&$1.07\pm0.40$&$1.25\pm0.40$&$118.9\pm9.4$\\
NGC~4085&0.16&0.31&$0.87\pm0.26$&$1.03\pm0.30$&$119.7\pm17.8$&$0.94\pm0.30$&$1.10\pm0.35$&$115.1\pm9.2$\\
NGC~4389&0.08&0.29&$0.35\pm0.15$&$0.43\pm0.20$&$88.8\pm19.2$&$0.28\pm0.30$&$0.36\pm0.25$&$87.0\pm15.3$\\
UGC~6399&0.11&0.48&$0.29\pm0.08$&$0.41\pm0.10$&$87.0\pm11.9$&$0.20\pm0.10$&$0.32\pm0.06$&$84.3\pm4.2$\\

UGC~6667&0.12&0.58&$0.40\pm0.09$&$0.52\pm0.10$&$94.5\pm10.0$&$0.26\pm0.20$&$0.38\pm0.22$&$88.5\pm12.7$\\
UGC~6818&0.15&0.36&$0.10\pm0.04$&$0.25\pm0.10$&$73.9\pm14.7$&$0.04\pm0.20$&$0.19\pm0.18$&$74.3\pm17.1$\\
UGC~6923&0.12&0.26&$0.17\pm0.04$&$0.29\pm0.10$&$77.8\pm8.0$&$0.11\pm0.10$&$0.23\pm0.13$&$77.7\pm10.9$\\
UGC~7089&0.19&0.62&$0.12\pm0.09$&$0.31\pm0.20$&$79.2\pm29.4$&$0.03\pm0.40$&$0.22\pm0.36$&$76.9\pm31.4$\\
UGC~6917&0.29&0.58&$0.58\pm0.13$&$0.87\pm0.20$&$112.7\pm13.0$&$0.52\pm0.30$&$0.81\pm0.33$&$106.6\pm11.0$\\
\cutinhead{\sc LSB Galaxies}
NGC~3917&0.27&0.62&$1.56\pm0.32$&$1.83\pm0.40$&$127.7\pm13.3$&$1.37\pm0.70$&$1.64\pm0.66$&$127.3\pm12.7$\\
UGC~6446&0.44&0.36&$0.22\pm0.17$&$0.66\pm0.50$&$91.9\pm36.3$&$0.08\pm0.50$&$0.52\pm0.52$&$95.5\pm23.9$\\
UGC~6983&0.42&0.53&$0.70\pm0.20$&$1.12\pm0.30$&$108.7\pm15.7$&$0.47\pm0.40$&$0.89\pm0.45$&$109.3\pm13.7$\\
NGC~4010&0.42&0.69&$1.40\pm0.44$&$1.82\pm0.60$&$127.4\pm20.2$&$1.06\pm0.40$&$1.48\pm0.38$&$123.9\pm8.0$\\
NGC~4183&0.53&0.64&$0.82\pm0.45$&$1.35\pm0.70$&$115.6\pm31.7$&$0.53\pm0.70$&$1.06\pm0.65$&$114.2\pm17.5$\\
\cutinhead{\sc HSB Galaxies}
NGC~3726&0.98&0.68&$2.62\pm0.87$&$3.59\pm1.20$&$161.3\pm26.9$&$2.59\pm1.40$&$3.56\pm1.40$&$154.5\pm15.2$\\
NGC~3769&0.68&0.36&$1.02\pm0.13$&$1.70\pm0.20$&$124.7\pm8.0$&$0.78\pm0.70$&$1.46\pm0.69$&$123.5\pm14.7$\\
NGC~3893&0.76&0.49&$3.67\pm0.66$&$4.43\pm0.80$&$173.9\pm15.7$&$4.24\pm1.50$&$5.00\pm1.52$&$168.2\pm12.8$\\
NGC~3953&0.43&0.77&$7.47\pm0.71$&$7.90\pm0.70$&$215.6\pm10.2$&$9.38\pm2.80$&$9.81\pm2.81$&$199.0\pm14.3$\\
NGC~3992&0.81&0.83&$12.57\pm1.58$&$13.39\pm1.70$&$264.6\pm16.6$&$16.15\pm5.00$&$16.96\pm4.96$&$228.2\pm16.7$\\
NGC~4013&0.45&0.41&$4.19\pm0.80$&$4.64\pm0.90$&$176.9\pm16.8$&$4.58\pm0.70$&$5.03\pm0.72$&$168.4\pm6.0$\\
NGC~4051&0.39&0.54&$2.57\pm0.29$&$2.96\pm0.30$&$150.7\pm8.4$&$2.65\pm1.10$&$3.04\pm1.06$&$148.5\pm12.9$\\
NGC~4088&1.12&0.67&$3.24\pm0.86$&$4.36\pm1.20$&$172.9\pm22.9$&$3.56\pm1.30$&$4.68\pm1.29$&$165.4\pm11.4$\\
NGC~4100&0.45&0.51&$3.96\pm0.82$&$4.41\pm0.90$&$173.7\pm18.1$&$4.43\pm1.20$&$4.88\pm1.25$&$167.1\pm10.7$\\
NGC~4138&0.21&0.28&$2.78\pm0.42$&$2.99\pm0.50$&$151.1\pm11.5$&$2.93\pm0.70$&$3.14\pm0.70$&$149.7\pm8.4$\\
NGC~4157&1.17&0.52&$4.38\pm0.71$&$5.55\pm0.90$&$188.8\pm15.3$&$4.96\pm0.90$&$6.13\pm0.88$&$176.9\pm6.3$\\
NGC~4217&0.37&0.58&$4.12\pm0.72$&$4.49\pm0.80$&$174.8\pm15.3$&$4.66\pm1.10$&$5.03\pm1.11$&$168.4\pm9.3$\\
UGC~6973&0.25&0.19&$2.38\pm0.72$&$2.63\pm0.80$&$144.6\pm21.7$&$2.66\pm0.20$&$2.91\pm0.16$&$146.9\pm2.0$\\
\enddata
\tablecomments{Best fit results of the UMa cluster of galaxies according to both MSTG and MOND via the single parameter
$(M/L)_{\rm{stars}}$ based on the $K$-band photometric data of the gaseous (HI plus He) and luminous stellar disks,
corresponding to the fits (labeled UMa) of Fig.\,\ref{photometricRotationCurves} and Fig.\,\ref{NGC4010}.  Column (1) is
the NGC/UGC galaxy number.  Column (2) is the mass of the infinitely thin gaseous disk (HI plus He).  Column (3) is the
$K$-band vertical scale height of the luminous stellar disk.  The MSTG best fit results are presented in Columns (4) -
(6), where Column (4) is the best fit mass of the luminous stellar disk; Column (5) is the MSTG predicted total mass of
the galaxy and is the sum of Column (2) and (4); and Column (6) is the predicted MSTG flat rotation velocity, $v_{0}$,
of equation (\ref{v0}).  The MOND best results
are presented in Columns (7) - (9), where Column (7) is the best fit mass of the luminous stellar disk; Column (8) is
the MOND predicted mass of the galaxy and is the sum of Column (2) and (7); and Column (9) is the MOND asymptotic
velocity, $v_{c}$ of equation (\ref{Milgromv}).}
\end{deluxetable}
\clearpage

\begin{deluxetable}{crrrrrr}
\tabletypesize{\footnotesize}
\tablecaption{\sc Parametric Rotation Curve Fit Results \label{parametricRotationCurveResults}}
\tablewidth{0pt}
\tablehead{&\colhead{\vector(-1,0){25}}&\colhead{MSTG}&\colhead{\vector(1,0){25}}&\colhead{\vector(-1,0){25}}&\colhead{MOND}&\colhead{\vector(1,0){25}}\\
\colhead{Galaxy}&\colhead{$M$}&\colhead{$r_{c}$}&\colhead{$v_{0}$}&\colhead{$M$}&\colhead{$r_{c}$}&\colhead{$v_{c}$}\\
&\colhead{\scriptsize $(10^{\tiny 10} M_{\sun})$}&\colhead{\scriptsize (kpc)}&\colhead{\scriptsize (km s$^{\tiny -1}$)}&\colhead{\scriptsize $(10^{\tiny 10} M_{\sun})$}&\colhead{\scriptsize (kpc)}&\colhead{\scriptsize (km s$^{\tiny -1}$)}\\
\colhead{\scriptsize (1)}&\colhead{\scriptsize (2)}&\colhead{\scriptsize (3)}&\colhead{\scriptsize (4)}&\colhead{\scriptsize (5)}&\colhead{\scriptsize (6)}&\colhead{\scriptsize (7)}}
\tablecolumns{7}
\startdata
\cutinhead{Dwarf (LSB \& HSB) Galaxies}
DDO~154\tablenotemark{a}&$0.13\pm0.02$&$0.53\pm0.07$&$48.9\pm2.4$&$0.07\pm0.01$&$0.95\pm0.12$&$57.9\pm2.3$\\
DDO~168&$0.42\pm0.09$&$0.66\pm0.08$&$67.1\pm4.7$&$0.23\pm0.06$&$0.89\pm0.14$&$77.9\pm5.5$\\
DDO~170&$0.40\pm0.04$&$0.82\pm0.07$&$61.9\pm2.3$&$0.26\pm0.03$&$1.38\pm0.10$&$80.7\pm1.9$\\
F583-4&$0.38\pm0.04$&$0.57\pm0.05$&$67.2\pm2.4$&$0.23\pm0.03$&$0.77\pm0.1$&$77.5\pm2.7$\\
NGC~55&$1.17\pm0.07$&$0.99\pm0.05$&$84.4\pm2$&$0.91\pm0.07$&$1.39\pm0.08$&$109.7\pm2.1$\\
NGC~1560&$0.79\pm0.05$&$0.93\pm0.04$&$74.9\pm1.7$&$0.59\pm0.05$&$1.43\pm0.08$&$98.5\pm2.2$\\
NGC~2708&$9.43\pm1.1$&$0.66\pm0.05$&$218.7\pm10.8$&$12.97\pm1.59$&$0.79\pm0.06$&$213.4\pm6.5$\\
NGC~3109&$0.78\pm0.04$&$1.15\pm0.04$&$68.6\pm1.3$&$0.62\pm0.04$&$2\pm0.06$&$99.8\pm1.5$\\
NGC~3877&$8.65\pm0.53$&$1.31\pm0.06$&$164.8\pm4.3$&$10.34\pm0.87$&$1.52\pm0.1$&$201.7\pm4.2$\\
NGC~3949&$6.51\pm0.3$&$0.99\pm0.03$&$164.5\pm3.2$&$7.77\pm0.41$&$1.12\pm0.04$&$187.7\pm2.5$\\
NGC~3972&$4.09\pm0.23$&$1.18\pm0.05$&$126.8\pm2.9$&$4.4\pm0.25$&$1.48\pm0.06$&$162.9\pm2.3$\\
NGC~4062&$2.98\pm0.17$&$0.43\pm0.02$&$149.4\pm3.4$&$4.07\pm0.27$&$0.53\pm0.03$&$159.7\pm2.7$\\
NGC~4085&$5.11\pm0.54$&$1.12\pm0.07$&$142\pm6.1$&$5.84\pm0.89$&$1.34\pm0.13$&$174.8\pm6.7$\\
NGC~4096&$1.07\pm0.07$&$0.24\pm0.01$&$110.1\pm2.8$&$1.21\pm0.09$&$0.26\pm0.01$&$117.9\pm2.1$\\
NGC~4389&$4.4\pm1.02$&$1.56\pm0.18$&$113.9\pm10.6$&$6.45\pm1.7$&$2.45\pm0.28$&$179.2\pm11.8$\\
NGC~4569&$6.23\pm0.51$&$0.39\pm0.03$&$205\pm7$&$11.72\pm1.18$&$0.67\pm0.06$&$208.1\pm5.2$\\
NGC~5585&$1.17\pm0.07$&$0.94\pm0.04$&$85.7\pm1.8$&$0.9\pm0.06$&$1.3\pm0.08$&$109.6\pm1.9$\\
UGC~2259&$0.77\pm0.02$&$0.48\pm0.01$&$88.8\pm1$&$0.55\pm0.02$&$0.49\pm0.02$&$96.7\pm0.9$\\
UGC~3691&$2.83\pm0.14$&$0.86\pm0.03$&$123.5\pm2.3$&$2.96\pm0.17$&$1.03\pm0.04$&$147.5\pm2.1$\\
UGC~6399&$1.34\pm0.08$&$1.05\pm0.04$&$86.7\pm2$&$1.08\pm0.09$&$1.48\pm0.08$&$114.8\pm2.5$\\
UGC~6446&$0.83\pm0.04$&$0.73\pm0.04$&$85.1\pm1.4$&$0.43\pm0.02$&$0.9\pm0.06$&$91.2\pm1.3$\\
UGC~6818&$1.31\pm0.53$&$1.5\pm0.32$&$73.1\pm10.8$&$1.37\pm0.75$&$2.78\pm0.68$&$121.7\pm16.7$\\
UGC~6917&$2.06\pm0.11$&$1.04\pm0.05$&$102.1\pm2.2$&$1.75\pm0.11$&$1.3\pm0.07$&$129.3\pm2$\\
UGC~6923&$0.96\pm0.17$&$0.74\pm0.1$&$86.5\pm5.6$&$0.69\pm0.16$&$0.91\pm0.15$&$102.5\pm5.9$\\
UGC~7089&$0.86\pm0.08$&$1.15\pm0.07$&$71.1\pm2.3$&$0.71\pm0.08$&$1.98\pm0.14$&$103.3\pm2.9$\\
\cutinhead{\sc LSB Galaxies}
F563-1&$2.26\pm0.16$&$1.06\pm0.07$&$110.4\pm2.7$&$1.48\pm0.15$&$1.24\pm0.14$&$124.1\pm3.1$\\
F568-3&$3.08\pm0.41$&$1.58\pm0.13$&$110.9\pm5.2$&$2.09\pm0.42$&$2.13\pm0.25$&$135.2\pm6.8$\\
F571-8&$5.46\pm0.84$&$1.40\pm0.14$&$141.2\pm8.0$&$4.86\pm0.73$&$1.69\pm0.17$&$167.0\pm6.3$\\
F583-1&$1.56\pm0.12$&$1.28\pm0.06$&$93.2\pm2.3$&$0.90\pm0.06$&$1.81\pm0.07$&$109.4\pm1.9$\\
NGC~247&$2.27\pm0.17$&$1.11\pm0.06$&$109.4\pm2.8$&$1.46\pm0.14$&$1.35\pm0.10$&$123.7\pm2.9$\\
NGC~598&$1.78\pm0.04$&$0.64\pm0.01$&$110.9\pm0.8$&$1.15\pm0.02$&$0.60\pm0.01$&$116.4\pm0.5$\\
NGC~1003&$1.64\pm0.03$&$0.08\pm0.00$&$121.5\pm0.8$&$1.66\pm0.03$&$0.08\pm0.00$&$127.6\pm0.6$\\
NGC~1417&$16.60\pm0.49$&$0.92\pm0.02$&$238.2\pm2.8$&$23.30\pm0.84$&$1.15\pm0.03$&$247.0\pm2.2$\\
NGC~3495&$4.16\pm0.27$&$0.87\pm0.04$&$142.1\pm3.3$&$4.01\pm0.28$&$0.95\pm0.04$&$159.2\pm2.7$\\
NGC~3672&$14.86\pm0.20$&$1.21\pm0.01$&$215.2\pm1.2$&$17.02\pm0.25$&$1.32\pm0.01$&$228.4\pm0.8$\\
NGC~3917&$6.25\pm0.45$&$1.60\pm0.09$&$142.8\pm3.8$&$5.18\pm0.52$&$1.89\pm0.15$&$169.7\pm4.2$\\
NGC~4010&$5.56\pm0.88$&$1.62\pm0.17$&$136.2\pm7.9$&$4.66\pm0.87$&$2.03\pm0.24$&$165.2\pm7.7$\\
NGC~4183&$2.04\pm0.11$&$0.85\pm0.05$&$111.3\pm2.0$&$1.35\pm0.07$&$0.91\pm0.06$&$121.2\pm1.5$\\
UGC~6446&$0.83\pm0.04$&$0.73\pm0.04$&$85.1\pm1.4$&$0.43\pm0.02$&$0.90\pm0.06$&$91.2\pm1.3$\\
UGC~6614&$11.36\pm1.79$&$1.24\pm0.22$&$192.3\pm11.9$&$9.90\pm1.18$&$1.14\pm0.16$&$199.5\pm5.9$\\
UGC~6930&$2.17\pm0.13$&$1.03\pm0.06$&$109.5\pm2.2$&$1.40\pm0.07$&$1.19\pm0.07$&$122.2\pm1.6$\\
UGC~6983&$2.12\pm0.16$&$0.90\pm0.07$&$111.5\pm2.8$&$1.34\pm0.10$&$0.95\pm0.09$&$121.0\pm2.3$\\
\cutinhead{\sc HSB Galaxies}
IC~342&$7.95\pm0.14$&$1.36\pm0.03$&$188.3\pm1.2$&$8.68\pm0.16$&$1.44\pm0.03$&$193.0\pm0.9$\\
Milky Way&$9.12\pm0.28$&$1.04\pm0.05$&$204.8\pm2.4$&$10.60\pm0.37$&$1.18\pm0.05$&$202.9\pm1.8$\\
NGC~224&$20.19\pm0.30$&$1.84\pm0.04$&$259.6\pm1.6$&$25.54\pm0.45$&$2.17\pm0.05$&$252.8\pm1.1$\\
NGC~253&$6.94\pm0.25$&$0.86\pm0.04$&$188.0\pm2.5$&$7.88\pm0.31$&$0.95\pm0.04$&$188.4\pm1.9$\\
NGC~300&$2.03\pm0.17$&$2.70\pm0.19$&$101.7\pm2.9$&$1.03\pm0.09$&$2.93\pm0.25$&$113.3\pm2.5$\\
NGC~660&$3.20\pm0.06$&$0.54\pm0.02$&$146.6\pm0.9$&$2.99\pm0.06$&$0.48\pm0.02$&$147.8\pm0.7$\\
NGC~801&$20.07\pm2.09$&$2.65\pm0.24$&$240.3\pm10.2$&$17.90\pm2.05$&$2.44\pm0.23$&$231.3\pm6.6$\\
NGC~891&$7.47\pm0.17$&$0.78\pm0.03$&$194.9\pm1.7$&$8.49\pm0.24$&$0.87\pm0.04$&$192.0\pm1.3$\\
NGC~1068&$9.42\pm0.54$&$1.11\pm0.07$&$205.9\pm4.5$&$12.96\pm0.87$&$1.46\pm0.09$&$213.4\pm3.6$\\
NGC~1097&$22.68\pm0.31$&$1.19\pm0.03$&$290.1\pm1.6$&$29.64\pm0.51$&$1.49\pm0.04$&$262.4\pm1.1$\\
NGC~1365&$14.96\pm0.25$&$1.29\pm0.03$&$242.6\pm1.6$&$18.51\pm0.36$&$1.52\pm0.04$&$233.2\pm1.1$\\
NGC~1808&$4.10\pm0.10$&$0.51\pm0.02$&$160.6\pm1.4$&$4.32\pm0.12$&$0.53\pm0.02$&$162.1\pm1.1$\\
NGC~2403&$3.80\pm0.13$&$2.09\pm0.07$&$133.7\pm1.6$&$2.77\pm0.09$&$1.95\pm0.07$&$145.1\pm1.1$\\
NGC~2590&$14.05\pm0.48$&$1.10\pm0.05$&$241.0\pm3.3$&$17.54\pm0.71$&$1.33\pm0.07$&$230.1\pm2.3$\\
NGC~2841&$33.04\pm1.31$&$2.19\pm0.14$&$308.3\pm5.2$&$41.07\pm0.80$&$2.46\pm0.07$&$284.7\pm1.4$\\
NGC~2903&$9.66\pm0.61$&$1.72\pm0.11$&$195.9\pm4.8$&$10.61\pm0.75$&$1.82\pm0.12$&$202.9\pm3.6$\\
NGC~2998&$15.13\pm1.20$&$2.52\pm0.19$&$216.7\pm6.8$&$14.05\pm1.25$&$2.39\pm0.19$&$217.7\pm4.8$\\
NGC~3031&$6.95\pm0.12$&$0.67\pm0.02$&$191.8\pm1.3$&$7.97\pm0.17$&$0.75\pm0.02$&$188.9\pm1.0$\\
NGC~3034&$0.52\pm0.03$&$0.08\pm0.01$&$85.0\pm1.6$&$0.59\pm0.05$&$0.09\pm0.01$&$98.4\pm2.0$\\
NGC~3079&$8.73\pm0.23$&$0.77\pm0.03$&$207.1\pm2.1$&$10.42\pm0.33$&$0.91\pm0.04$&$202.0\pm1.6$\\
NGC~3198&$5.55\pm0.28$&$2.18\pm0.12$&$152.1\pm2.8$&$4.49\pm0.26$&$2.05\pm0.15$&$163.7\pm2.4$\\
NGC~3379&$6.99\pm0.06$&$0.45\pm0.01$&$196.7\pm0.6$&$8.13\pm0.08$&$0.51\pm0.01$&$189.9\pm0.5$\\
NGC~3379&$6.99\pm0.06$&$0.45\pm0.01$&$196.7\pm0.6$&$8.13\pm0.08$&$0.51\pm0.01$&$189.9\pm0.5$\\
NGC~3521&$7.89\pm0.10$&$0.80\pm0.02$&$198.7\pm1.0$&$9.12\pm0.14$&$0.89\pm0.02$&$195.4\pm0.8$\\
NGC~3628&$9.13\pm0.31$&$1.17\pm0.05$&$202.3\pm2.6$&$10.67\pm0.41$&$1.32\pm0.06$&$203.2\pm2.0$\\
NGC~3726&$9.60\pm1.37$&$4.07\pm0.58$&$158.4\pm8.8$&$7.22\pm1.06$&$4.00\pm0.72$&$184.3\pm6.8$\\
NGC~3769&$2.59\pm0.24$&$1.66\pm0.20$&$121.7\pm3.8$&$1.61\pm0.16$&$1.34\pm0.24$&$126.6\pm3.2$\\
NGC~3893&$7.70\pm1.00$&$1.74\pm0.29$&$179.3\pm8.9$&$7.85\pm1.17$&$1.78\pm0.34$&$188.2\pm7.0$\\
NGC~3953&$20.47\pm1.65$&$3.46\pm0.28$&$225.5\pm7.4$&$24.30\pm1.94$&$3.92\pm0.32$&$249.7\pm5.0$\\
NGC~3992&$25.16\pm2.32$&$2.77\pm0.44$&$260.9\pm10.0$&$28.35\pm3.90$&$2.74\pm0.72$&$259.5\pm8.9$\\
NGC~4013&$6.01\pm0.35$&$0.70\pm0.19$&$181.1\pm3.9$&$5.52\pm0.26$&$0.16\pm0.17$&$172.4\pm2.1$\\
NGC~4051&$7.21\pm1.31$&$2.58\pm0.43$&$161.7\pm11.1$&$6.20\pm1.22$&$2.50\pm0.47$&$177.4\pm8.8$\\
NGC~4088&$9.74\pm1.52$&$3.15\pm0.51$&$172.4\pm10.4$&$8.87\pm1.46$&$3.23\pm0.60$&$194.0\pm8.0$\\
NGC~4100&$10.30\pm1.59$&$2.89\pm0.49$&$180.2\pm10.8$&$9.91\pm1.64$&$3.00\pm0.58$&$199.5\pm8.2$\\
NGC~4138&$4.31\pm0.90$&$0.68\pm0.39$&$160.7\pm12.1$&$4.25\pm1.03$&$0.62\pm0.45$&$161.4\pm9.8$\\
NGC~4157&$11.64\pm1.21$&$2.92\pm0.36$&$188.5\pm7.7$&$11.05\pm1.28$&$2.93\pm0.43$&$205.0\pm5.9$\\
NGC~4217&$12.92\pm1.54$&$3.31\pm0.36$&$189.7\pm8.9$&$13.01\pm1.66$&$3.49\pm0.42$&$213.6\pm6.8$\\
NGC~4258&$7.29\pm0.14$&$0.84\pm0.03$&$191.9\pm1.4$&$7.79\pm0.16$&$0.86\pm0.03$&$187.8\pm1.0$\\
NGC~4303&$3.08\pm0.08$&$0.59\pm0.02$&$143.8\pm1.4$&$2.92\pm0.08$&$0.55\pm0.02$&$146.9\pm1.0$\\
NGC~4321&$21.67\pm0.45$&$2.12\pm0.06$&$260.2\pm2.2$&$28.20\pm0.68$&$2.56\pm0.07$&$259.1\pm1.6$\\
NGC~4448&$1.98\pm0.08$&$0.27\pm0.01$&$127.8\pm1.7$&$2.10\pm0.09$&$0.28\pm0.01$&$135.4\pm1.4$\\
NGC~4527&$5.55\pm0.23$&$0.79\pm0.05$&$174.3\pm2.7$&$6.02\pm0.28$&$0.84\pm0.05$&$176.2\pm2.1$\\
NGC~4565&$18.11\pm0.21$&$1.72\pm0.03$&$251.2\pm1.2$&$21.74\pm0.32$&$1.93\pm0.04$&$242.8\pm0.9$\\
NGC~4631&$6.15\pm0.10$&$1.34\pm0.03$&$171.4\pm1.0$&$6.36\pm0.12$&$1.37\pm0.03$&$178.6\pm0.8$\\
NGC~4736&$3.15\pm0.08$&$0.47\pm0.02$&$146.8\pm1.3$&$3.21\pm0.09$&$0.47\pm0.02$&$150.5\pm1.1$\\
NGC~4945&$4.58\pm0.12$&$0.63\pm0.03$&$165.1\pm1.6$&$4.65\pm0.14$&$0.63\pm0.03$&$165.1\pm1.2$\\
NGC~5033&$9.90\pm0.51$&$1.10\pm0.08$&$210.2\pm4.2$&$10.80\pm0.70$&$1.15\pm0.10$&$203.9\pm3.3$\\
NGC~5055&$8.38\pm0.06$&$1.11\pm0.01$&$196.9\pm0.5$&$8.44\pm0.08$&$1.07\pm0.02$&$191.6\pm0.5$\\
NGC~5194&$7.29\pm0.23$&$0.61\pm0.03$&$196.6\pm2.3$&$8.72\pm0.30$&$0.71\pm0.03$&$193.2\pm1.7$\\
NGC~5236&$6.16\pm0.12$&$1.10\pm0.04$&$175.5\pm1.3$&$5.56\pm0.13$&$0.96\pm0.04$&$172.6\pm1.0$\\
NGC~5457&$10.20\pm0.27$&$1.39\pm0.04$&$206.5\pm2.1$&$12.03\pm0.36$&$1.57\pm0.05$&$209.4\pm1.6$\\
NGC~5533&$28.81\pm1.92$&$2.11\pm0.23$&$293.2\pm8.2$&$25.61\pm1.68$&$1.68\pm0.18$&$253.0\pm4.1$\\
NGC~5907&$4.59\pm0.26$&$0.40\pm0.05$&$169.3\pm3.5$&$23.16\pm0.46$&$2.35\pm0.06$&$246.7\pm1.2$\\
NGC~6503&$1.98\pm0.06$&$1.10\pm0.05$&$117.4\pm1.3$&$1.38\pm0.05$&$0.91\pm0.05$&$122.0\pm1.0$\\
NGC~6674&$32.48\pm2.38$&$3.27\pm0.33$&$277.7\pm8.6$&$28.04\pm2.43$&$2.70\pm0.33$&$258.7\pm5.6$\\
NGC~6946&$8.95\pm0.65$&$3.54\pm0.27$&$161.2\pm4.5$&$7.29\pm0.61$&$3.57\pm0.37$&$184.8\pm3.8$\\
NGC~6951&$6.22\pm0.22$&$0.58\pm0.03$&$185.8\pm2.5$&$6.92\pm0.28$&$0.63\pm0.03$&$182.4\pm1.9$\\
NGC~7331&$21.47\pm0.76$&$2.56\pm0.10$&$248.9\pm3.6$&$24.73\pm0.83$&$2.74\pm0.10$&$250.8\pm2.1$\\
UGC~6973&$6.41\pm0.45$&$1.43\pm0.12$&$172.5\pm4.5$&$6.63\pm0.46$&$1.46\pm0.12$&$180.4\pm3.1$\\
\enddata
\tablecomments{Best fit results according to both MSTG and MOND
via a parametric mass distribution (independent of luminosity
observations) --- corresponding to the galaxy rotation curves of
Figs.\,\ref{parametricRotationCurves}--\ref{NGC3379}. Column (1)
is the NGC/UGC galaxy number.  The MSTG best fit results are
presented in Columns (2) - (4), where Column (2) is the MSTG
predicted total mass of the galaxy, $M$; Column (3) is the MSTG
predicted core radius, $r_{c}$; and Column (4) is the predicted
MSTG flat rotation velocity, $v_{0}$, of equation (\ref{v0}).  The
MOND best results are presented in Columns (5) - (7), where Column
(5) is the MOND predicted mass of the galaxy, $M$; Column (6) is
the MOND predicted core radius, $r_{c}$; and Column (7) is the
MOND asymptotic velocity, $v_{c}$ of equation (\ref{Milgromv}).}
\tablenotetext{a}{A MOND best fit was not possible due to
${\langle M/L \rangle}_{stars} < 0$.  The MOND fit shown for
DDO~154 neglects $M_{disk}$, whereas all the data was used for the
MSTG result.}
\end{deluxetable}
\clearpage

\begin{deluxetable}{crrrr}
\tablecaption{\sc Observed \& Actual Tully-Fisher Relation Results \label{TullyFisherResults}}
\tablewidth{0pt}
\tablehead{& \colhead{$B$-band} & \colhead{$K$-band} & \colhead{MSTG} & \colhead{MOND}\\
&\colhead{\footnotesize (1)}&\colhead{\footnotesize (2)}&\colhead{\footnotesize (3)}&\colhead{\footnotesize (4)}}
\tablecolumns{5}
\startdata
\cutinhead{UMa Cluster of Galaxies --- Photometric Fits}
$a$ & $2.67\pm0.25$ & $4.73\pm0.44$ & $3.19\pm0.10$ & $4.00\pm0.10$ \\
$b$ & $-5.73\pm0.54$ & $-9.82\pm0.95$ & $-6.54\pm0.21$ & $-8.20\pm0.01$ \\
\cutinhead{{\em Complete Sample} of Galaxies --- Parametric Fits}
$a$ & $2.89\pm0.22$ & $2.85\pm0.33$ & $2.68\pm0.07$ & $4.00\pm0.00$ \\
$b$ & $-6.30\pm0.47$ & $-5.73\pm0.74$ & $-5.18\pm0.16$ & $-8.21\pm0.00$
\enddata

\tablecomments{The
$B$-band luminosity data are taken from the original references except for~\citet{sof96} and~\citet{rom03}
which are taken from~\citet{tul88}.  The $K$-band luminosity data are taken from the 2MASS except for the Schombert
F-type galaxies, which are taken from the original reference.  The values of $a$ and $b$ are
determined using a nonlinear least-squares fitting routine including estimated errors.  Columns (1) and (2) are the best fit solutions to
the observed Tully-Fisher relation, $\log(L) = a \log(v_{\rm out}) +b$ in the $B$- and $K$-bands, respectively.  Columns (3)
and (4) are the best fit solutions to the
actual Tully-Fisher relation, $\log(M) = a \log(v) +b$ for MSTG and MOND, where v is determined by
equations (\ref{TullyFisherMSTG}) and (\ref{TullyFisherMOND}), respectively.  The observed \& actual Tully-Fisher
relation presented here are shown graphically in Figs.\,\ref{observedTullyFisher} \& \ref{actualTullyFisher},
respectively.}
\end{deluxetable}
\clearpage

\end{document}